\documentclass{aastex62}

\usepackage{amsmath}
\usepackage{mathrsfs}
\usepackage[caption=false]{subfig}

\received{\today}
\revised{\today}
\accepted{\today}

\shorttitle{GRBs with a X-ray Plateau}
\shortauthors{Tang et al.}

\begin{document}

\title{Statistical Study of Gamma-Ray Bursts with a Plateau Phase in the X-ray Afterglow}

\correspondingauthor{Yong-Feng Huang}
\email{hyf@nju.edu.cn, gengjinjun@nju.edu.cn}

\author{Chen-Han Tang}
\affiliation{School of Astronomy and Space Science, Nanjing University \\
	Nanjing 210023, P. R. China}

\author{Yong-Feng Huang}
\affiliation{School of Astronomy and Space Science, Nanjing University \\
	Nanjing 210023, P. R. China}
\affiliation{Key Laboratory of Modern Astronomy and Astrophysics (Nanjing University) \\
	Ministry of Education, P. R. China}

\author{Jin-Jun Geng}
\affiliation{School of Astronomy and Space Science, Nanjing University \\
	Nanjing 210023, P. R. China}

\author{Zhi-Bin Zhang}
\affiliation{College of Physics and Engineering, Qufu Normal University \\ Qufu 273165, P. R. China}

\begin{abstract}
A plateau phase in the X-ray afterglow is observed in a significant fraction of gamma-ray bursts (GRBs). Previously, it has been found that there exists a correlation among three key parameters concerning the plateau phase, i.e., the end time of the plateau phase in the GRB rest frame ($T_{a}$), the corresponding X-ray luminosity at the end time ($L_{X}$) and the isotropic energy of the prompt GRB ($E_{\gamma,\rm{iso}}$). In this study, we systematically search through all the \emph{Swift} GRBs with a plateau phase that occurred between 2005 May and 2018 August. We collect 174 GRBs, with redshifts available for all of them.
For the whole sample, the correlation between $L_{X}$, $T_{a}$ and $E_{\gamma,\rm{iso}}$ is confirmed, with the best fit relation being
$L_{X}\propto T_{a}^{-1.01}E_{\gamma,\rm{iso}}^{0.84}$. Such an updated three-parameter correlation still supports that the central leftover after GRBs is probably a millisecond magnetar. It is interesting to note that short GRBs with duration less than 2 s in our sample also follow the same correlation, which hints that the merger production of two neutron stars could be a high mass magnetar, but not necessarily a black hole. Moreover, GRBs having an ``internal'' plateau (i.e., with a following decay index being generally smaller than -3) also obey this correlation. It further strengthens the idea that the internal plateau is due to the delayed collapse of a high mass neutron star into a black hole. The updated three-parameter correlation indicates that GRBs with a plateau phase may act as a standard candle for cosmology study.
\end{abstract}

\keywords{gamma-ray burst: general -- methods: statistical}

\section{Introduction} \label{sec:intro}

Gamma-ray bursts (GRBs) are erratic $\gamma$-ray flashes in the universe, lasting from milliseconds
to as long as thousands of seconds\footnote{Some ultra-long GRBs even last for tens of thousands of seconds \citep{Levan2014}.}, usually with a non-thermal spectrum, isotropically distributed
on the sky \citep{Band1993, Kouveliotou1993, Meegan1992}. The isotropic energy release of the prompt
GRB emission ranges from $10^{48}$ to $10^{55}$ erg \citep{Kumar2015}, making GRBs the most energetic
stellar explosions in our universe. In the era of BeppoSAX \citep{Fishman1995}, multi-wavelength
(from X-ray to radio) afterglows following the prompt emission were detected, which help to localize
GRBs precisely. The first redshift measurement was made for GRB 970508 \citep{Metzger1997}.
Its spectrum revealed a redshift of 0.835, which formally confirms its cosmological origin. GRBs can
even be observed up to $z \sim$ 10 with current detectors \citep{Cucchiara2011}. After decades of
observations and researches, people have obtained a general picture of GRB physics. The widely accepted
model to explain the origin of GRBs is the so called ``fireball"
model \citep{Rees1992, Piran1993, Wijers1997, Meszaros2006}. The prompt emission is considered to be
produced by internal shocks due to the interaction of ejecta in the fireball, while the broadband
afterglow is produced by the interaction between the GRB ejecta and the circumburst
medium \citep{Meszaros1997, Vietri1997, Tavani1997, Waxman1997, Sari1997, Huang1999, Huang2000}.

It has long been found that GRBs can be grouped into two distinct classes by considering
their durations and spectral hardness  \citep{Kouveliotou1993, Bromberg2013}. Long GRBs typically
last for 20 -- 30 s, while short GRBs typically last for 0.2 -- 0.3 s. Besides, short GRBs are on
average harder than long GRBs in hardness ratio.
Long GRBs, with a good number being identified to be associated with core-collapse
supernovae \citep{Galama1998, Hjorth2003, Stanek2003, Campana2006, Xu2013}, are generally
believed to originate from the deaths of massive stars. In contrast, the favored scenario
for short GRBs is the coalescence of two compact stars, especially two neutron
stars \citep[NS-NS]{Paczynski1986, Eichler1989}, or a neutron star and a black
hole \citep[NS-BH]{Paczynski1991}. In both the collapsar models and the coalescence models,
the violent explosion would generate a stellar-size, hyper-accreting black hole, or a
rapidly-spinning, strongly-magnetized neutron
star \citep{Usov1992, Thompson1994, Dai1998a, Popham1999, Rosswog2003, Lei2013}.
Recently, the detection of gravitational waves with the advanced Laser
Interferometer Gravitational Wave Observatory (LIGO) opened a ``multi-messenger''
era in astronomy studies\citep{Abbott2016b, Abbott2016a, Abbott2017b, Abbott2017a}.
The association of the gravitational wave event GW170817 and the short event of GRB 170817A
confirms the binary neutron star coalescence scenario for short GRBs \citep{Abbott2017b, Goldstein2017}.

Although important progresses have been made toward understanding GRBs, some crucial
issues still remain unsolved. With a cosmological origin, GRBs could be used as a powerful
tool to probe the early universe. Nevertheless, compared to type Ia supernovae (SNe Ia), GRBs
are not perfect standard candles. The luminosities of GRBs cover several orders of magnitude.
Their diversity, together with the complex classifications and triggering mechanisms, prevents
direct applications of GRBs in cosmology. Interestingly, a few correlations have been found
between various GRB parameters, either on GRB prompt
emission \citep{Amati2002,Norris2000,Ghirlanda2004a,Ghirlanda2009,Yonetoku2004,Tsutsui2009,Willingale2010, Geng2013, Zhang2018},
or on the afterglow \citep{Liang2005,Oates2009,Oates2012,Dainotti2010,Dainotti2016,Dainotti2017,Xu2012,Liang2015}.
These correlations can provide important clues for studying the physical mechanisms of the
GRB central engines. They can also hopefully make GRBs as some kinds of standard candles
so that these energetic events can be used to limit cosmological
parameters \citep{Fenimore2000,Schaefer2003,Schaefer2007,Dai2004,Ghirlanda2004a,Ghirlanda2006,Wang2006,Amati2008,Wang2011,Wang2015}.
In this aspect, a lot of work still needs to be done because GRBs originate from various mechanisms.

There is a special subclass of GRBs, i.e., GRBs with a plateau phase in the X-ray light curve.
They are characterized by a shallow decay phase followed by a normal decay in X-ray afterglow.
The X-ray light curve of the shallow decay phase is usually very flat so that it typically shows up as
a plateau, while the power-lay index of the subsequent normal decay is usually $\sim -1$.
It is usually believed that the plateau phase is due to some kinds of extra energy injection
into the external shock, possibly from a long-lasting central
engine \citep{Dai1998a,Zhang2001,Rowlinson2010,Rowlinson2013,Bucciantini2012,Gompertz2013, Geng2016}.
Thus the X-ray plateau observations can help constrain various central engine models.
For example, \cite{Li2018} analyzed the \emph{Swift}/XRT light curves of 101 GRBs that have a
plateau phase (and with the redshift being available). They compared the energetics with the
maximum energy budget of magnetars ($\sim 2\times 10^{52}$ erg), trying to determine whether
the central engine is a magnetar or a black hole.
More interestingly, it is also found that there exists another special kind of plateau,
known as the ``internal plateau'', characterized by a plateau followed by an extremely rapid decay
(decay index usually $< -3$). Such a rapid decay at the end of the ``internal plateau'' is much
steeper than that due to the evolution of the synchrotron emission from a decelerating forward shock,
but has to invoke some internal dissipation processes.
One promising interpretation within the magnetar framework is that the ``internal plateau'' is due to
the continuous energy supply from a ``supra-massive'' neutron star. The neutron star will finally collapse
into a black hole, leading to a sudden switching off of the central engine, which manifests as the
subsequent steep decay phase. The collapse of the supra-massive neutron star may either be triggered
by spinning down, or by fall-back accretion, e.g., GRB 070110 as studied by \cite{Chen2017}
and GRB 170714A as studied by \cite{Hou2018}. Comparing with the ``internal plateau'', a plateau
followed by a normal decay is usually called an ``external plateau''
due to the fact that it can be well explained by the deceleration of the external shock.

Several interesting correlations are obtained for GRBs with a plateau phase. \cite{Dainotti2008}
first found an anti-correlation between the end time of the plateau phase in the GRB rest
frame ($T_{a}$) and the corresponding X-ray luminosity at that moment ($L_{X}$).  By analyzing a sample
of 77 GRBs, they derived the relation as $L_{X} \propto T_{a}^{-1.06\pm 0.27}$\citep{Dainotti2010}.
\citet[hereafter, XH2012]{Xu2012} further found a much tighter three-parameter correlation among
$T_{a}$, $L_{X}$ and the isotropic energy in the prompt emission ($E_{\gamma,\rm{iso}}$, here we denote
it as the L-T-E correlation), e.g., $L_{X} \propto T_{a}^{-0.87\pm 0.09}E_{\gamma,\rm{iso}}^{0.88\pm 0.08}$.
\cite{Dainotti2017} later also presented a $L_{X}-T_{a}-L_{\gamma,\rm{peak}}$ correlation between
$T_{a}$, $L_{X}$, and the peak luminosity in the prompt emission ($L_{\gamma,\rm{peak}}$),
$L_{X} \propto T_{a}^{-0.83\pm 0.10} L_{\rm peak}^{0.64 \pm 0.11}$. Interestingly, \cite{Si2018} recently
analyzed 50 GRBs which have a plateau phase in the optical afterglow light curve. They also found a
similar three-parameter correlation for their sample. Their analysis
indicates $L_{b,z}\propto T_{b,z}^{-0.9}E_{\gamma, {\rm iso}}^{0.4}$
and $L_{b,z}\propto T_{b,z}^{-0.9}E_{\rm p,i}^{0.5}$, where $T_{b,z}$ is the break time of
the plateau phase in the optical band, $L_{b,z}$ is the corresponding optical luminosity during the plateau phase,
and $E_{\rm p,i}$ is the peak energy of the prompt emission.
For a comprehensive overview of various GRB correlations, one may refer to \cite{Wang2019}, or \cite{Zhao2019}.

In this study, we have collected a sample of 174 \emph{Swift} GRBs that have a plateau phase
in the X-ray afterglow light curve. These GRBs, all with the redshift being measured, occurred
between March 2005 and August 2018. This sample provides a good opportunity for various statistical analysis.
We fit the light curve of each GRB to get relevant parameters of the plateau phase, and then
explore potential correlations between various pairs of parameters. Especially, the
L-T-E correlation is extensively examined with this enlarged sample, and the physics behind
the L-T-E correlation is investigated. The structure of our article is organized as follows.
We describe the sample selection and data reduction processes in Section \ref{sec:sample}.
The data are statistically analyzed and the results are presented in Section 3. In Section 4,
we briefly summarize our conclusions and discuss the implications.

\section{Sample Selection and Data Analysis} \label{sec:sample}

After the successful launch of \emph{Swift} in 2004 \citep{Gehrels2004}, many interesting features
were soon discovered in the X-ray afterglows of some GRBs, including X-ray flares and the plateau
phase \citep{Zhang2006,Nousek2006}. In this paper, we mainly analyze \emph{Swift} GRBs with a plateau
phase in the X-ray light curve. Our sample are selected from the GRBs occurred between March 2005 and
August 2018, all with the redshift being measured. The observed X-ray data are taken
from the \emph{Swift} GRB light curve repository \citep{Evans2007, Evans2009}, with the spectrum
range of the light curve being 0.3 --- 10 keV. We select our GRBs by the
following four criteria: (1) There should be an obvious flat segment in the X-ray light curve that could
be reliably identified as a plateau phase. To be more specific, we require that the power-law index of
the plateau phase should be in a range of $-1.0$ --- $+1.0$. We exclude any GRBs that do not meet this
requirement. (2) There are abundant observational data points during the plateau phase. This is to ensure that the
light curve can be well defined during the fitting process so that we could correctly extract the
key parameters relevant to the plateau phase without any difficulties. (3) There are no flares observed
during the plateau phase. Again it is to ensure a smooth fit of the plateau phase. (4) The redshift
should be available for the event, so that we could calculate the isotropic $\gamma$-ray energy
release ($E_{\rm \gamma,iso}$). Also, with the redshift, we will be able to derive the intrinsic $T_a$
parameter by considering the time dilation effect. With these four criteria, we finally obtained a
sample of 174 GRBs.  The redshift range of our sample is 0.04 --- 8.0.
We note that seven short GRBs ($T_{90} < 2$ s) are interestingly included in this sample, as listed in
Table \ref{tab:sGRBname}.

\begin{table}[h!]
	\renewcommand{\thetable}{\arabic{table}}
	\centering
	\caption{Seven Short GRBs Included in Our Sample}
	\label{tab:sGRBname}
	\begin{tabular}{ccc}
		\tablewidth{0pt}
		\hline
		GRB name & $T_{90}$ (s) & $z$ \\
		\hline
		GRB 051221A & 1.4 & 0.55 \\
		GRB 061201 & 0.76 & 0.11 \\
		GRB 070809 & 1.3 & 0.22 \\
		GRB 090510 & 0.3 & 0.90 \\
		GRB 130603B & 0.18 & 0.36 \\
		GRB 140903A & 0.3 & 0.35 \\
		GRB 150423A & 0.22 & 1.40 \\
		\hline
	\end{tabular}
\end{table}

Having selected the sample, we then fit the X-ray light curve of each GRB with a smoothly broken
power-law function, which takes the form of \citep{Evans2009,Li2012,Yi2016}
\begin{equation}
\label{eq:smo}
F_{X}(t)=F_{X0}\left[ \left( \frac{t}{T_{0}}\right)^{\alpha_{1}\omega}+\left( \frac{t}{T_{0}}\right)^{\alpha_{2}\omega}\right]^{-1/\omega}.
\end{equation}
Here, $\alpha_{1}$ is the power-law index during the plateau phase, which should generally be close
to zero. $\alpha_{2}$ describes the decay in the following phase. $T_{0}$ is the observed end time
of the plateau phase. The end time in the GRB rest frame can then be obtained by $T_{a} = T_{0}/(1+z)$.
$F_{X0}\times 2^{-1/\omega}$ is the corresponding flux at the end of the plateau,
and $\omega$ is a smoothness parameter, characterizing the sharpness of the transition from the
plateau phase to the subsequent decay phase.

Fitting the observed X-ray light curves of the GRBs in our sample with Equation (1), we can get the
main parameters relevant to the plateau phase, i.e., the end time of the plateau phase ($T_{a}$),
the flux at the break time ($F_{X0}$), the power-law timing index during the plateau phase ($\alpha_{1}$),
and the power-law timing index in the following decaying phase ($\alpha_{2}$).
In order to get the best fit, the Markov chain Monte Carlo (MCMC) algorithm is applied.
$10^{6}$ samples are generated to fit the light curves.

Our best fit results for the X-ray light curves of all the 174 GRBs are presented in Figure \ref{fig:1},
which can give us a general idea on what the plateau is like and how good the fit is. We obtain $\alpha_{1}$, $\alpha_{2}$, $T_{a}$ and $F_{X0}$ of all GRBs in our sample after the light curve fittings.
The range of $\alpha_{1}$ is (-0.79, 0.79), and the range of $\alpha_{2}$ is (0.80, 15.11).
It is interesting to note that several
GRBs have very large $\alpha_{2}$ values, which means their plateau phase is followed by a very steep decay.
In other words, the plateau phase in these GRBs should be  ``internal plateau''. Here we use a quantified
criteria of $\alpha_{2} > 2+\beta_{X}$ to define ``internal plateaus'', where the convention
of $F_{\nu} \propto t^{-\alpha} \nu^{-\beta_{X}}$ is used and $\beta_{X}$ corresponds to the spectral
index of the X-ray afterglow \citep{Evans2009}. According to this criteria, 11 GRBs in our sample
have ``internal plateaus", which are listed in Table \ref{tab:IGRBname}.

\begin{table}[h!]
	\renewcommand{\thetable}{\arabic{table}}
	\centering
	\caption{GRBs with an``Internal Plateau" in Our Sample}
	\label{tab:IGRBname}
	\begin{tabular}{ccc}
		\tablewidth{0pt}
		\hline
		GRB name & $\alpha_{2}$ & $z$ \\
		\hline
		GRB 050730  & 2.77 & 3.97 \\
		GRB 060607A & 3.47 & 3.08 \\	
		GRB 070110  & 8.95 & 2.35 \\
		GRB 100219A & 4.64 & 4.7 \\
		GRB 100902A & 4.69 & 4.5 \\
		GRB 111209A & 15.11 & 0.68 \\
		GRB 111229A & 3.15 & 1.38 \\
		GRB 120521C & 3.03 & 6.0 \\
		GRB 120712A & 3.26 & 4.17 \\
		GRB 130408A & 3.79 & 3.76 \\
		GRB 170714A & 4.97 & 0.79 \\
		\hline
	\end{tabular}
\end{table}

Our sample is one of the largest sample of GRBs with a plateau phase so far. The number of GRBs in
our sample is about three times that of XH2012, which only includes 55 ``golden" events. Comparing with
XH2012's study, we have removed several GRBs from XH2012 sample because they do not meet the
selection criteria applied here, e.g. GRBs 050724 and 070802. \cite{Li2018} recently composed
a sample of 101 \emph{Swift}  GRBs (up to 2017 May), which seems to be quite incomplete as compared
with our sample. It is interesting to note that the sample of \cite{Dainotti2017} includes 183 GRBs
with X-ray plateaus (up to 2016 August). However, in their selection process, they did not impose
any compulsory conditions to identify the plateau phase clearly. In our study, we require that
the power-law timing index during the plateau phase should be in the range of $-1.0$ --- $+1.0$.
This can effectively avoid confusing those GRBs with a jet break as a GRB with a plateau phase.
For those GRBs with a jet break, the timing index is usually $\sim -1$, then followed by a
steeper decay with a slope of $\sim -2$). They might be confused as a GRB with a plateau phase
in some cases.

With $F_{X0}$ being derived from the fitting, the X-ray luminosity at the end time of the
plateau phase ($L_{X}$) can be calculated as \citep{Willingale2007},
\begin{equation}
\label{eq:lx}
L_{X}=\frac{4\pi D_{\rm L}^{2}(z)F_{X0}}{(1+z)^{1-\beta_{X}}},
\end{equation}
where $z$ is the redshift and $D_{\rm L}(z)$ is the luminosity distance. The values of $z$
and $\beta_{X}$ are obtained from the \emph{Swift} GRB
Table\footnote{\url{https://swift.gsfc.nasa.gov/archive/grb\_table.html/}}.
Note that in the \emph{Swift} table, two redshifts are available for GRB 151027A: XRT suggests a
redshift of 0.38, while Keck and GMG give a redshift of 0.81. As noted on the website, the XRT
measurement is only a tentative value. Thus we chose GRB 151027A's redshift as 0.81.
Another special case is GRB 101225A, for which Keck I gives a redshift of 3.8 while Gemini GMOS gives a redshift
of 0.847. In this case, we adopted the larger z value (3.8) since the smaller redshift of 0.847 may
be due to absorption features of the foreground galaxies. 
All through this study, we take a fla t $\Lambda$CDM cosmology
with $H_{0}$ = 70.0 km s$^{-1}$ Mpc$^{-1}$ and $\Omega_{\rm M}$ = 0.286 to calculate $D_{\rm L}(z)$.

The observed isotropic $\gamma$-ray energy ($E_{\gamma,\rm{iso}}'$) of the prompt emission is calculated from
\begin{equation}
\label{eq:ex}
E_{\gamma,\rm{iso}}'=\frac{4\pi D_{\rm L}^{2}(z)S}{1+z},
\end{equation}
where $S$ is the BAT fluence (15 --- 150 keV, in units of $\rm{erg/cm^{2}}$), also taken from the \emph{Swift} GRB table.

The so called k-correction should be considered in our calculations. Due to the cosmological
time dilation, the observer frame 15-150 keV bandpass of Swift GRBs is different from the
rest-frame bandpass with redshift \citep{Bloom2001}. Considering this effect, the corrected
isotropic energy should be
\begin{equation}
\label{eq:corr_ex}
E_{\gamma,\rm{iso}}=E_{\gamma,\rm{iso}}'\times \frac{\int_{15/1+z}^{150/1+z} \ E\Phi(E)\,dE}{\int_{15}^{150} \ E\Phi(E)\,dE},
\end{equation}
where $\Phi (E)$ is the energy spectrum. Typically, the peak energy of the observed GRB prompt emission is
$\sim$ 200 --- 300 keV \citep{Preece2000, Goldstein2012}. Since Swift/BAT observations are carried out in a
relatively narrow energy range (15 --- 150 keV), we adopt a simple power-law spectral model,
i.e., $\Phi (E)=E^{\alpha_{\gamma}}$, to calculate the k-correction in this study.
Here $\alpha_{\gamma}$ is the photon spectral index, taken from the \emph{Swift} GRB table.
The derived k-corrections for the GRBs of our sample are mainly in the range of 0.07 --- 1.7. 


The various parameters relevant to our study are listed in Table \ref{tab:parameter},
which includes the GRB name, $T_{90}$, redshift ($z$), BAT fluence ($S$), power-law index of the plateau phase ($\alpha_{1}$), power-law index of the subsequent segment ($\alpha_{2}$), the end time of the plateau phase ($T_{a}$), the corresponding luminosity at the end time ($L_{X}$), and the isotropic energy ($E_{\gamma,\rm{iso}}$). After getting all the relevant data, we then can explore the correlations between various parameters. The L-T-E correlation can also be thoroughly examined. We present the results in the
next section.

\section{Results} \label{sec:results}

\subsection{Parameter Distributions}
In Figure \ref{fig:hist1}, we plot the distributions of $L_{X}$, $T_{a}$, and
$E_{\gamma,\rm{iso}}$. The typical value of $L_{\rm x}$
is $2.6\times 10^{47}$ erg$\cdot$s$^{-1}$ with a 50\% distribution range of
(0.4, 12)$\times 10^{47}$ erg$\cdot$s$^{-1}$. The typical value of $T_{a}$ is $2.4\times 10^{3}$ s
with a 50\% distribution range of (0.8, 7)$\times 10^{3}$ s. The typical value
of $E_{\gamma,\rm{iso}}$ is $0.13\times 10^{53}$ s, with a 50\% distribution range
of (0.04, 0.3)$\times 10^{53}$ erg. We notice that the distributions of these three parameters
are very similar to those of XH2012 sample. For XH2012 sample, the typical value of $L_{\rm x}$
is $3.9\times 10^{47}$ erg$\cdot$s$^{-1}$ with a 50\% distribution range
of (0.4, 17)$\times 10^{47}$ erg$\cdot$s$^{-1}$; the typical value of $T_{a}$ is $2.5\times 10^{3}$ s,
with a 50\% distribution range of (0.9, 10)$\times 10^{3}$ s; and the typical value of
$E_{\gamma,\rm{iso}}$ (bolometric isotropic energy, in XH2012) is $0.62\times 10^{53}$ erg,  with a 50\% distribution range
of (0.07, 1.1)$\times 10^{53}$ erg.

The distributions of $S$, $z$, and $T_{90}$ are displayed in Figure \ref{fig:hist2}. The typical
value of $S$ is $20\times 10^{-7}$ erg$\cdot$cm$^{-2}$, with a 50\% distribution range
of (8, 50)$\times 10^{-7}$ erg$\cdot$cm$^{-2}$. The typical value of $z$ is 2.0, with
a 50\% distribution range of (1.0, 3.2). As for $T_{90}$, the typical value is 58 s, with a 50\% distribution range
of (18, 123) s. The distribution of $T_{90}$ indicates that most GRBs in our sample are long GRBs.

$\alpha_{1}$ and $\alpha_{2}$ are two important parameters derived from our light curve fitting. Although
we have applied a criteria of $-1.0 \leq \alpha_1 \leq +1.0$ in selecting the GRBs with a plateau phase,
$\alpha_{1}$ actually has a typical value of 0.11 for our whole sample, with a 50\% distribution range of (0.03, 0.3). It means
that in most cases, the timing index of the plateau is in the range of $0.03$ --- $0.3$, so that
the plateau phase is really very flat in the light curve. Again, it supports the idea that the plateau
phase is a special phenomenon in GRB afterglows. In our sample, $\alpha_{2}$ has a typical value 1.5,
with a 50\% distribution range of 1.3 --- 1.9 (see Figure \ref{fig:hist3}).
For the internal plateau sub-sample, we find that the typical value of $\alpha_{1, {\rm int}}$ is 0.23,
with a 50\% distribution range of 0.06 --- 0.5; and the typical value of $\alpha_{2, {\rm int}}$ is 3.8, with
a 50\% distribution range of 3.2 --- 5. We have also calculated the slope variation
of $\Delta \alpha = \alpha_{2}-\alpha_{1}$.  The typical value of $\Delta \alpha$ is 1.4 for the whole sample,
with a 50\% distribution range of 1.1 --- 1.7;  and it is $\Delta \alpha_{\rm int} = 3.5$ for the internal plateau
sub-sample, with a 50\% distribution range of 3.0 --- 4.4. We have also applied the K-S test on $\Delta \alpha$ for
the internal plateau sub-sample and the rest of the whole sample. The derived p-value is as small as
$5\times 10^{-10}$, which means the internal plateau GRBs are very likely to form a distinct subclass.

\subsection{Two-Parameter Correlations}

Here, we investigate whether there are any significant correlations between various parameter pairs.
In Figure \ref{fig:2D}, we plot $L_{X}$ versus $T_{a}$, $L_{X}$ versus $E_{\gamma,\rm{iso}}$, $S$ versus $T_{90}$,
and $E_{\gamma,\rm{iso}}$ versus $T_{90}$, respectively. It could be clearly seen that these parameter pairs
are somewhat correlated. For the $L_{X}$ -- $T_{a}$ correlation, the best fit result is
$\log(L_{X}/10^{47}{\rm erg\cdot s^{-1}}) = (0.76\pm 0.13) + (-1.12\pm 0.18)\log(T_{a}/10^{3}{\rm s})$,
with the Spearmaan correlation coefficient $\rho = -0.74$ and a chance probability $p=2\times 10^{-31}$.
For the $L_{X}$ -- $E_{\gamma,\rm{iso}}$ correlation,
the best fit result is $\log(L_{X}/10^{47}{\rm erg\cdot s^{-1}})
= (1.55\pm 0.20) + (0.97\pm 0.13) \log(E_{\rm \gamma,iso}/10^{53}{\rm erg})$. The Spearman correlation
coefficient $\rho$ is 0.60, with a chance probability $p=1\times 10^{-18}$. Generally speaking,
these two-parameter correlations are still quite dispersive.  In the
next subsection, we will see that when we use the three parameters of $L_{X}$, $T_{a}$ and
$E_{\gamma,\rm{iso}}$ to get a three-parameter correlation, it would be much tighter.
We have also used the bivariate Bayesian linear regression procedure \citep{Kelly2007} to
further check these two parameter correlations. This method can conveniently take into account the
heteroscedastic measurement errors. The fitting results
are $\log(L_{X}/10^{47}{\rm erg\cdot s^{-1}}) = (0.81\pm 0.07) + (-1.33\pm 0.10)\log(T_{a}/10^{3}{\rm s})$
with an intrinsic scatter of 0.59,
and $\log(L_{X}/10^{47}{\rm erg\cdot s^{-1}})= (1.37\pm 0.10) + (1.01\pm 0.07)\log(E_{\rm \gamma,iso}/10^{53}{\rm erg})$
with an intrinsic scatter of 0.64. We see that the results are consistent with the previous least square linear fits.

Additionally, we find that both $S$ and $E_{\gamma,\rm{iso}}$ are positively correlated with $T_{90}$.
The best fit results are $\log (S/10^{-7}{\rm erg\cdot cm^{-2}}) = (0.72\pm 0.06)
+ (0.34\pm 0.03) \log (T_{90}/{\rm s})$ with a Spearman correlation
coefficient of $\rho = 0.54$ and a chance probability of $p = 2\times 10^{-14}$,
and $\log(E_{\rm \gamma,iso}/10^{53}{\rm erg}) = (-1.71\pm 0.22)
+ (0.42\pm 0.11) \log (T_{90}/{\rm s})$ with a Spearman correlation coefficient
of $\rho = 0.55$ and a chance probability of $p = 1\times 10^{-14}$,
respectively. The Spearman correlation coefficients support the existence of
the $S$ - $T_{90}$ and $E_{\gamma,\rm{iso}}$ - $T_{90}$ correlations, but note
that these correlations again are relatively dispersive.
As a reference, the fitting results by using the bivariate Bayesian linear
regression procedure are: $\log(S/10^{-7}{\rm erg\cdot cm^{-2}}) = (0.52\pm 0.10) + (0.48\pm 0.05)
\log (T_{90}/{\rm s})$ with an intrinsic scatter of 0.22,
and $\log(E_{\rm \gamma,iso}/10^{53}{\rm erg}) = (-2.15\pm 0.17) + (0.67\pm 0.10)\log (T_{90}/{\rm s})$
with an intrinsic scatter of 0.46. 
Note that $T_{90}$ might be detector dependent \citep{Bromberg2013,Resmi2017}, but
this effect should not be significant in our sample, since all our bursts are \emph{Swift} GRBs.
Generally, the positive correlations of $S$ -- $T_{90}$ and $E_{\gamma,\rm{iso}}$ -- $T_{90}$  are
not difficult to understand. It means the observed intensities of GRBs are somewhat in a limited range.

In these plots, we have also marked the 7 short GRBs in our sample with green squares, and
the 11 GRBs of ``internal plateaus'' with red triangles, to see whether there are any
systematic difference in these sub-samples. It is interesting to note that in all the four
panels of Figure 5, the internal plateau sub-sample is largely consistent with the whole
sample. We speculate that maybe the reason is that all the internal plateau GRBs
in our sample are essentially long GRBs. In order to further investigate the nature of
the internal plateau sub-sample, we have applied the Bromberg criteria test on these GRBs,
following \cite{Bromberg2013}. The Bromberg criteria is an improved classification method
for distinguishing long GRBs (or more precisely, collapsars) and short GRBs (or non-collapsars)
by considering the duration as well as the hardness.
More specifically, for \emph{Swift} GRBs, if the harness raito is hard (Power-Law index PL $>$ -1.13),
then the critical duration for long and short GRBs is $2.8^{+1.5}_{-1.0}$ s;
If the harness raito is intermediate (-1.65 $<$ PL $<$ -1.13), then the critical duration
is $0.6^{+0.2}_{-0.3}$ s; If the harness raito is soft (PL $<$ -1.65), then the separation
duration is $0.3^{+0.4}_{-0.2}$ s. In addition to $T_{90}$ and the hardness ratio, we further
calculated the probability for the GRB to be a non-collapar, i.e. the non-collapsar probability
($f_{NC}$) in \citet{Bromberg2013}. The results are listed in Table \ref{tab:fNC}.
Note that for two GRBs (111209A and 170714A), the $T_{90}$ data are unavailable so that the
test cannot be applied. Table \ref{tab:fNC} clearly shows that all other
``internal plateau'' GRBs in our sample are long GRBs based on the Bromberg criteria test.

Contrastively, although the short GRB sub-sample is also generally consistent with
the whole sample, there is still some systematic deviation. It is most significant in
Figure 5(a). Maybe it is due to the fact that long GRBs and short GRBs have quite different
origins.

In Figure \ref{fig:2D_seleceff}, we plot our sample on the $L_{X}$ - $z$,
$E_{\gamma,\rm{iso}}$ - $z$, and $T_{a}$ - $z$ planes.
Figure 6(a) and 6(b) show that GRBs at high redshifts usually have a relatively larger $L_{X}$
and $E_{\gamma,\rm{iso}}$. This may be due to the selection effect: at a large distance, GRBs
with a small $L_{X}$ and $E_{\gamma,\rm{iso}}$ will generally be too weak to be observed, since
our detectors have a limited threshold. In Figure 6(a), we have plotted the XRT detection limit
with the dashed line, which corresponds to a minimum flux
of $f_{\rm lim}\sim 2\times 10^{-14}$ erg cm$^{-2}$ s$^{-1}$. Note that all the observed data
points are reasonably above the line. Similarly, in Figure 6(b), the dashed line represents the
BAT detection limit for $f_{\rm lim}\sim 10^{-8}$ erg cm$^{-2}$ s$^{-1}$ by assuming a typical
duration of 2 s. Again, all the observed data points are above the limit line.

In Figure \ref{fig:2D_T-z}, we see that no correlation exists between $T_{a}$ and $z$.
It means that the plateau phase is somewhat redshift-independent. This tendency is especially
clear for the internal plateau sub-sample.

In Figure 7, we plot our sample on a few other parameter planes.
We have applied the Kendell's tau method to do the non-parametric test for any
potential correlations between the parameters (see Table \ref{tab:kendalltau} for numerical
results). The $\tau$ statistics are all rather small, indicating that there is no
significant correlation between any two of the parameters.

\subsection{L-T-E Correlation}

In this section, we examine whether the L-T-E correlation still exists for our significantly expanded sample.
Using the fitted $T_{a}$, $L_{X}$ and $E_{\gamma,\rm{iso}}$ data, we fit the possible L-T-E correlation by
the following expression (XH2012):
\begin{equation}
\log(L_{X}/10^{47}{\rm erg\cdot s^{-1}})=a+b \log(T_{a}/10^{3}{\rm s})+c \log(E_{\rm \gamma,iso}/10^{53}{\rm erg}),
\end{equation}
where the coefficients $a$, $b$, and $c$ are constants to be determined.

In order to get the best fit, a Markov chain Monte Carlo (MCMC) algorithm is used.
$10^{7}$ samples are generated to fit the L-T-E correlation. During our fitting, the joint likelihood
function for $a$, $b$, and $c$ is \citep{DAgostini2005}:
\begin{equation}
\mathscr{L}\left( a,b,c,\sigma_{\rm int}\right)\propto \prod_{i}\frac{{\rm exp}\left[ -( y_{i}-a-bx_{{\rm 1},i}-cx_{{\rm 2},i})^{2}/(2(\sigma_{\rm int}^{2}+\sigma_{y_{i}}^{2}+b^{2}\sigma_{x_{1, i}}^{2}+c^{2}\sigma_{x_{2, i}}^{2}))\right]}{\sqrt{\sigma_{\rm int}^{2}+\sigma_{y_{i}}^{2}+b^{2}\sigma_{x_{1, i}}^{2}+c^{2}\sigma_{x_{2, i}}^{2}}},
\end{equation}
where $i$ is the corresponding serial number of GRBs in our sample, $\sigma_{\rm int}$ represents the
intrinsic scattering. Here, $x_{1}=\log(L_{X}/10^{47}\rm{erg\cdot s^{-1}})$,
$x_{2}=\log(E_{\gamma,\rm{iso}}/10^{53}\rm{erg})$ and $y=\log(T_{a}/10^{3}\rm{s})$.
$\sigma_{x_{1, i}}, \sigma_{x_{2, i}}$, and $\sigma_{y_{i}}$ are errors of $x_{1}, x_{2}$, and $y$ respectively.

We have performed the three-parameter fitting for all the 174 GRBs in our sample. The best fit result is
\begin{equation}
\log(L_{X}/10^{47}{\rm erg\cdot s^{-1}})= (1.11\pm 0.04) + (-1.01\pm 0.05) \log(T_{a}/10^{3}{\rm s}) + (0.84\pm 0.04) \log(E_{\rm \gamma,iso}/10^{53}{\rm erg}).
\end{equation}
The Spearman correlation coefficient is $\rho = 0.90$, with a chance probability of $p = 2\times 10^{-67}$.
Our fitting result is illustrated in Figure \ref{fig:result_fig}, where the Y-axis is the X-ray
luminosity at the end time of the plateau phase ($L_{X}$, in units of $10^{47}$ erg/s), and the
X-axis is a combination of $T_{a}$ (in units of $10^{3}$ s) and $E_{\gamma,\rm{iso}}$ (in units
of $10^{53}$ erg). We see that the observed data points are distributed tightly along the
best fit line.

Comparing with the two-parameter correlations presented in the above subsection, such as
the $L_{X}$ - $T_{a}$ correlation ($L_{X} \propto T_{a}^{-1.10\pm 0.20}$, see Figure \ref{fig:2D_L-T})
and the $L_{X}$ - $E_{\gamma,\rm{iso}}$ correlation ($L_{X} \propto E_{\gamma,\rm{iso}}^{0.97\pm 0.13}$,
see Figure \ref{fig:2D_L-E}), the L-T-E correlation is much tighter. 
The correlation coefficient is very close to 1.
The error bars of the parameters
derived here are also much smaller, with the intrinsic scatter of the L-T-E correlation being
as small as $\sigma_{\rm int} = 0.39\pm 0.03$.

Our result indicates that $L_{X} \propto T_{a}^{-1.01\pm 0.05} E_{\gamma,\rm{iso}}^{0.84\pm 0.04}$. 
The relation is generally consistent with the result of XH2012, which
reads $L_{X} \propto T_{a}^{-0.87\pm 0.09} E_{\gamma,\rm{iso}}^{0.88\pm 0.08}$.
The error bars of our derived indices are roughly one half of those in XH2012, showing
that the correlation is even tighter here. It is striking to see that the L-T-E correlation still
exists for such a significantly enlarged sample, with an even smaller dispersion.

Several different models have been suggested to interpret the plateau phase. First, it may be due
to a structured jet. In this case, emission from the high latitude ejecta will peak at a later stage as
compared with the central energetic ejecta, leading to a slower decay of the afterglow. Second, it may be
produced by a stratified fireball. In this case, the material in different shells have different velocities.
Slower shells will finally catch up with the preceding faster shells and supply energy into the external
shock to show up as the plateau phase. Generally, in the above two kinds of models, some arbitrary
assumptions have to be made to depict the jet structure or the velocity distribution of the ejecta, and no
simple conclusions could be drawn for the plateau phase.

In a third model, it is argued that the plateau phase is produced by energy injection from a
rapidly spinning millisecond magnetar that resides at the center of the GRB
remnant \citep{Zhang2001,Dai2004,Troja2007,Dall2011,Metzger2011,Rowlinson2013,Rowlinson2014,Siegel2014,Yi2014,Rea2015,Lv2018}.
In this model, after the prompt burst, the millisecond magnetar loses its rotational energy due to spinning down,
generating a Poynting-flux or an electron-positron wind and injecting energy into the external shock.
Xu \& Huang (2012) suggested that the L-T-E correlation supports this magnetar model.
It has been shown that the luminosity of the energy injection is roughly constant within the characteristic
spinning down time of the magnetar \citep{Dai2004}, naturally leading to a plateau phase. After the magnetar
spins down, the energy injection ceases. So, Xu \& Huang argued that the product of the observed $L_X$ and
$T_a$ gives a good measure for the rotational energy of the magnetar. On the other hand, the isotropic
$\gamma$-ray energy release ($E_{\gamma,\rm{iso}}$) is also closely related to the rotational energy.
As a result, we roughly have $L_X \times T_a \propto E_{\gamma,\rm{iso}}$,
i.e., $L_X \propto  E_{\gamma,\rm{iso}} {T_a}^{-1} $.
It is interesting to note that our fitting result
of $L_{X} \propto T_{a}^{-1.01\pm 0.05} E_{\gamma,\rm{iso}}^{0.84\pm 0.04}$ is again consistent 
with the theoretical expectation. The fitted power-law index of $T_{a}$ is
$-1.01\pm 0.05$, very close to $-1$, and the fitted power-law index of $E_{\gamma,\rm{iso}}$ is
$0.84\pm 0.04$, also close to $1$. We thus suggest that our updated L-T-E correlation is
a support for the magnetar explanation of the X-ray plateaus.

In Figure 8, we have marked the short GRB sub-sample (7 events) with green squares. Interestingly,
these short GRBs follow the same L-T-E correlation. They are generally at the low luminosity end of the
plot, but they are distributed along the best fit line almost as tightly as other GRBs. Theoretically,
while a millisecond magnetar may be born during the hypernova that gives birth to a long
GRB \citep{Usov1992, Bucciantini2009, Metzger2011}, it has also been argued that the central
engine of short GRBs could also be magnetars \citep{Dai2006, Fan2006, Metzger2008, Kiuchi2012, Lv2015}.
The distribution of the short GRB sub-sample on Figure 8 suggests that
the leftover of the double compact star merger might be a magnetar in these cases.

In Figure 8, we have also specially marked the ``internal plateau'' sub-sample (11 GRBs) with red
triangles. As mentioned before, for these internal plateau GRBs, the most popular explanation is that
a fast-rotating supra-massive neutron star is initially born associated with the GRB. It then collapses
into a black hole after significantly spinning down. So, the internal plateau sub-sample are valuable
events for checking our magnetar explanation of the plateau phase. In Figure 8, we see that these
GRBs do follow the same L-T-E correlation as expected. However, we also note that the scattering of the
red triangles are slightly larger than other data points. It may be due to the fact that the condition
that determines the end time of the internal plateau phase is somewhat different from other GRBs.
For normal plateaus, the end time is reached when the magnetar loses half of its rotational energy,
while for internal plateaus, the end time is achieved when the magnetar rotates slowly enough so that
it cannot resist the gravity. So, in the internal plateau case, the end time is dependent on
both the initial spinning period and the initial mass.

\section{Discussion} \label{sec:discussion}

In this study, we focus on those GRBs with a plateau phase in the X-ray afterglow light curve.
We define the plateau phase by requiring that the power-law timing index should be in a range of
$-1.0$ --- $+1.0$. We finally extracted a large sample of 174 events from the \emph{Swift} GRBs
that happened between March 2005 and August 2018, all with the redshift being measured. Statistical
analysis is then carried out based on the sample. It is found that some parameter pairs are more or
less correlated, such as $L_{X}$ vs. $T_{a}$, $L_{X}$ vs. $E_{\gamma,\rm{iso}}$, $S$ vs. $T_{90}$,
and $E_{\gamma,\rm{iso}}$ vs. $T_{90}$. However, the most striking finding about our sample is that
the three-parameter L-T-E correlation still exists when the capacity of our sample has been significantly
expanded as compared with that of XH2012. The best fitting result gives
$L_{X} \propto T_{a}^{-1.01} E_{\gamma,\rm{iso}}^{0.84}$, also largely consistent with
the result of XH2012. It is argued that the L-T-E correlation supports
the magnetar explanation for the plateau phase.
In our study, two sub-samples have been identified, i.e. short GRBs with a plateau phase and
those GRBs with a so called ``internal plateau''. It is interesting to note that the same L-T-E
correlation is followed by these two sub-samples, indicating that a rapidly rotating millisecond
magnetar might have been born during the GRB.

Short GRBs are generally believed to be produced by
the merger of binary compact stars, such as by a double neutron star system or by a black
hole-neutron star binary. For the merger of a black hole and a neutron star, the aftermath
is undoubtedly a black hole. However, for the merger of a double neutron star system, the case
is much more complicated. The leftover could be either a black hole or a massive neutron
star. Here, for the short GRBs with an X-ray plateau as in our sample, the L-T-E correlation
strongly indicates that the leftover of the merger should be a neutron star that has a strong
magnetic field and spins at millisecond period. However, we should also notice that only a
small fraction of short GRBs has the plateau phase. For other short GRBs, we cannot accurately
determine the aftermath of the merger. Anyway, the merit of our study is that it proves that
at least some of the short GRBs are produced by double neutron star mergers, and that a fraction
of them lead to the birth of massive millisecond magnetars.

For GRBs that have an internal plateau, the most natural interpretation is that a rapidly rotating
supra-massive magnetar is born at the center. The plateau phase is then followed by a steep
decay when the magnetar collapse to form a black hole after it spins down significantly so that
it cannot resist the gravity. As a result, it is naturally expected that internal plateau 
GRBs should also follow the L-T-E correlation, as shown in Figure 8. Though some authors recently have 
suggested that it is possible to explain the internal plateaus by some other mechanisms such as 
the deceleration of the ejecta by external medium \citep{Beniamini2017a, Beniamini2017b, Metzger2018}, 
our results support the magnetar interpretation of the L-T-E correlation.

In our study, the sub-sample of short GRBs is consisted of 7 events while the sub-sample of
internal plateau GRBs is consisted of 11 events. Interestingly, we find no overlapping between
these two sub-samples. In other words, no internal plateau has been found in a short GRB in
our sample. However, we notice that internal plateaus attributed to neutron star collapse actually
have been claimed to exist in short GRBs at least by \cite{Rowlinson2013}.
In their study, by defining internal plateaus as $\alpha_{2} > \beta_{X}$ + 2,
four short GRBs were found to have internal plateaus, e.g. GRBs 060801, 080919, 100702A, and 120305A.
The former three GRBs are even listed in the \emph{Swift} short GRB cataloge in \cite{Bromberg2013}.
But nearly 80\% of the short GRBs in \cite{Rowlinson2013} are lacking of redshift measurements.
Unfortunately, redshifts are not available for the four ``internal plateau'' GRBs listed above, thus
they are not included in our sample. In fact, theoretically, it is more likely that the merger events
would result in a supra-massive NS than the core collapse process. Before the final merge, the binary
neutron stars rotate around each other at a speed close to the Kepler limit. This makes the merger
remnant more likely to have greater rotational energy to generate a supra-massive NS than the core collapse.
It will then result in a delayed collapse to form a black hole, and thus lead to an internal plateau.
Indeed, evidence supporting the existence of supra-massive NSs after NS - NS mergers has been
suggested by a few authors \citep{Lasky2014, Lv2015, Gao2016, Li2016}.
In our sample, the total number of internal plateau GRBs is still very small. We cannot exclude
the possibility that short GRBs may also have internal plateaus. In the future, if short GRBs with clear
internal plateaus are observed and if redshifts are also measured, then the L-T-E correlation will
be a useful tool to probe their nature, and the correlation itself will also be meaningfully examined. 

Finally, it should be noted that in recent years, some authors have suggested that the
central engine of GRBs with a plateau phase may also be newly formed black holes with a
neutrino-cooling-dominated accretion flow (NDAF) \citep{Cannizzo2009,Cannizzo2011,Lv2014,Lei2017}.
Two popular mechanisms are considered to launch a continuous jet to supply energy into the external shock
and produce the plateau. One is the Blandford-Znajek (BZ) mechanism, which extracts the rotational
energy from a Kerr black hole \citep{Blandford1977}. The other is the neutrino-antineutrino
annihilation mechanism, releasing gravitational energy from the accretion
disk \citep{Popham1999, Di2002, Gu2006, Liu2007, Chen2007, Janiuk2007, Lei2009, Liu2015}.
In reality, it is quite possible that both black holes and magnetars can somewhat act as the central
engine of energy injection. Actually, \cite{Li2018} even argued that about 20\% of the central engines
are magnetars, and others are black holes. A mixed origin of the plateau phase may make the L-T-E
correlation complicated and dispersive.


\acknowledgments
We thank the anonymous referee for valuable suggestions. This study is partially supported by the National Natural Science Foundation of
China (Grants No. 11873030, and 11833003), and by the Strategic Priority Research
Program of the Chinese Academy of Sciences ``Multi-waveband Gravitational Wave Universe''
(Grant No. XDB23040400). ZBZ acknowledges the support from two Chinese Provincial
Natural Science Foundations (Grant No. 20171125 and 20165660). This work made use of data
supplied by the UK \emph{Swift} Science Data Center at the University of Leicester.

\bibliographystyle{aasjournal}
\bibliography{bibtex}

\clearpage

\begin{figure}[ht!]
	\centering
	\includegraphics[width=7.0in]{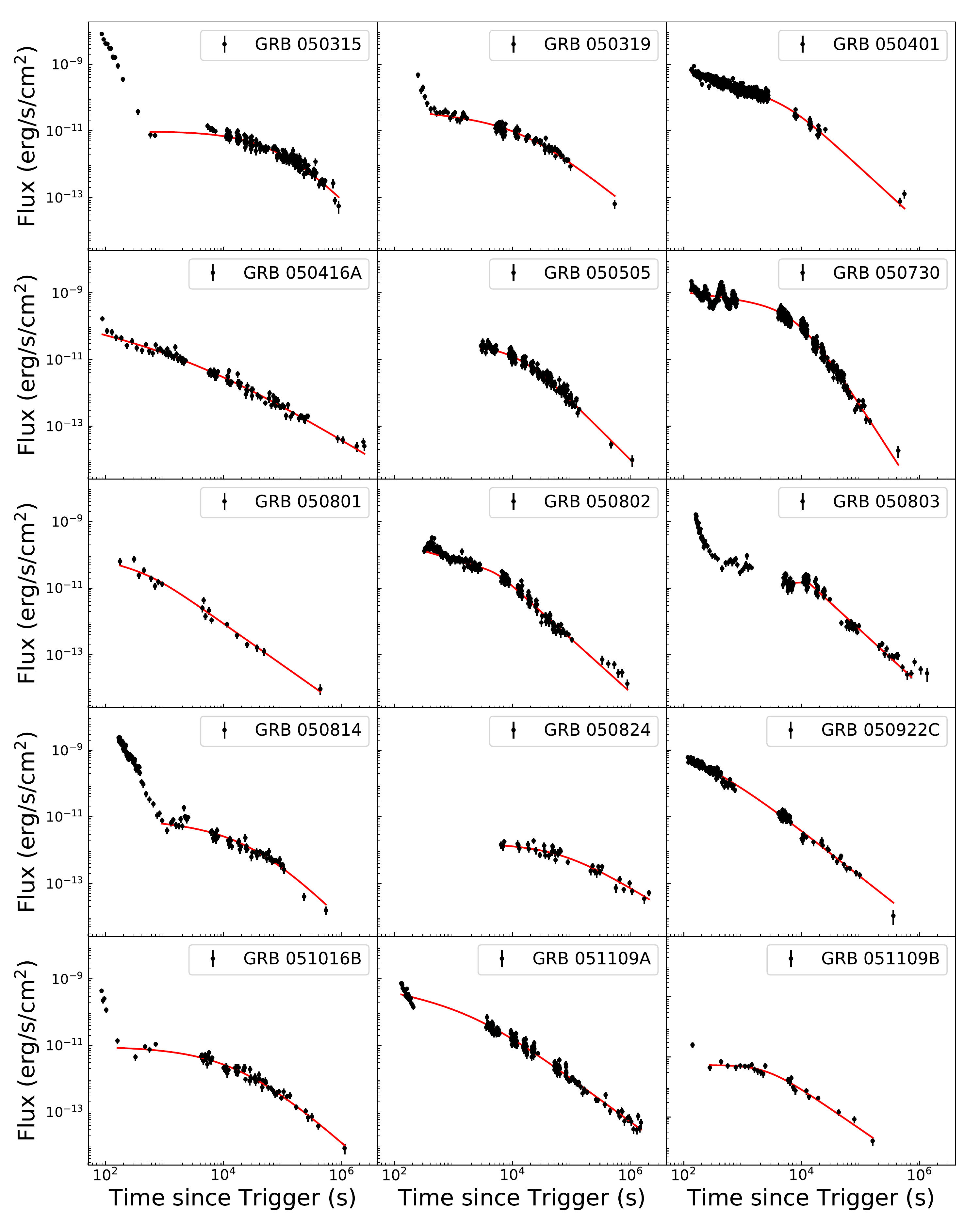}
	\caption{Observed X-ray afterglows of all the 174 GRBs in our sample, and our best fit to the light curves.
             Black dots correspond to the observational data by \emph{Swift}/XRT \citep{Evans2007, Evans2009}.
             The data fit is performed with the Markov chain Monte Carlo (MCMC) algorithm.\label{fig:1}}
\end{figure}

\begin{figure}[ht!]
	\centering
	\addtocounter{figure}{-1}
	\includegraphics[width=7.0in]{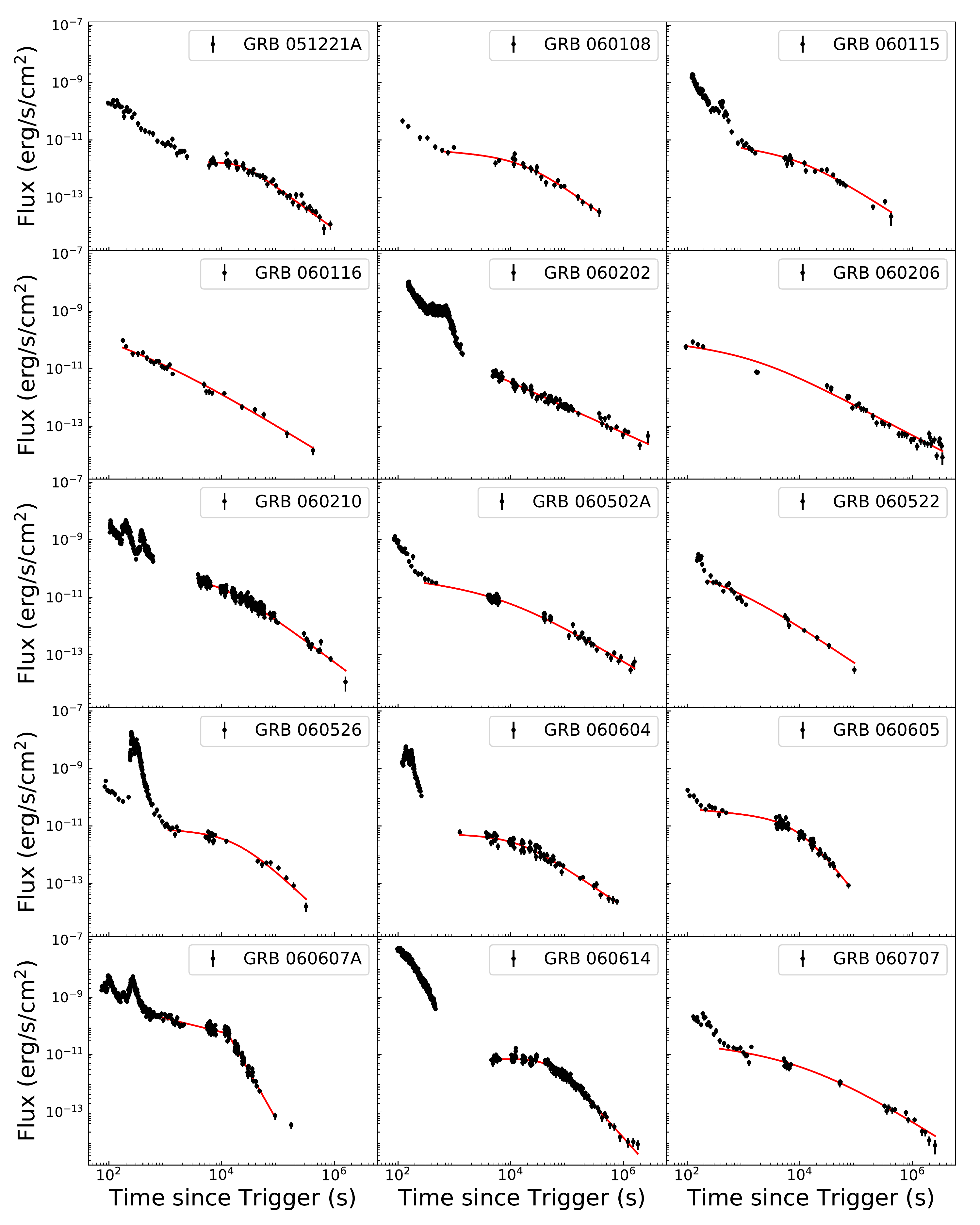}
	\caption{--- Continued}
\end{figure}

\begin{figure}[ht!]
	\centering
	\addtocounter{figure}{-1}
	\includegraphics[width=7.0in]{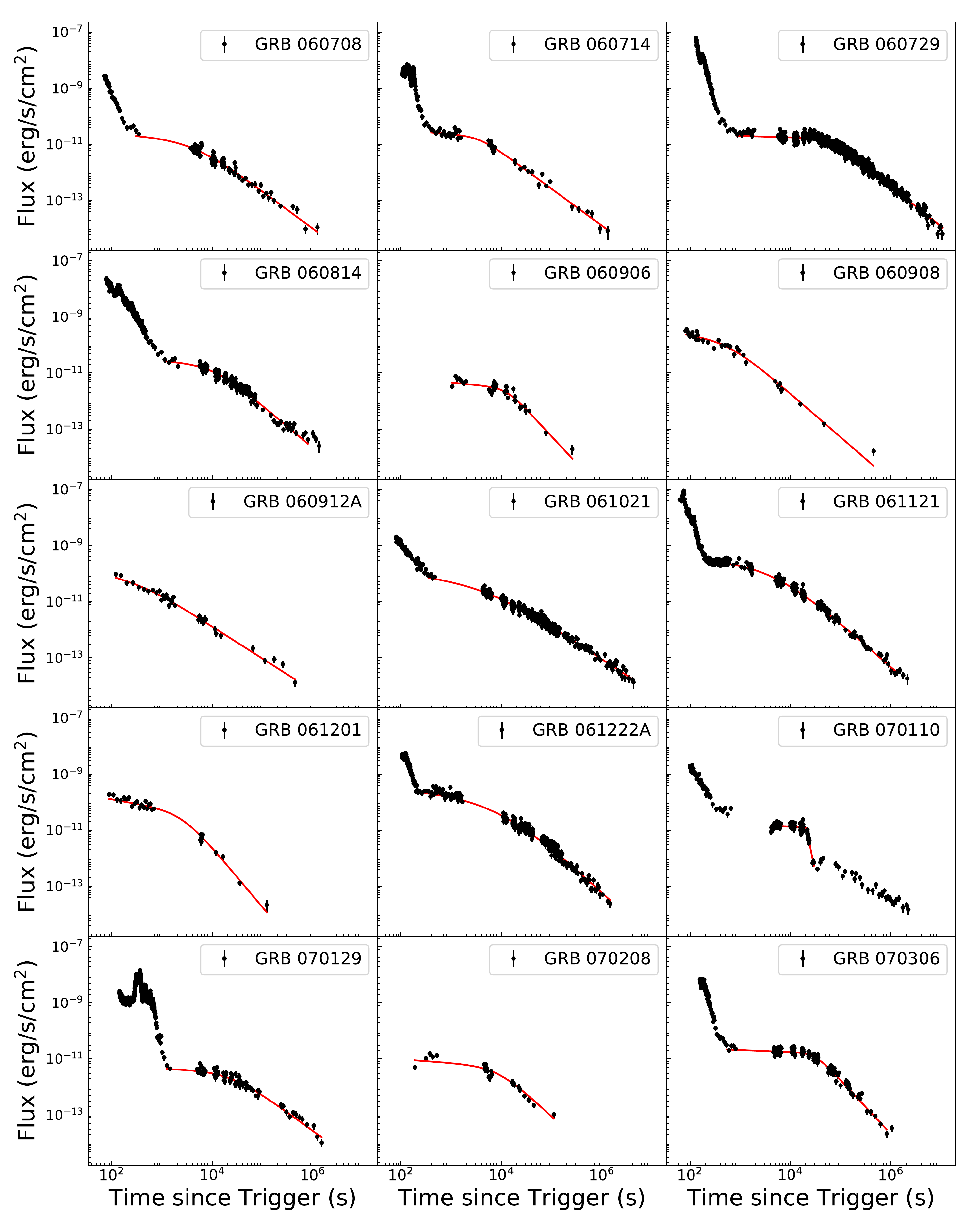}
	\caption{--- Continued}
\end{figure}

\begin{figure}[ht!]
	\centering
	\addtocounter{figure}{-1}
	\includegraphics[width=7.0in]{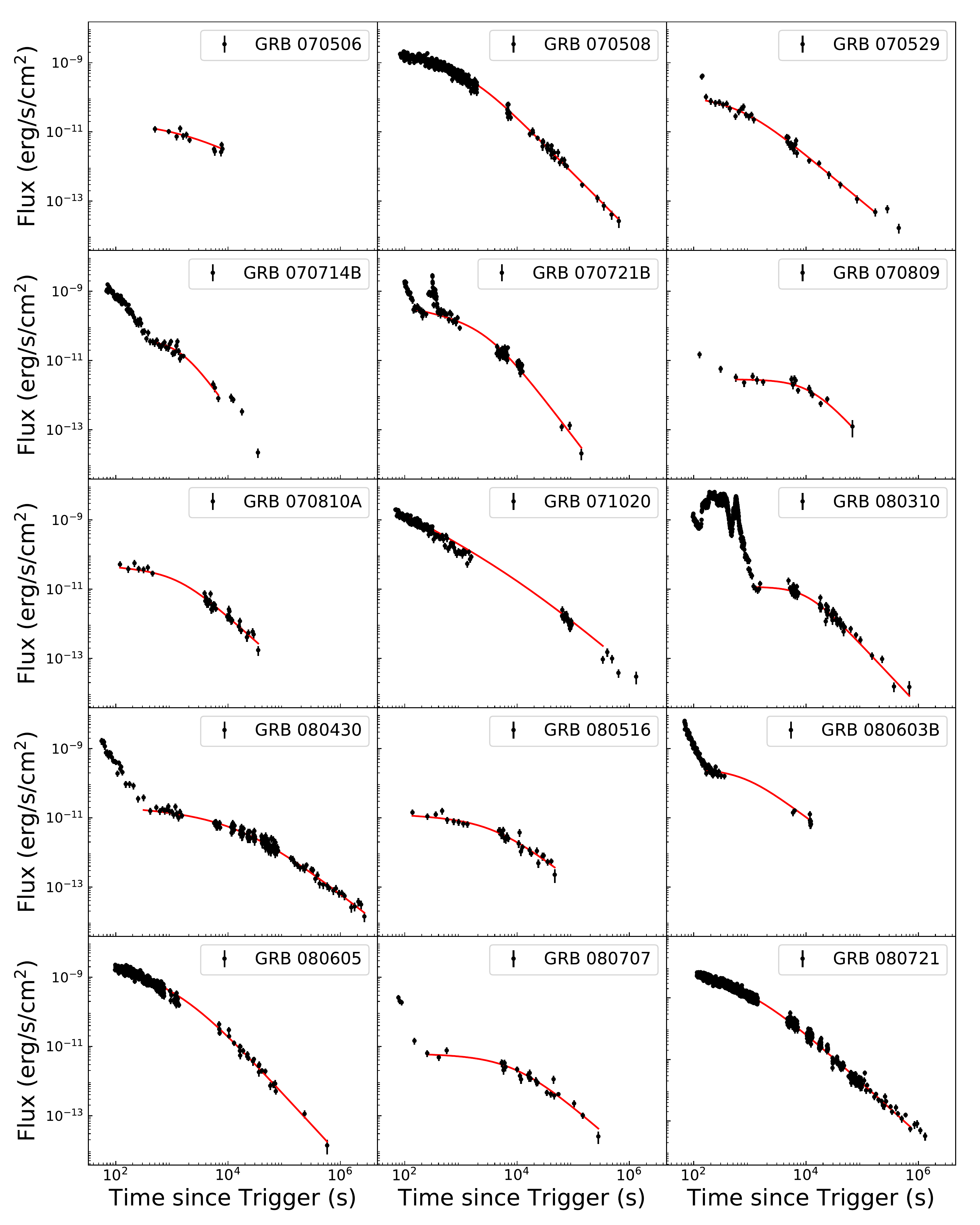}
	\caption{--- Continued}
\end{figure}

\begin{figure}[ht!]
	\centering
	\addtocounter{figure}{-1}
	\includegraphics[width=7.0in]{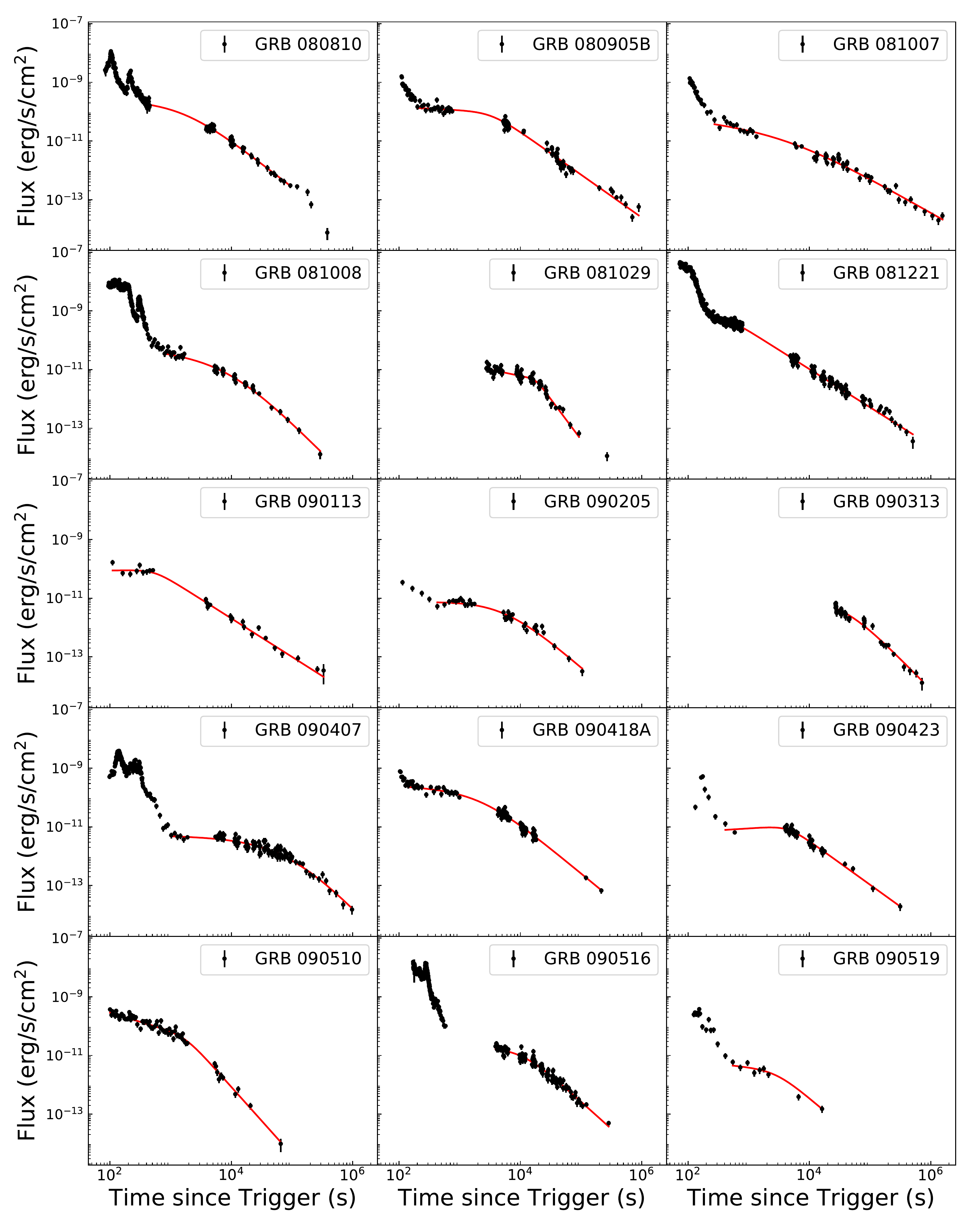}
	\caption{--- Continued}
\end{figure}

\begin{figure}[ht!]
	\centering
	\addtocounter{figure}{-1}
	\includegraphics[width=7.0in]{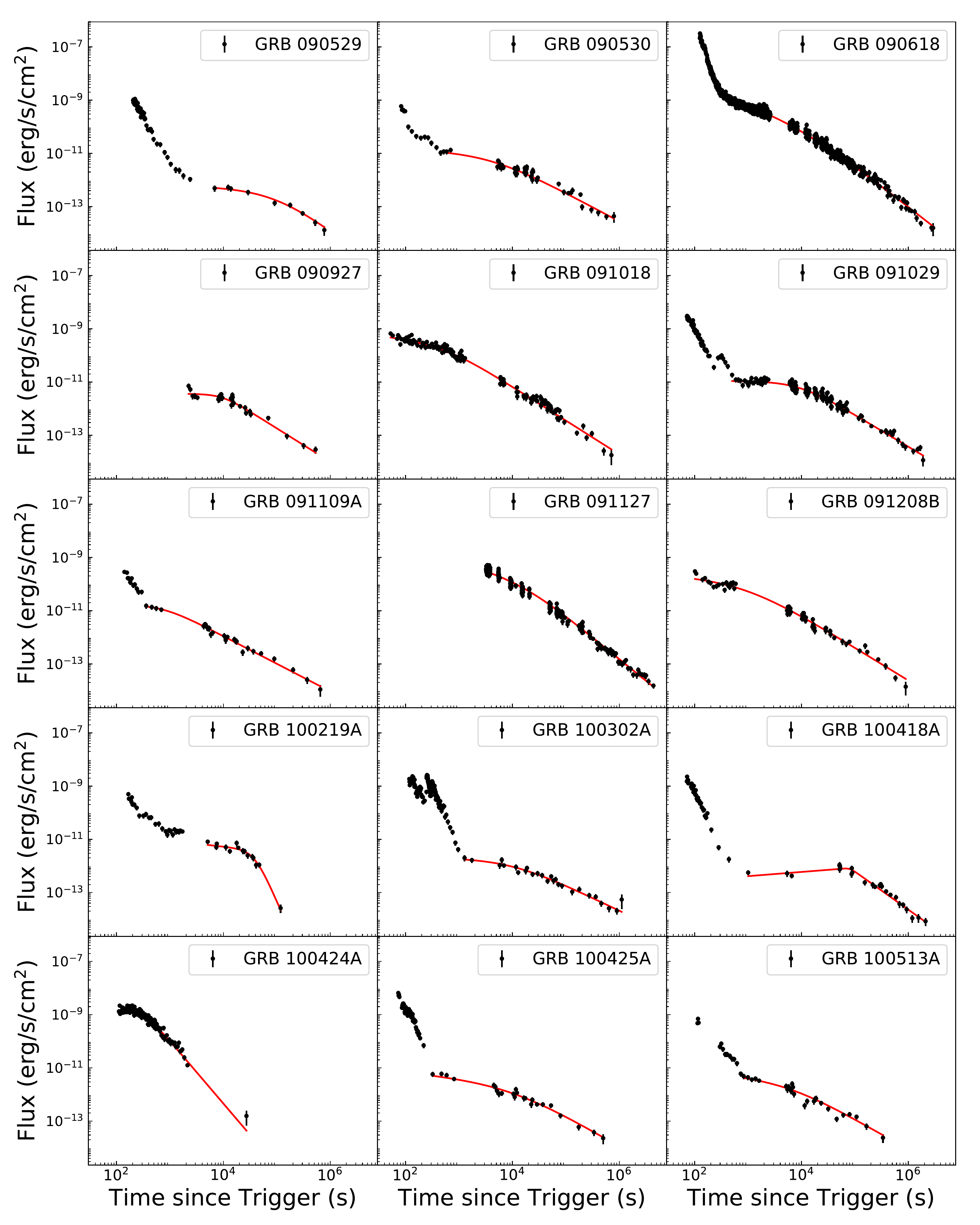}
	\caption{--- Continued}
\end{figure}

\begin{figure}[ht!]
	\centering
	\addtocounter{figure}{-1}
	\includegraphics[width=7.0in]{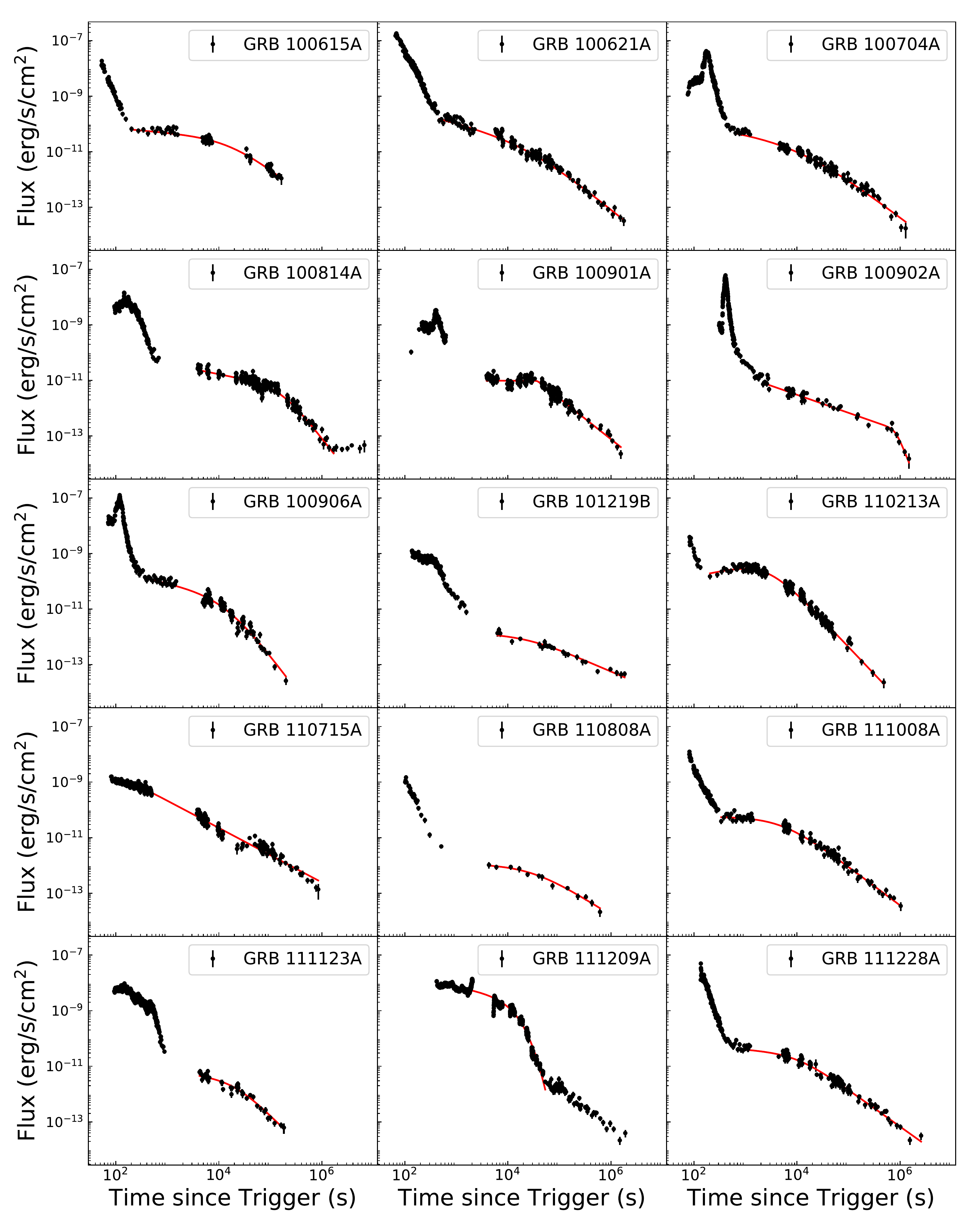}
	\caption{--- Continued}
\end{figure}

\begin{figure}[ht!]
	\centering
	\addtocounter{figure}{-1}
	\includegraphics[width=7.0in]{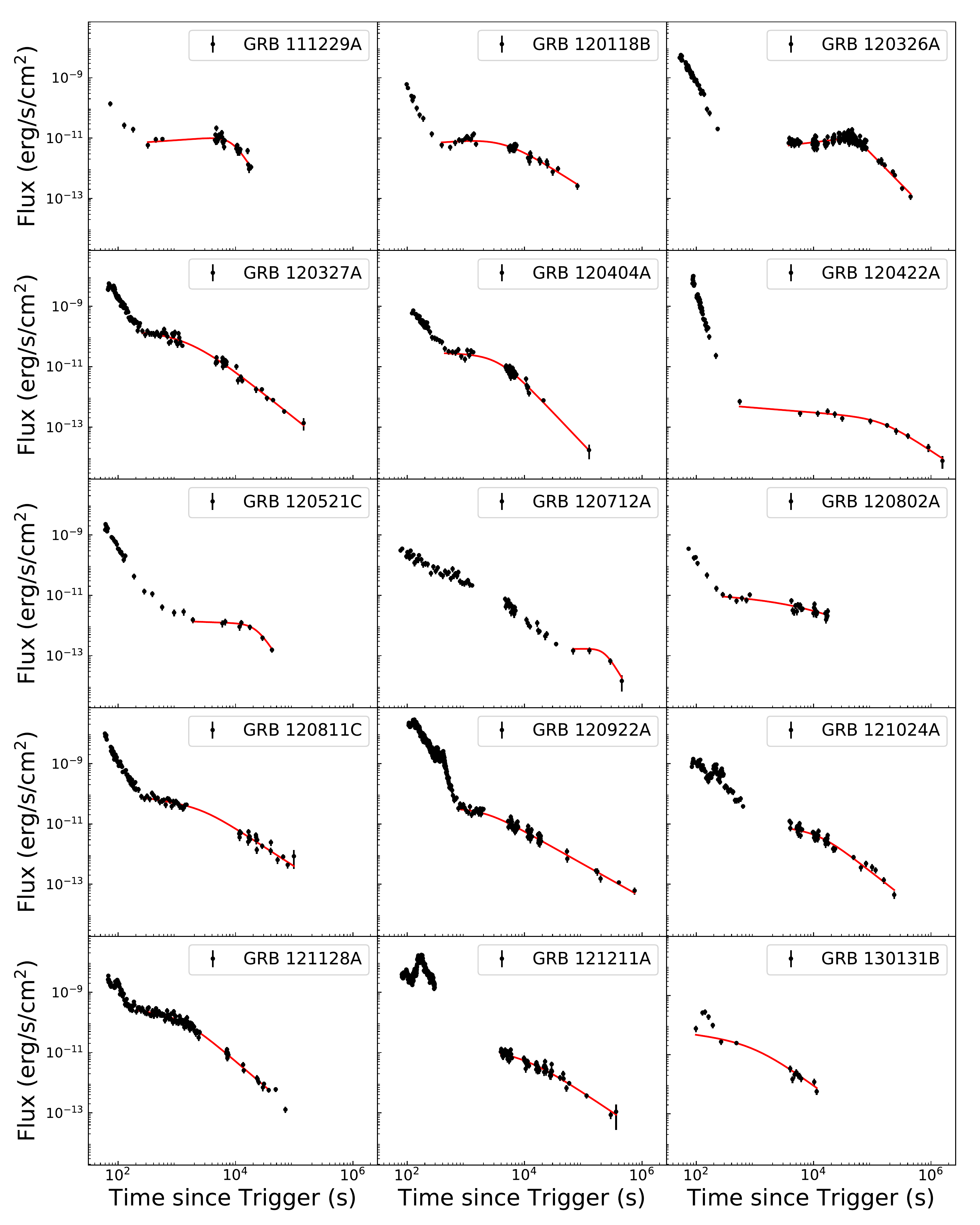}
	\caption{--- Continued}
\end{figure}

\begin{figure}[ht!]
	\centering
	\addtocounter{figure}{-1}
	\includegraphics[width=7.0in]{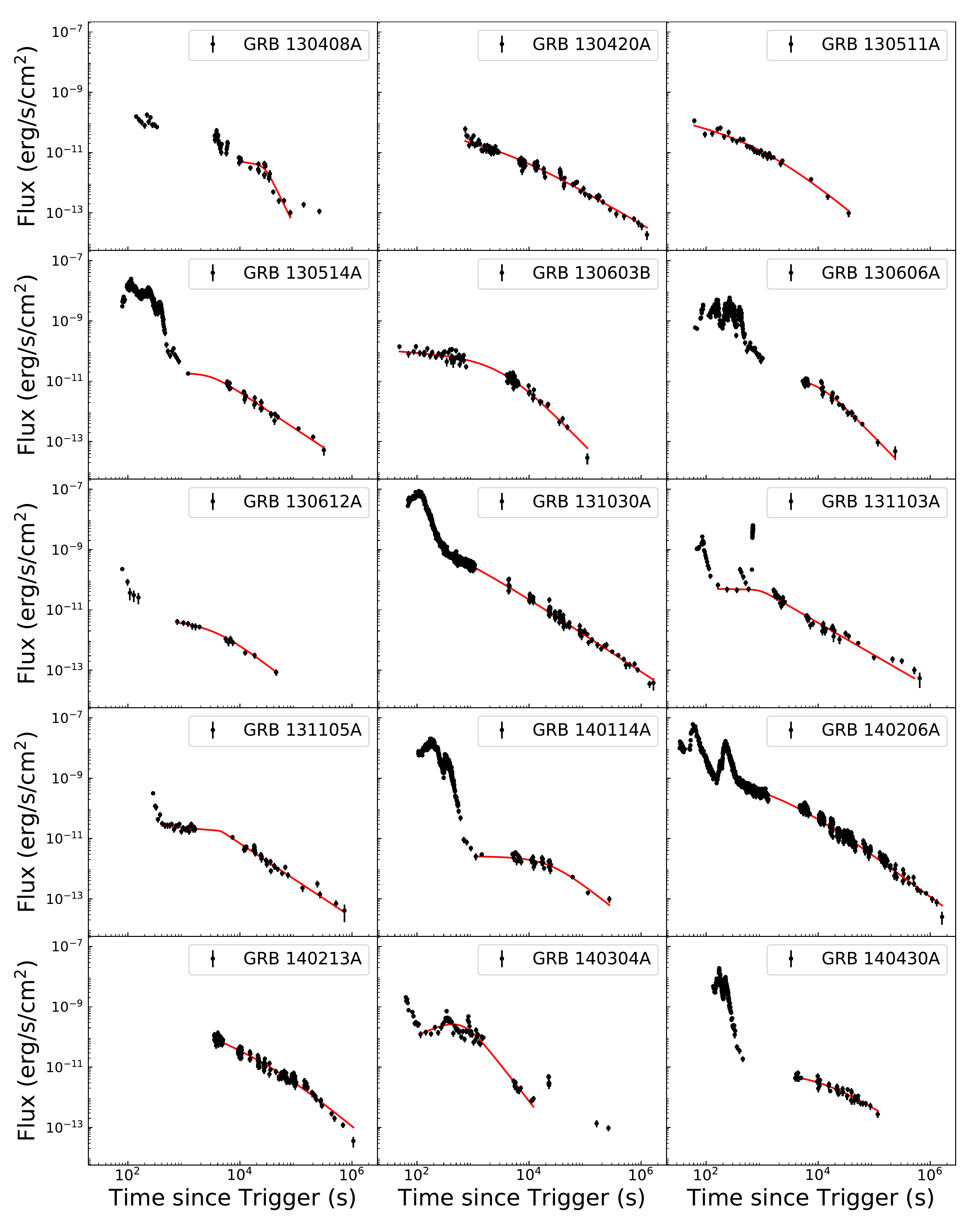}
	\caption{--- Continued}
\end{figure}

\begin{figure}[ht!]
	\centering
	\addtocounter{figure}{-1}
	\includegraphics[width=7.0in]{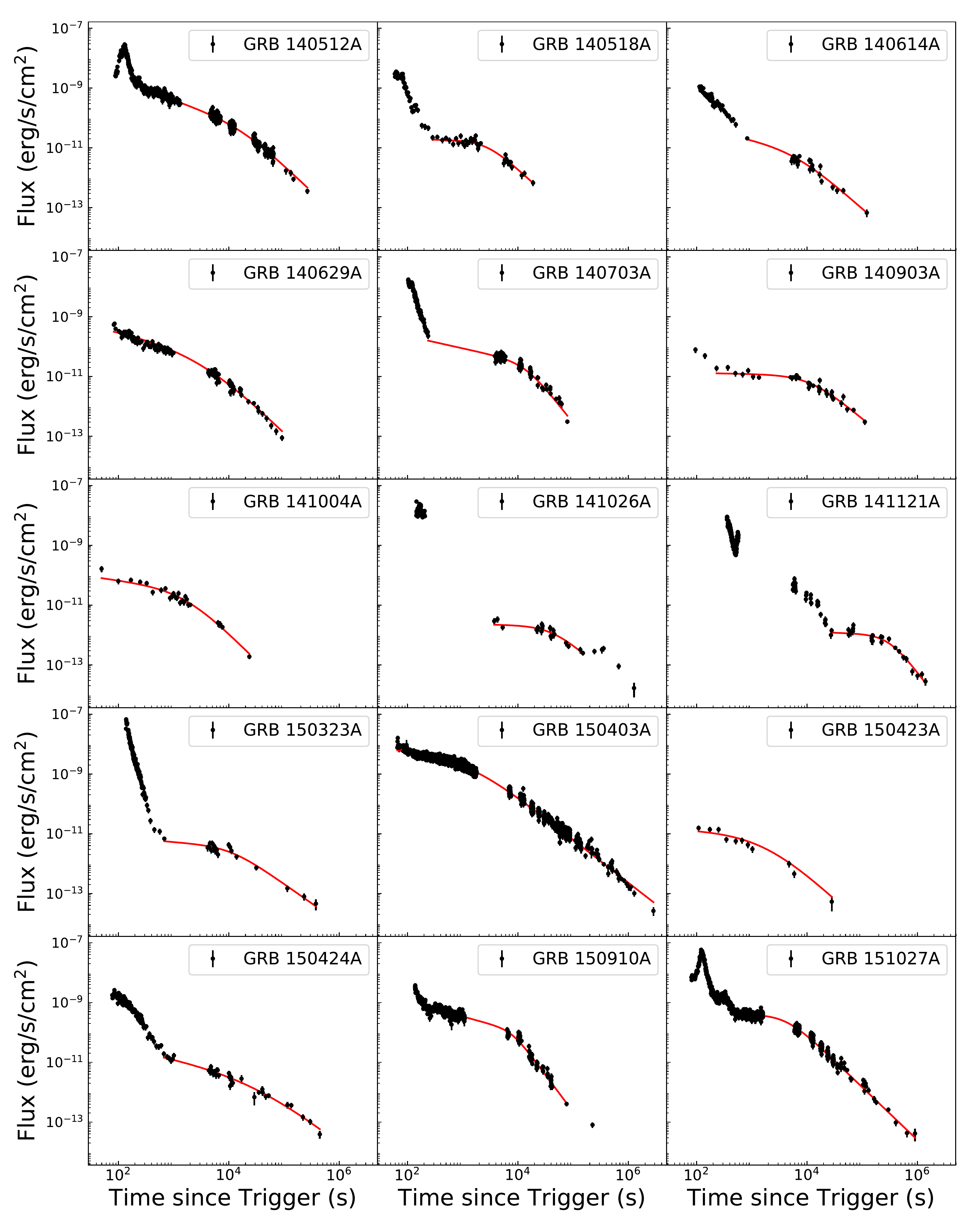}
	\caption{--- Continued}
\end{figure}

\begin{figure}[ht!]
	\centering
	\addtocounter{figure}{-1}
	\includegraphics[width=7.0in]{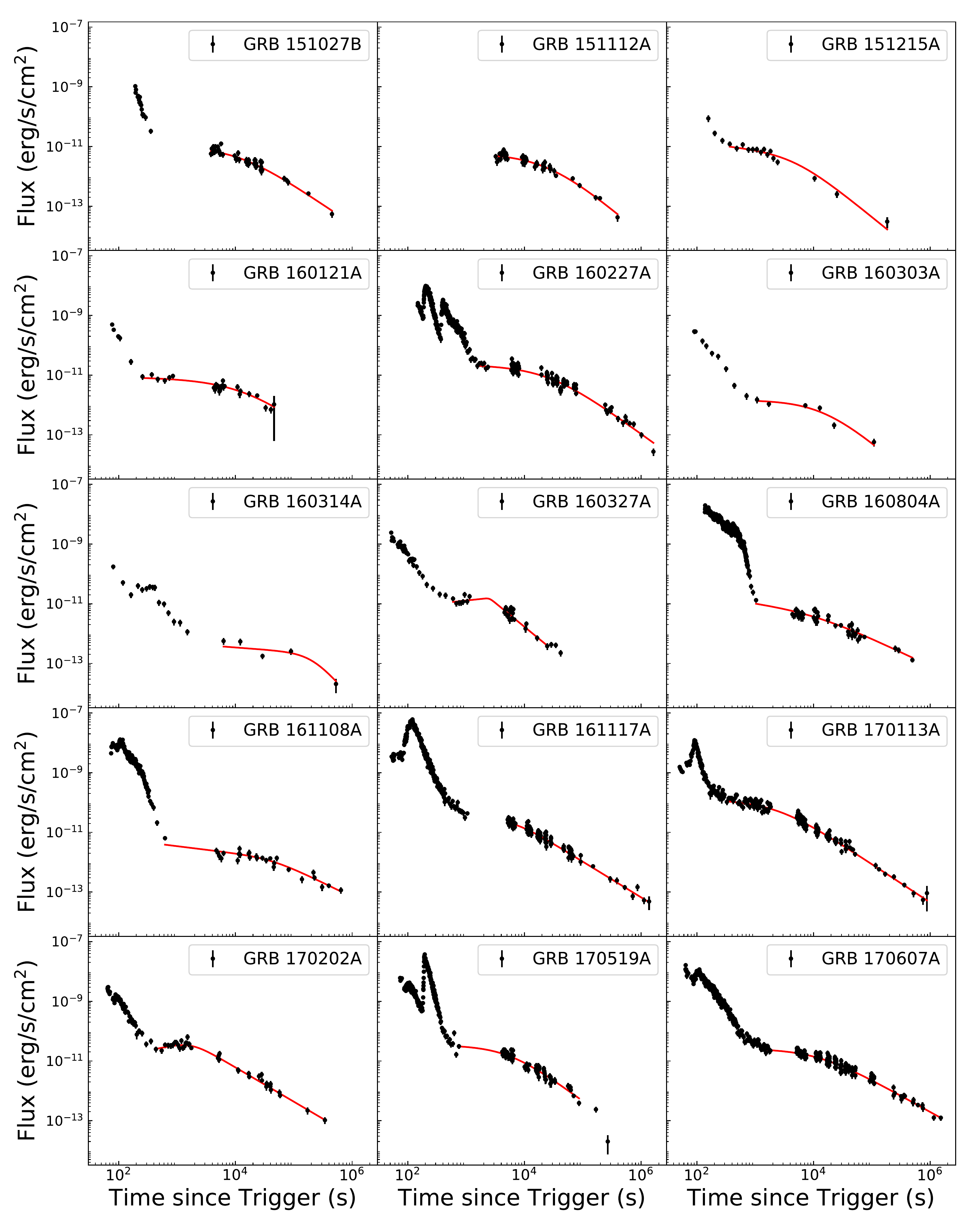}
	\caption{--- Continued}
\end{figure}

\begin{figure}[ht!]
	\centering
	\addtocounter{figure}{-1}
	\includegraphics[width=7.0in]{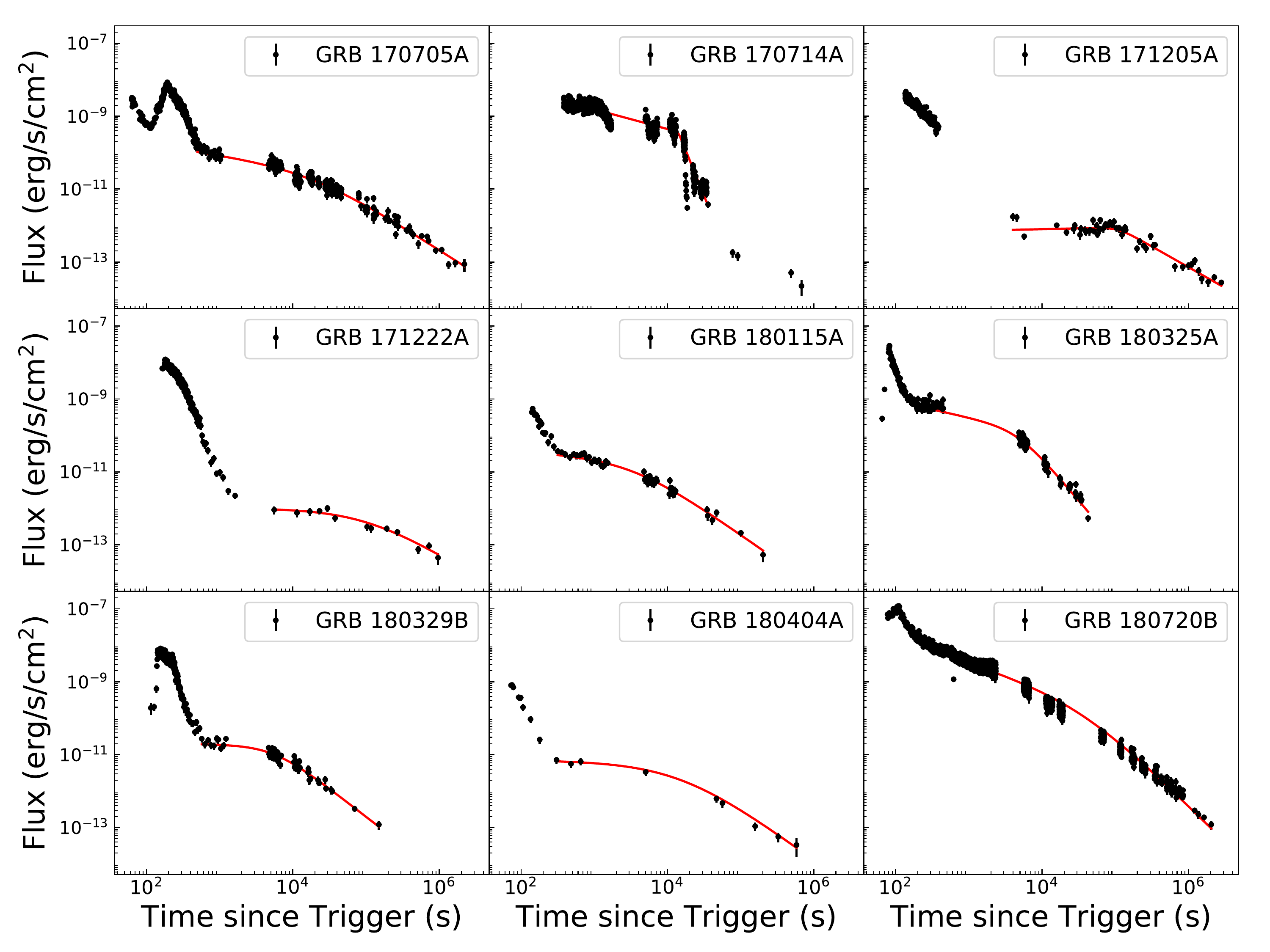}
	\caption{--- Continued}
\end{figure}

\begin{figure}[htbp]
	\centering
	\subfloat{
		\label{fig:hist_logT}
		\includegraphics[width=0.5\linewidth]{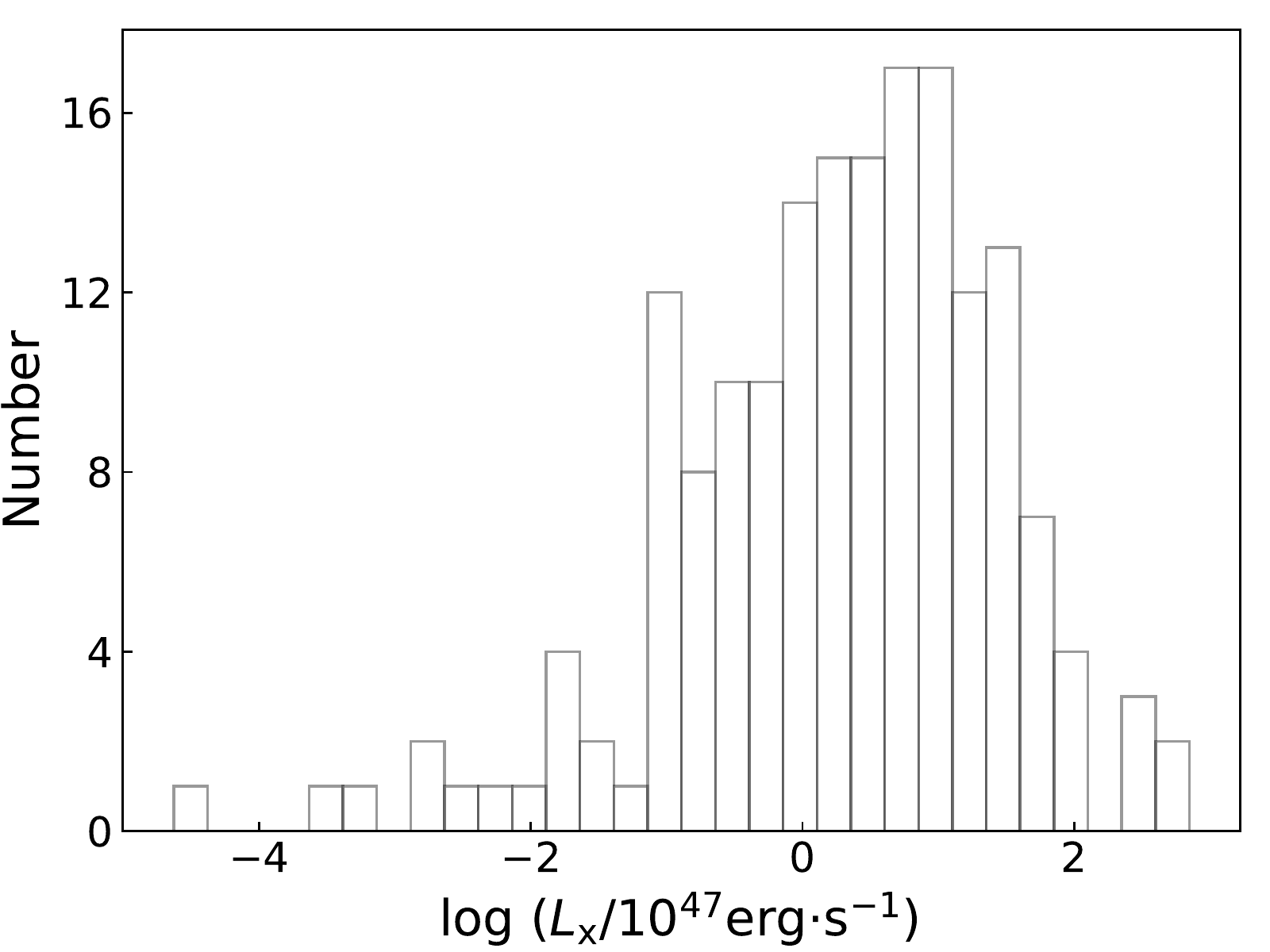}
	}
	
	\subfloat{
		\label{fig:hist_logL}
		\includegraphics[width=0.5\linewidth]{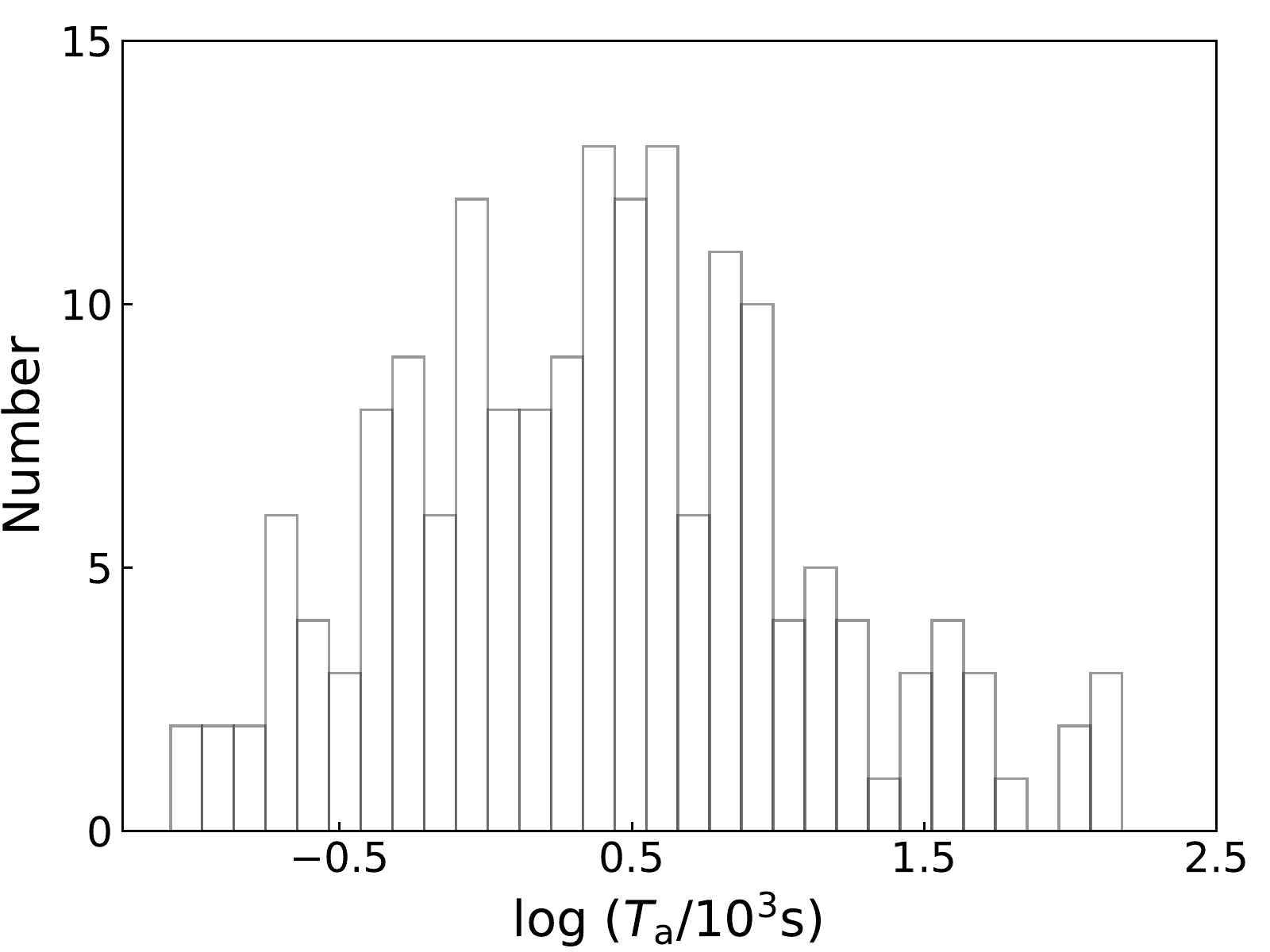}
	}
	
	\subfloat{
		\label{fig:hist_logE}
		\includegraphics[width=0.5\linewidth]{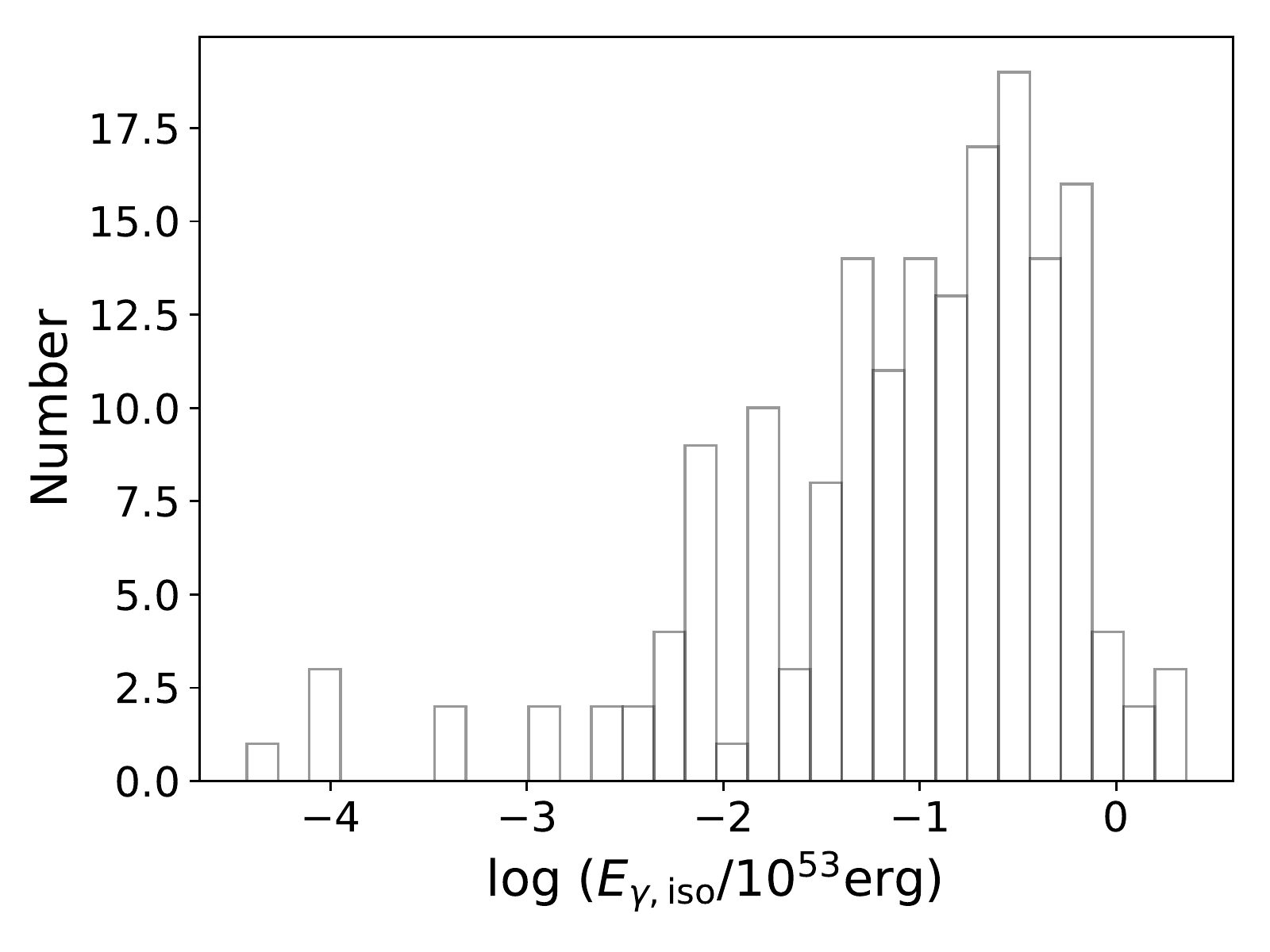}
	}
	\caption{Histograms of $L_{X}$, $T_{a}$, and $E_{\gamma,\rm{iso}}$ of our sample.
             The typical value of $L_{\rm x}$ is $2.6\times 10^{47}$ erg$\cdot$s$^{-1}$ with a 50\%
             distribution range of (0.4, 12)$\times 10^{47}$ erg$\cdot$s$^{-1}$. The typical value of
             $T_{a}$ is $2.4\times 10^{3}$ s with a 50\% distribution range of (0.8, 7)$\times 10^{3}$ s.
             The typical value of $E_{\gamma,\rm{iso}}$ is $0.13\times 10^{53}$ s, with a 50\%
             distribution range of (0.04, 0.3)$\times 10^{53}$ erg.}
	\label{fig:hist1}
\end{figure}

\begin{figure}[htbp]
	\centering
	\subfloat{
		\label{fig:hist_logS}
		\includegraphics[width=0.5\linewidth]{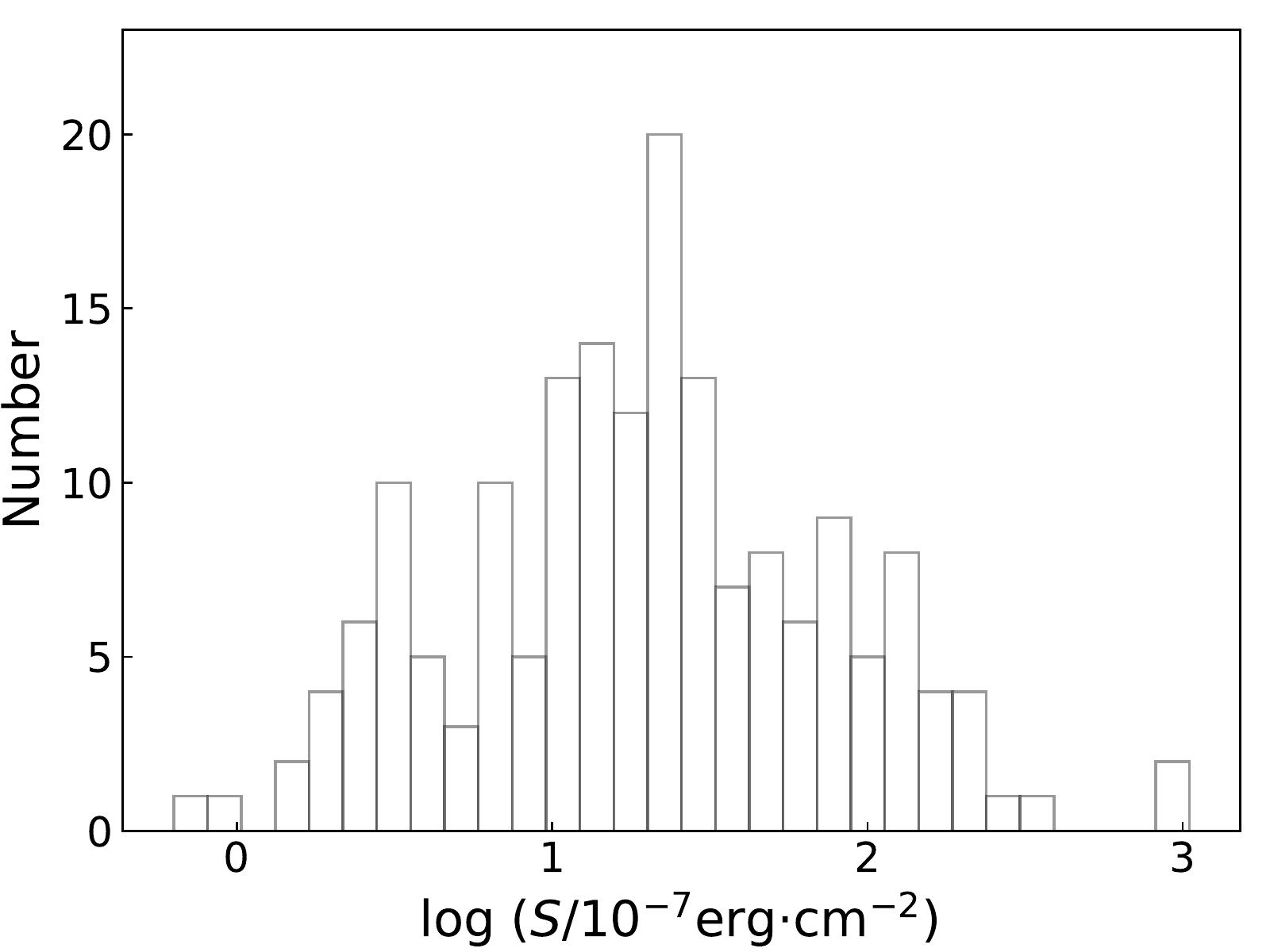}
	}
	
	\subfloat{
		\label{fig:hist_z}
		\includegraphics[width=0.5\linewidth]{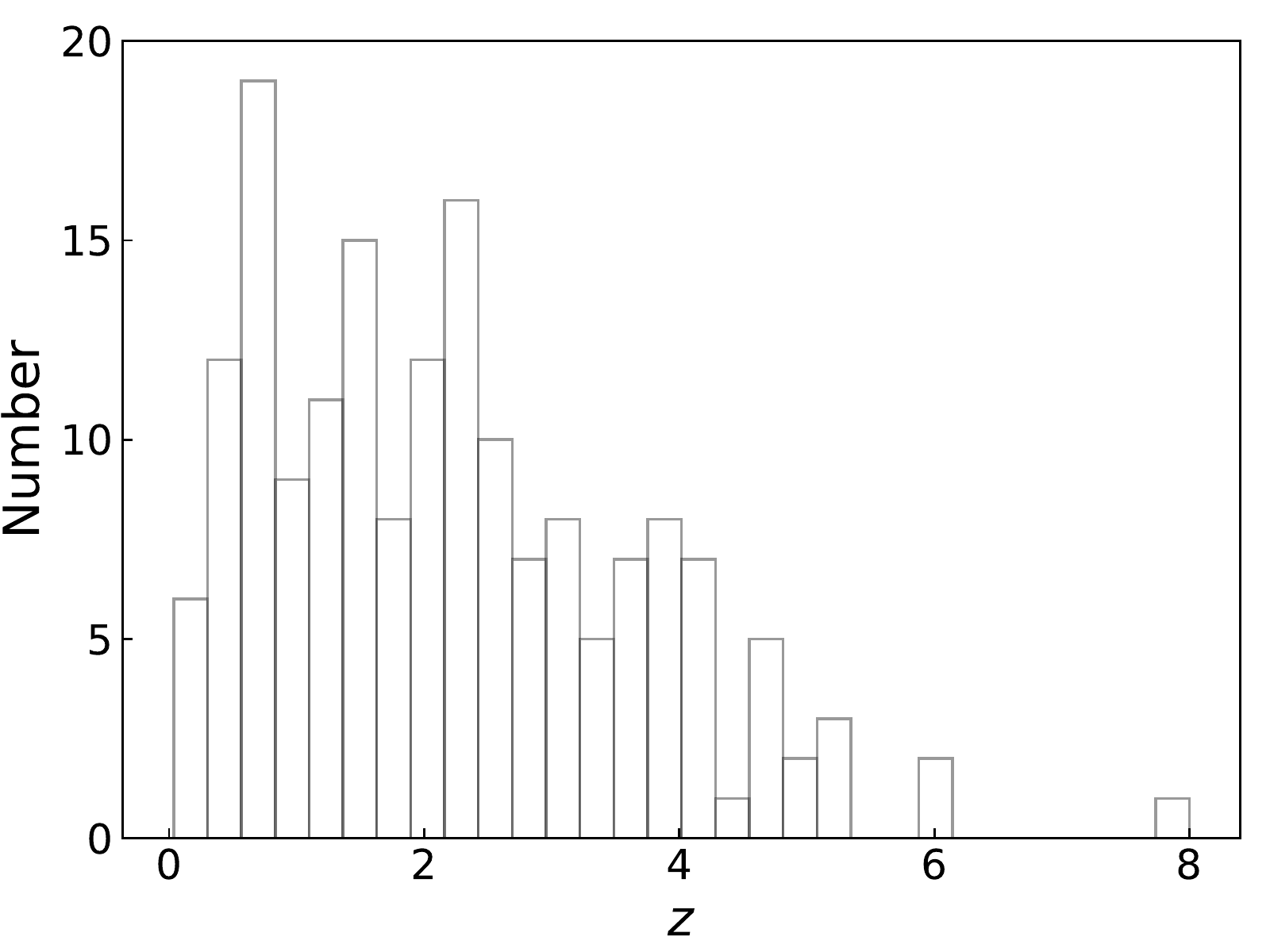}
	}
	
	\subfloat{
		\label{fig:hist_T90}
		\includegraphics[width=0.5\linewidth]{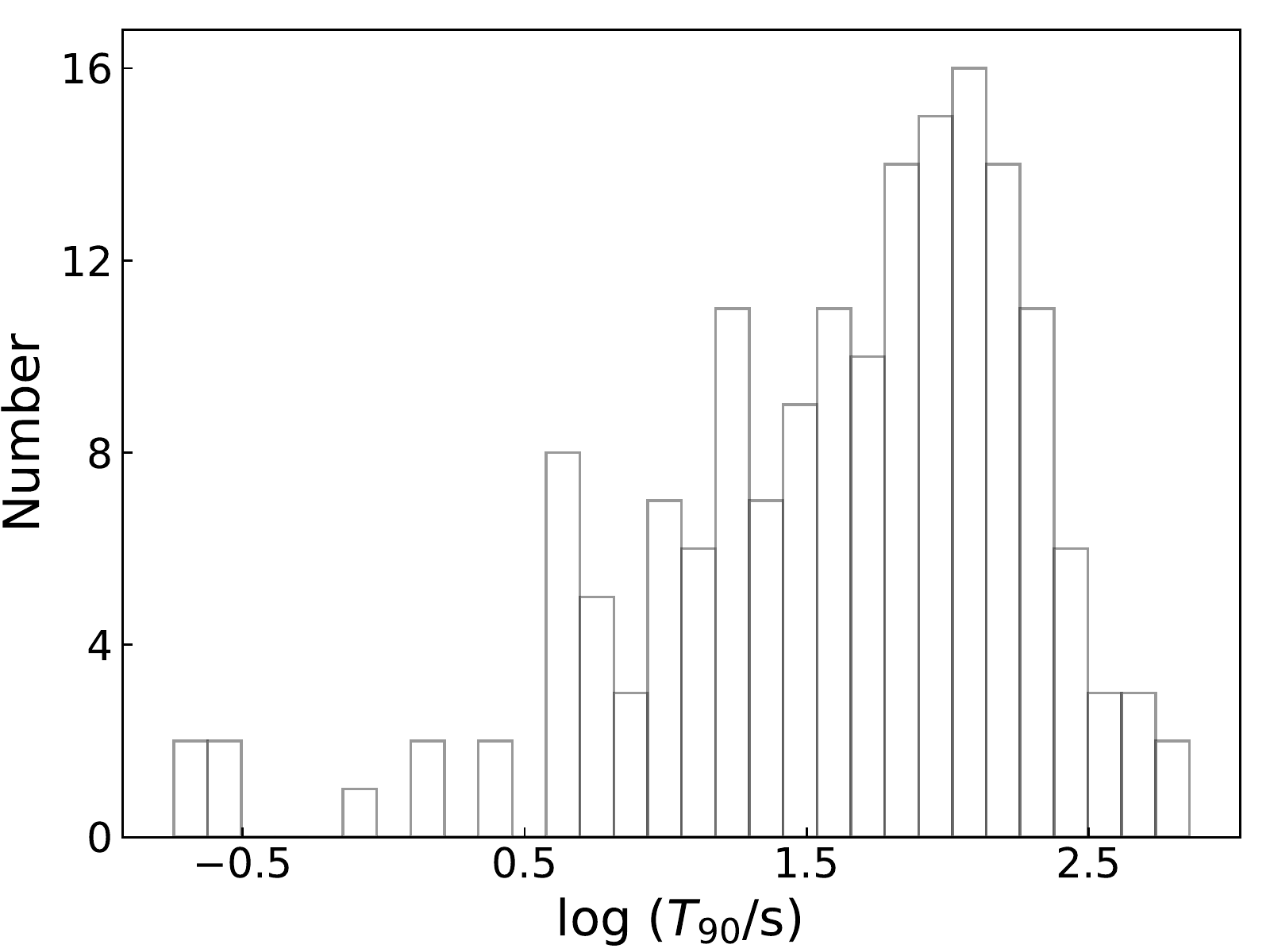}
	}
	\caption{Histograms of $S$, $z$, and $T_{90}$ of our sample. The typical value of $S$ is
             $20\times 10^{-7}$ erg$\cdot$cm$^{-2}$, with a 50\% distribution range of
             (8, 50)$\times 10^{-7}$ erg$\cdot$cm$^{-2}$. The typical value of $z$ is 2.0,
             with a 50\% distribution range of (1.0, 3.2). The typical value of $T_{90}$ is 58 s,
             with a 50\% distribution range of (18, 123) s.}
	\label{fig:hist2}
\end{figure}

\begin{figure}[htbp]
	\centering
	\subfloat{
		\label{fig:hist_alpha1}
		\includegraphics[width=0.5\linewidth]{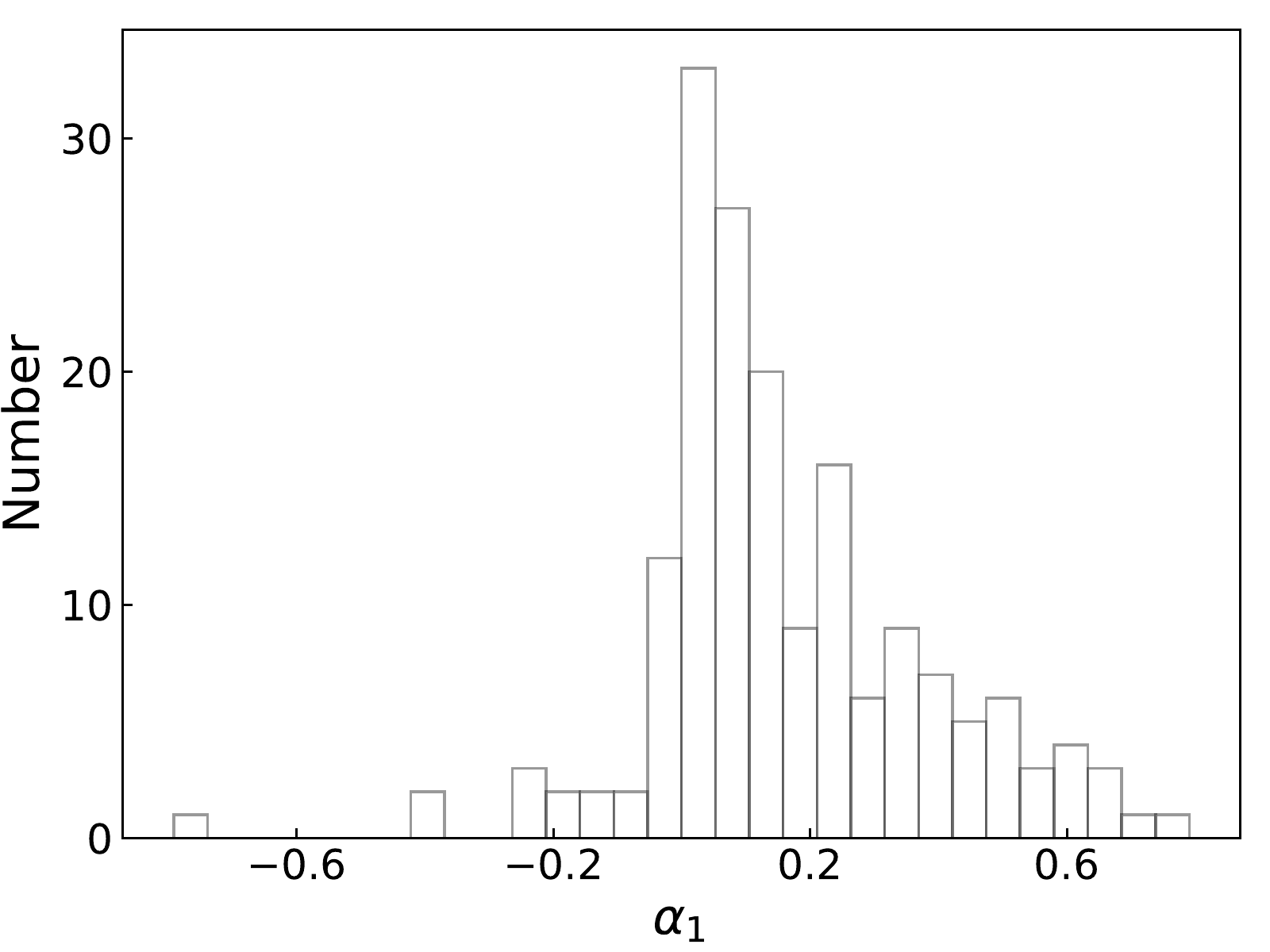}
	}
	
	\subfloat{
		\label{fig:hist_alpha2}
		\includegraphics[width=0.5\linewidth]{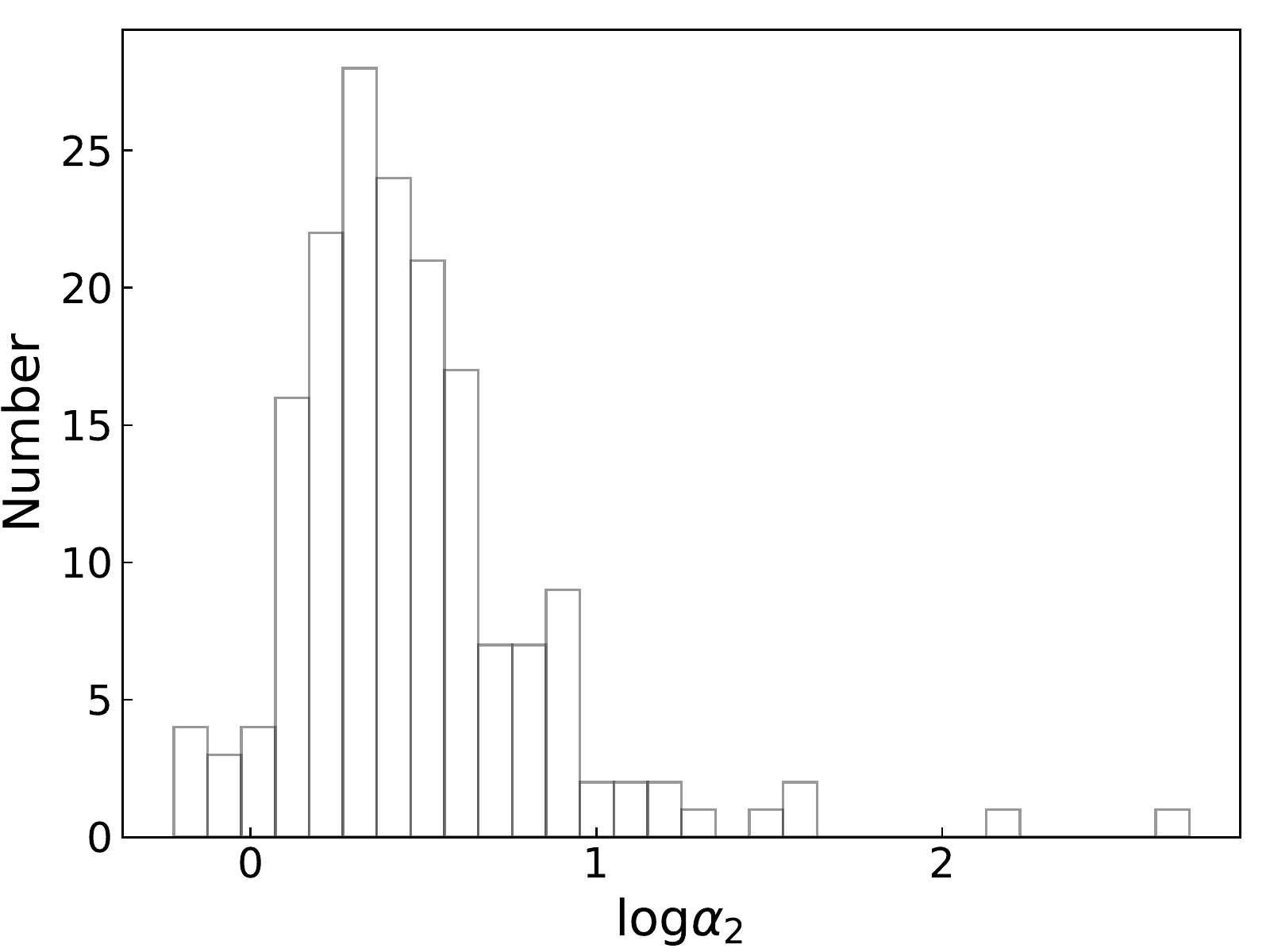}
	}
	
	\subfloat{
		\label{fig:hist_deltalpha}
		\includegraphics[width=0.5\linewidth]{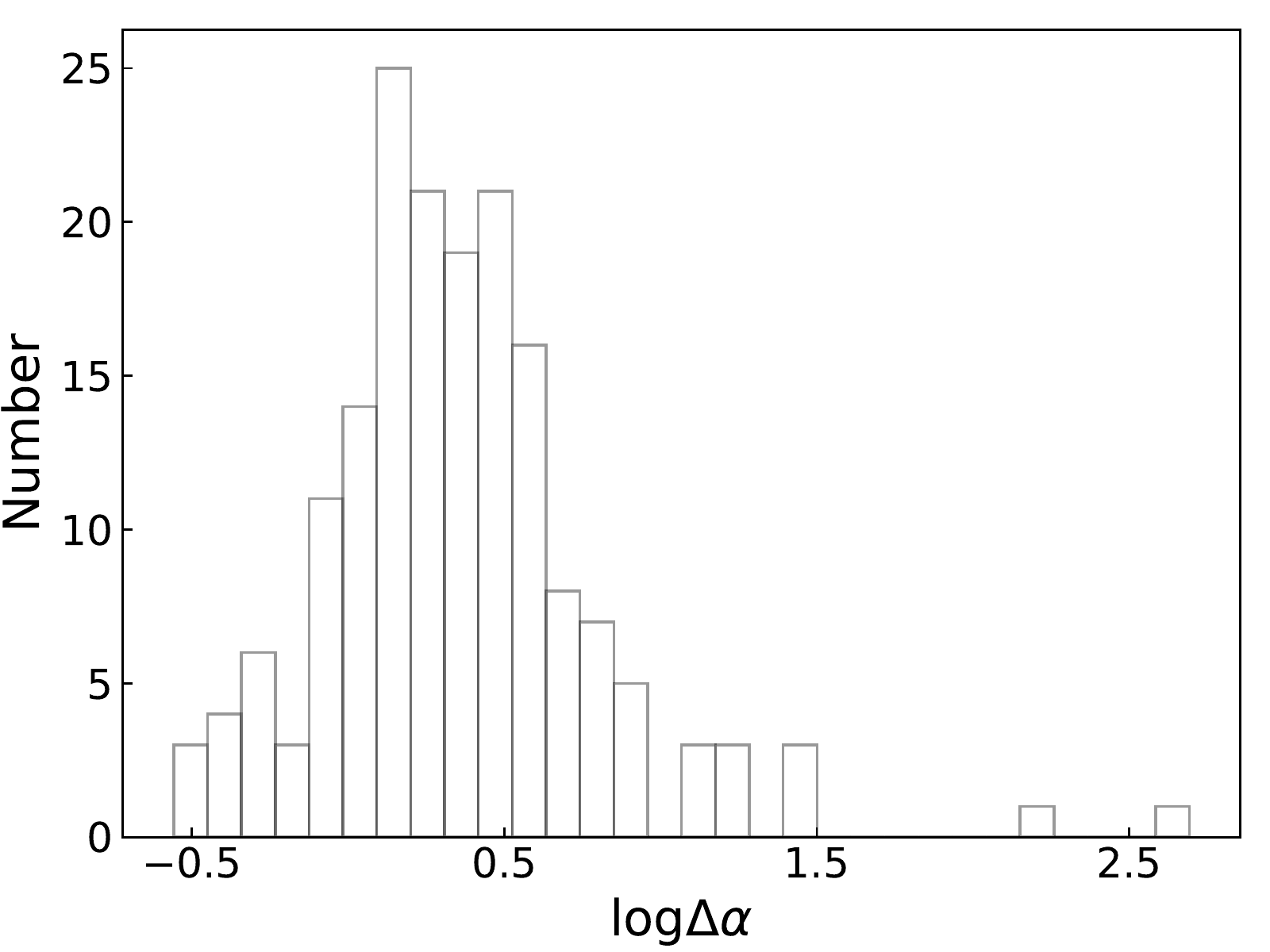}
	}
	\caption{Histograms of $\alpha_{1}$, $\alpha_{2}$, and $\Delta \alpha$ of our sample. The typical
             value of $\alpha_{1}$ is 0.11, with a 50\% distribution range of (0.03, 0.3). The typical
             value of $\alpha_{2}$ is 1.5, with a 50\% distribution range of (1.3, 1.9). The typical
             value of $\Delta \alpha$ is 1.4, with a 50\% distribution range of (1.1, 1.7).}
	\label{fig:hist3}
\end{figure}

\begin{figure}[htbp]
	\centering
	\subfloat[]{
		\label{fig:2D_L-T}
		\includegraphics[width=0.5\linewidth]{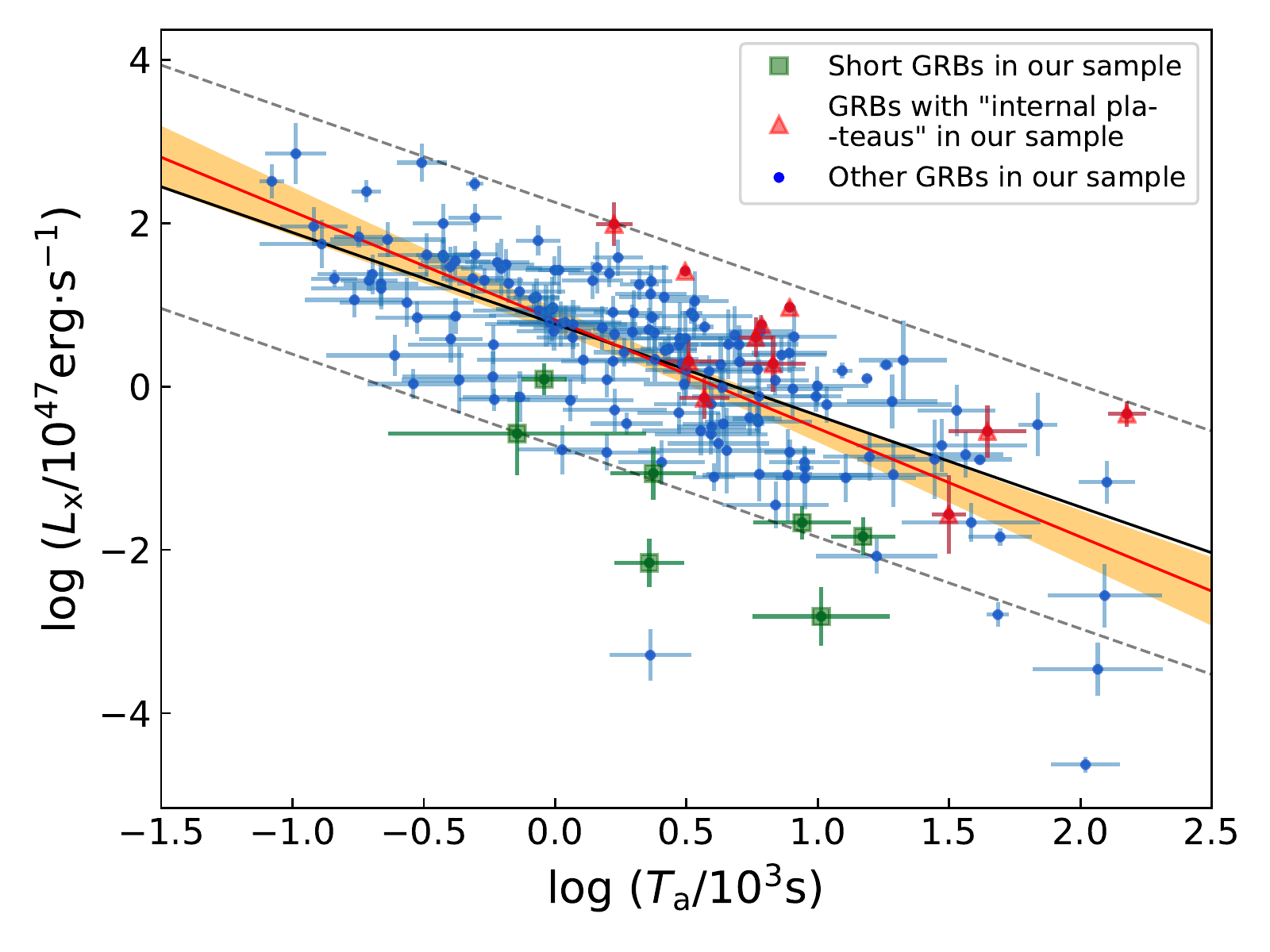}
	}
	\subfloat[]{
		\label{fig:2D_L-E}
		\includegraphics[width=0.5\linewidth]{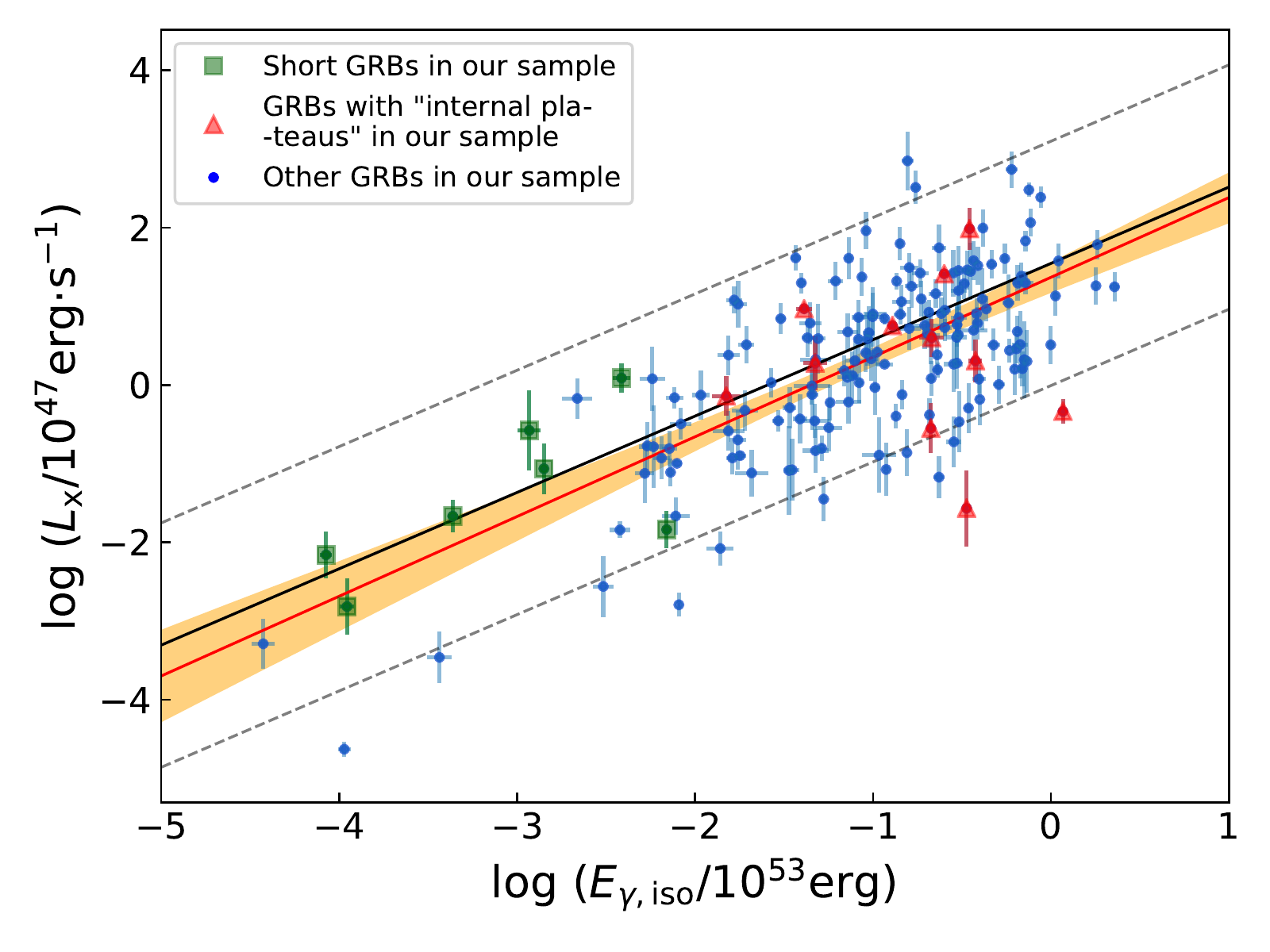}
	}
	
	\subfloat[]{
		\label{fig:2D_S-T90}
		\includegraphics[width=0.5\linewidth]{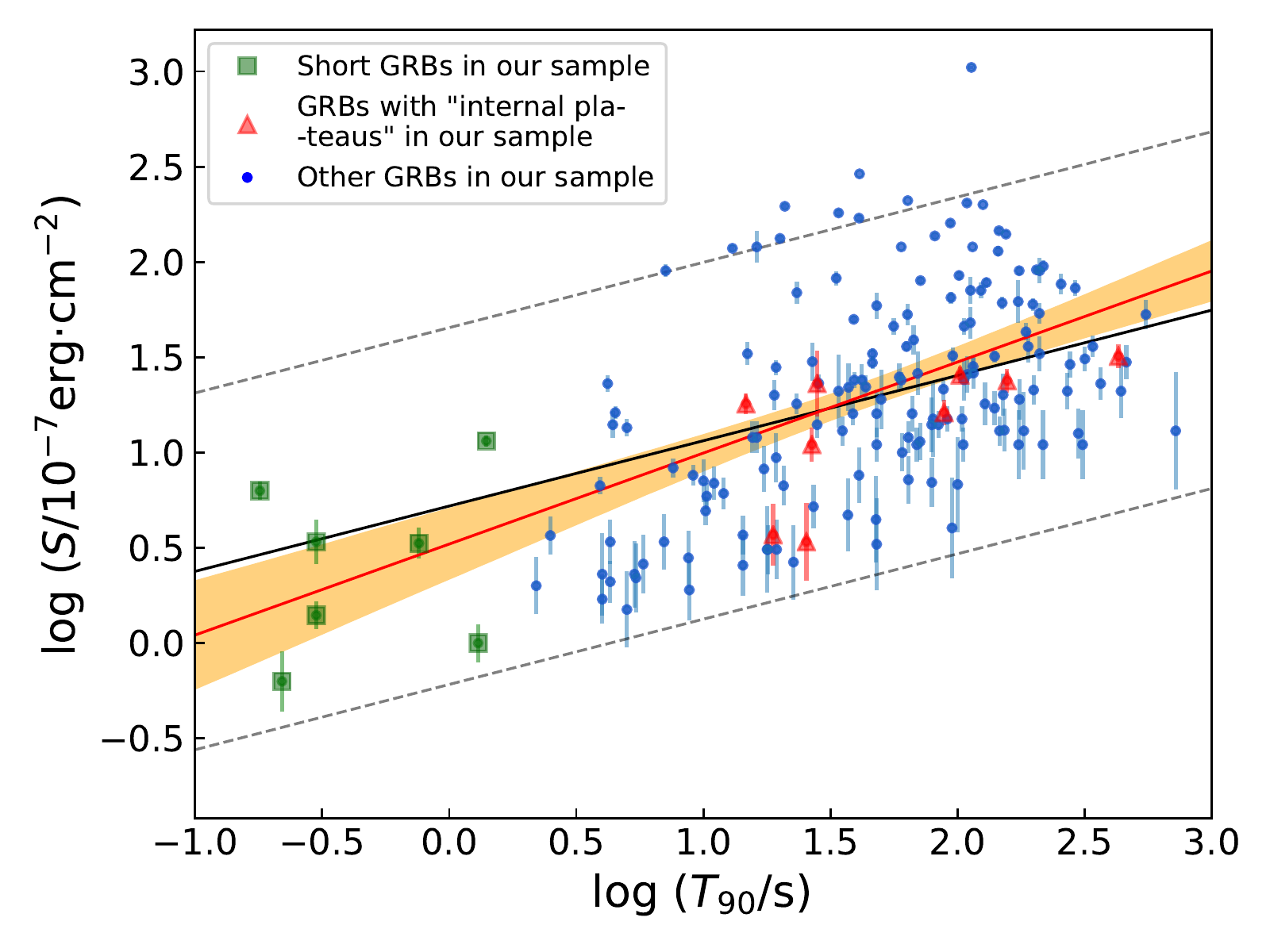}
	}
	\subfloat[]{
		\label{fig:2D_E-T90}
		\includegraphics[width=0.5\linewidth]{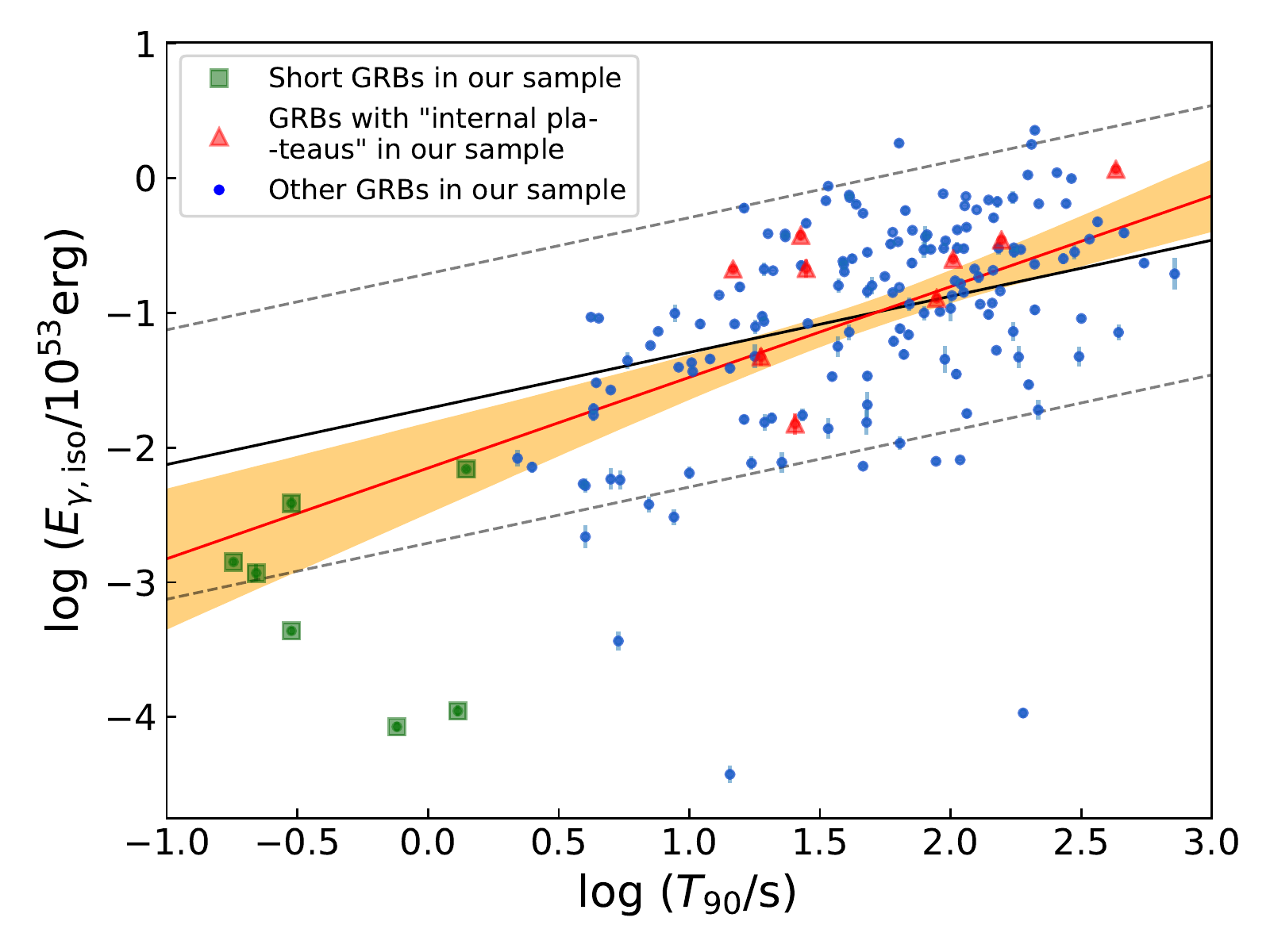}
	}
	\caption{Logarithmic plots of various two-parameter correlations:
            (a) $L_{X}$ - $T_{a}$, (b) $L_{X}$ - $E_{\gamma,\rm{iso}}$, (c) $S$ - $T_{90}$, and
            (d) $E_{\gamma,\rm{iso}}$ - $T_{90}$. 7 short GRBs are marked with green squares,
            while 11 GRBs with ``internal plateaus'' are marked with red triangles. The black solid
            lines are the best fit for the observational data points by the least square linear
            regression. The black dashed lines are 3$\sigma$ confidence
            range. The red solid lines are the best fits by using the bivariate Bayesian linear
            regression, and the orange regions show the 3$\sigma$ ranges correspondingly.}
	\label{fig:2D}
\end{figure}

\begin{figure}[htbp]
	\centering
	\subfloat[]{
		\label{fig:2D_L-z}
		\includegraphics[width=0.5\linewidth]{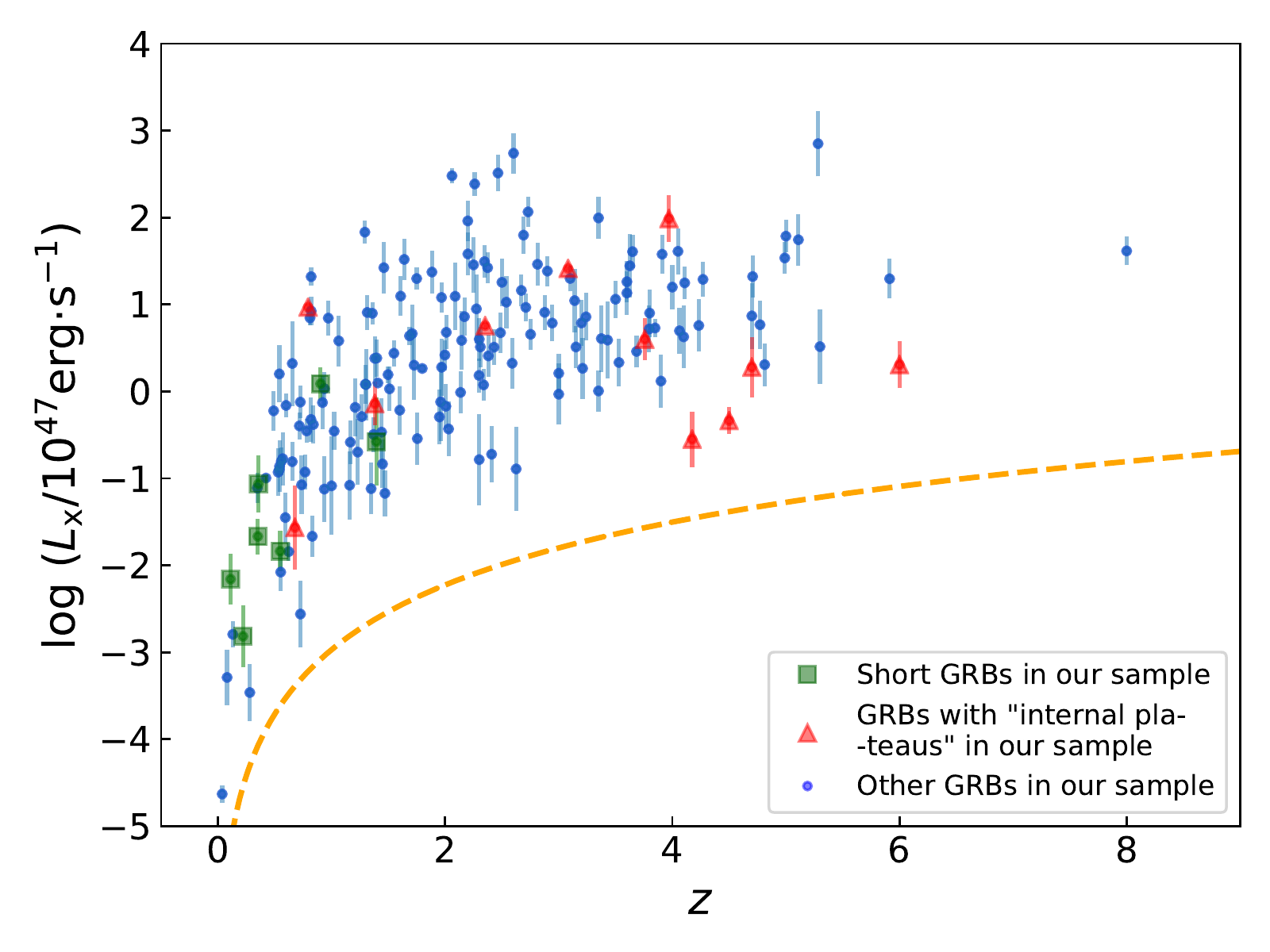}
	}

	\subfloat[]{
		\label{fig:2D_E-z}
		\includegraphics[width=0.5\linewidth]{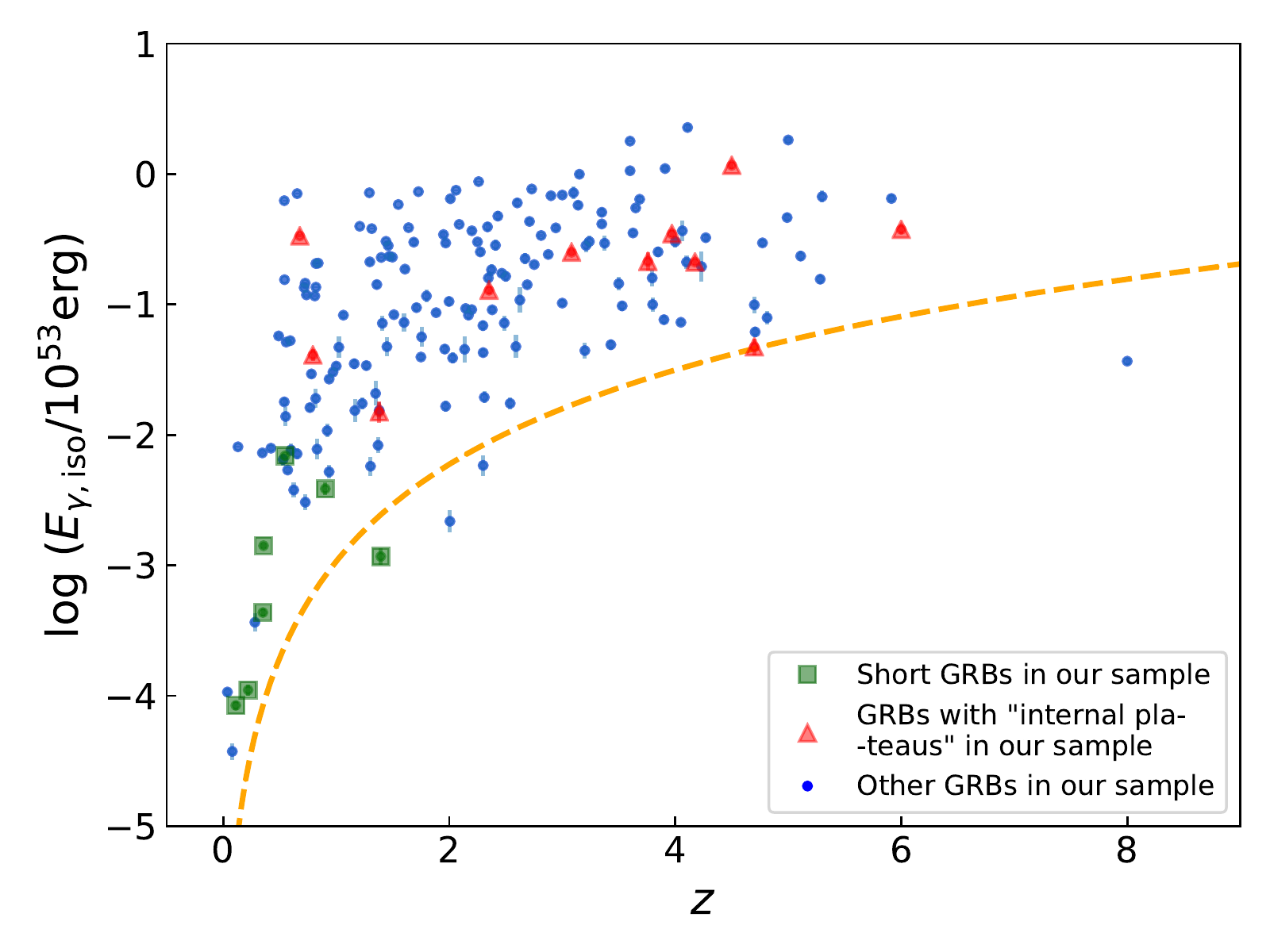}
	}

	\subfloat[]{
		\label{fig:2D_T-z}
		\includegraphics[width=0.5\linewidth]{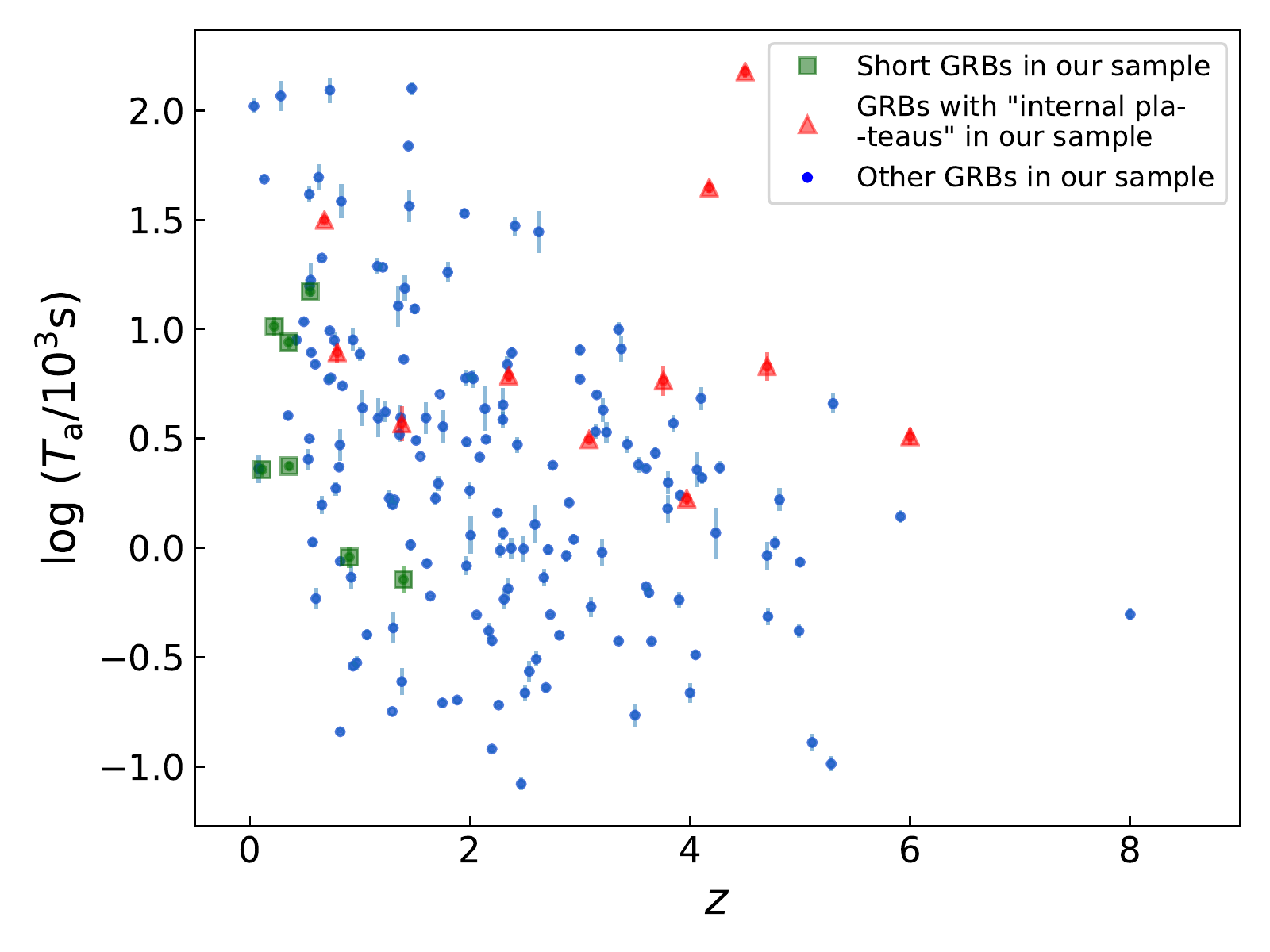}
	}	
	\caption{Dependence of $L_{X}$, $E_{\gamma, \rm{iso}}$, and $T_{a}$ on redshifts. Panel (a) shows $L_{X}$ versus $z$. The dashed line represents the XRT detection limit of $f_{\rm lim}\sim 2\times 10^{-14}$ erg cm$^{-2}$ s$^{-1}$. Panel (b) shows $E_{\gamma, \rm{iso}}$ versus $z$. The dashed line represents the BAT detection limit of $f_{\rm lim}\sim 10^{-8}$ erg cm$^{-2}$ s$^{-1}$ by assuming a GRB duration of 2 s. Panel (c) illustrates $T_{a}$ versus $z$.}
	\label{fig:2D_seleceff}
\end{figure}

\begin{figure}[htbp]
	\centering
	\subfloat[]{
		\label{fig:2D_E-T}
		\includegraphics[width=0.5\linewidth]{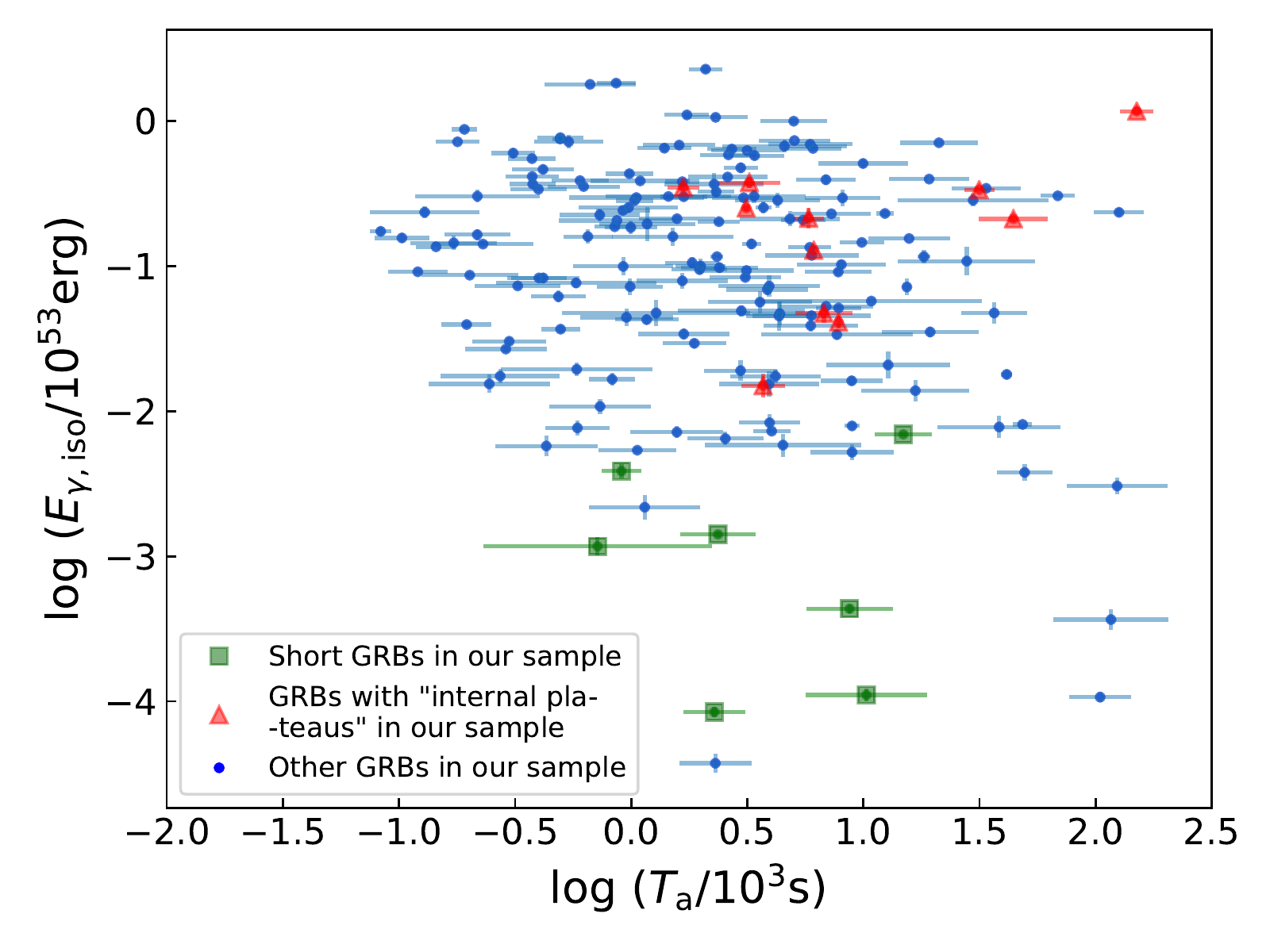}
	}
	\subfloat[]{
		\label{fig:2D_alpha1-alpha2}
		\includegraphics[width=0.5\linewidth]{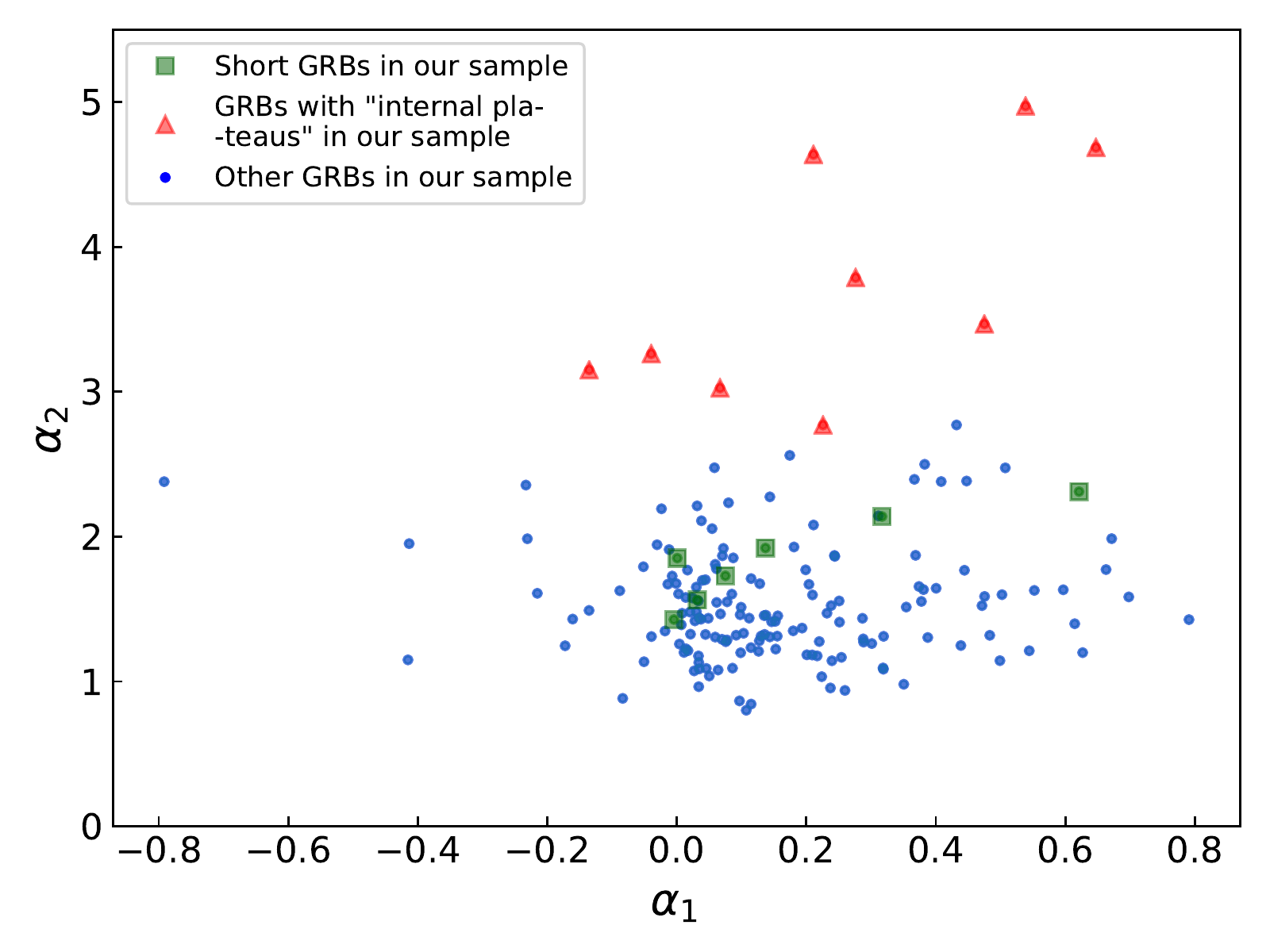}
	}
	
	\subfloat[]{
		\label{fig:2D_logL-alpha1}
		\includegraphics[width=0.5\linewidth]{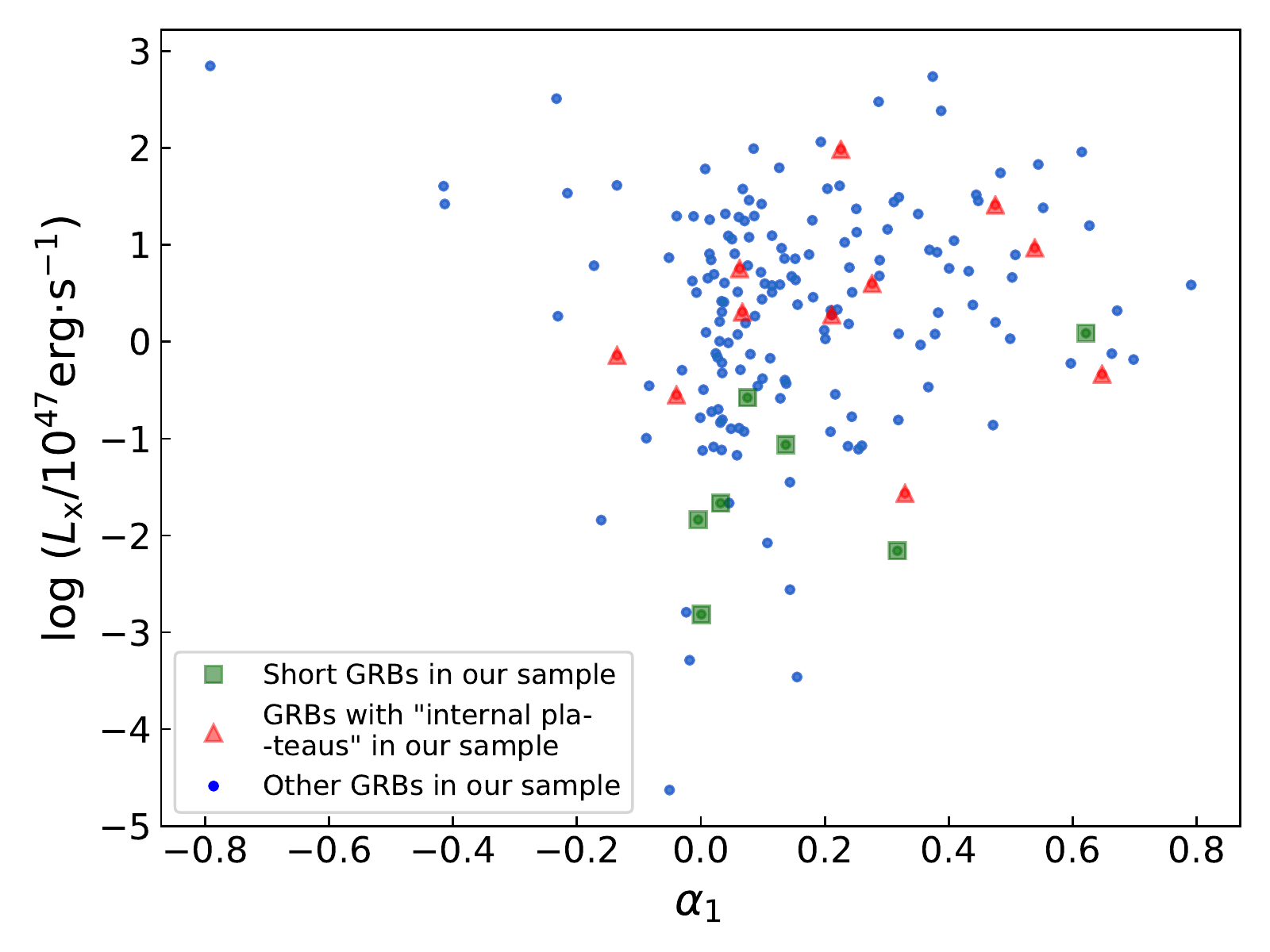}
	}
	\subfloat[]{
		\label{fig:2D_logT-alpha1}
		\includegraphics[width=0.5\linewidth]{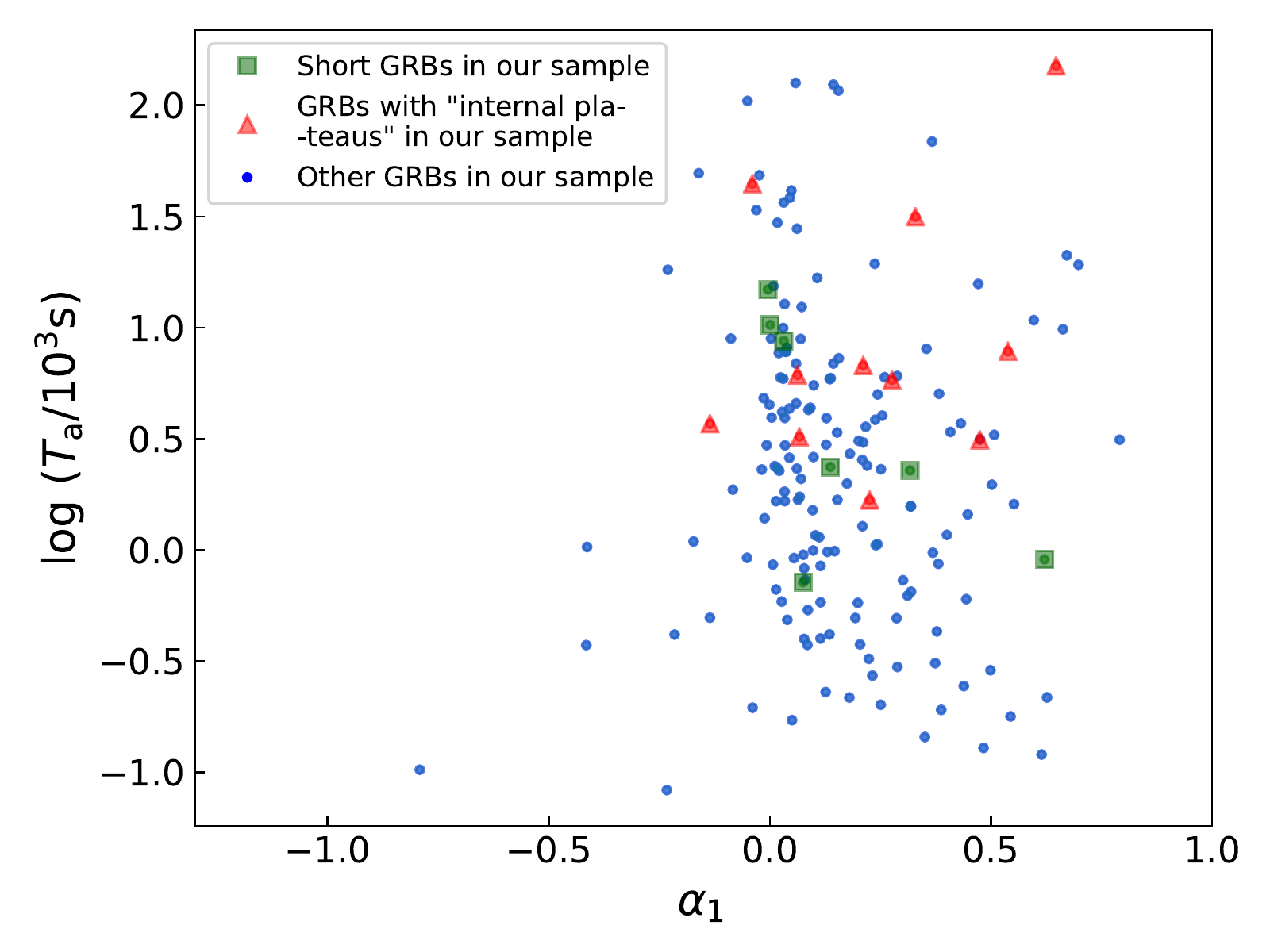}
	}
	
	\subfloat[]{
		\label{fig:2D_logE-alpha1}
		\includegraphics[width=0.5\linewidth]{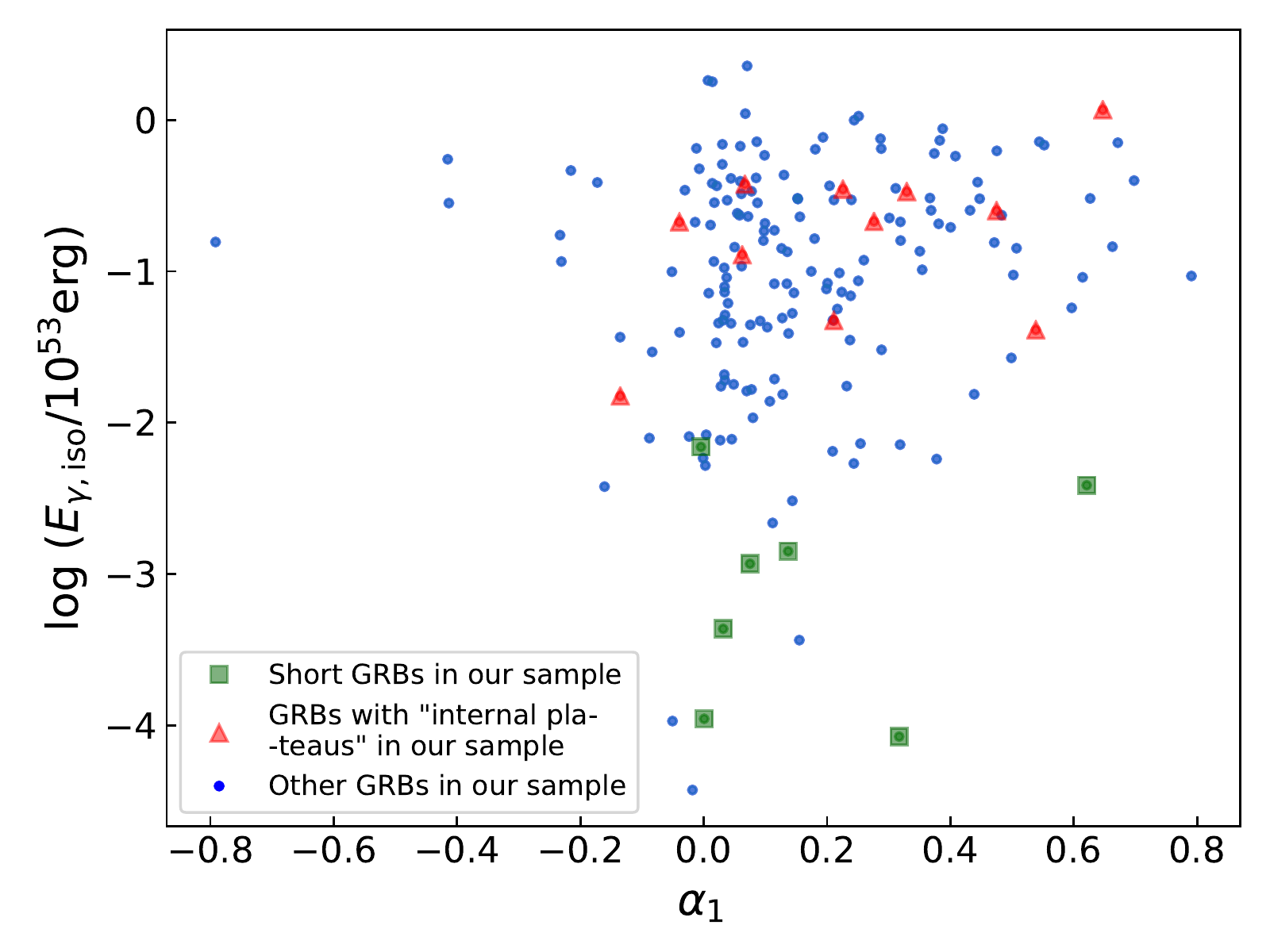}
	}
	\subfloat[]{
		\label{fig:2D_logL-alpha2}
		\includegraphics[width=0.5\linewidth]{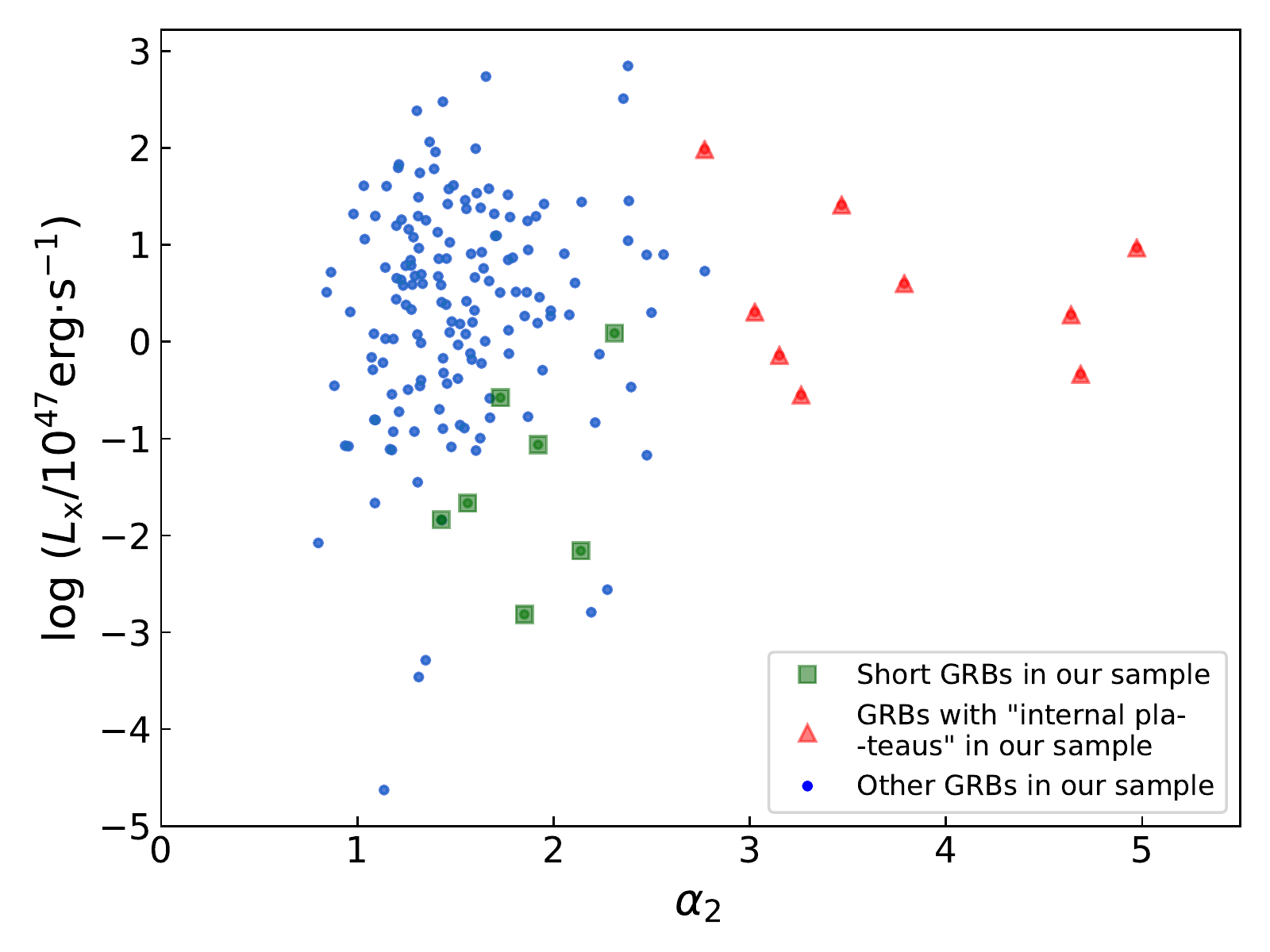}
	}
	\caption{Logarithmic plots between various parameter pairs. Here, $T_{a}$ is
the end time of the plateau, $L_{X}$ is the luminosity at the end time, $E_{\gamma,\rm{iso}}$ is
the isotropic energy of the prompt emission, $\alpha_{1}$ is the power-law timing index during the plateau phase
and $\alpha_{2}$ is the power-law timing index in the following decay phase, and finally $T_{90}$ is the
GRB duration. In order to clearly demonstrate the distributions, we do not show the error bars in 7 of
the plots (Panels \ref{fig:2D_alpha1-alpha2}, \ref{fig:2D_logL-alpha1}, \ref{fig:2D_logT-alpha1}, \ref{fig:2D_logE-alpha1}, \ref{fig:2D_logL-alpha2}, \ref{fig:2D_logT-alpha2}, and \ref{fig:2D_logE-alpha2}).
}
	\label{fig:2D_grp2}
\end{figure}

\begin{figure}[htbp]
	\centering
	\addtocounter{figure}{-1}
	\subfloat[]{
		\addtocounter{subfigure}{6}
		\label{fig:2D_logT-alpha2}
		\includegraphics[width=0.5\linewidth]{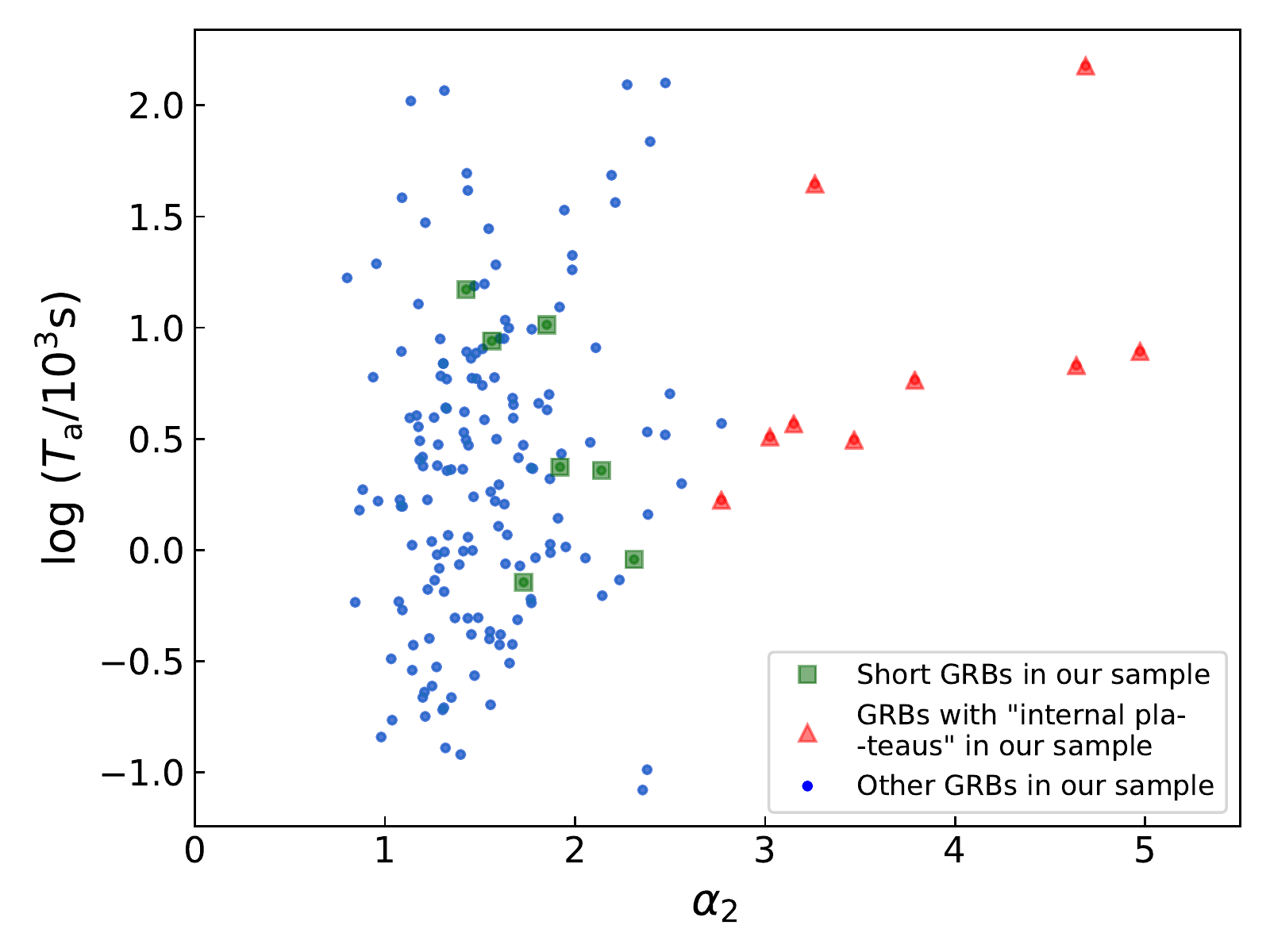}
	}
	\subfloat[]{
		\label{fig:2D_logE-alpha2}
		\includegraphics[width=0.5\linewidth]{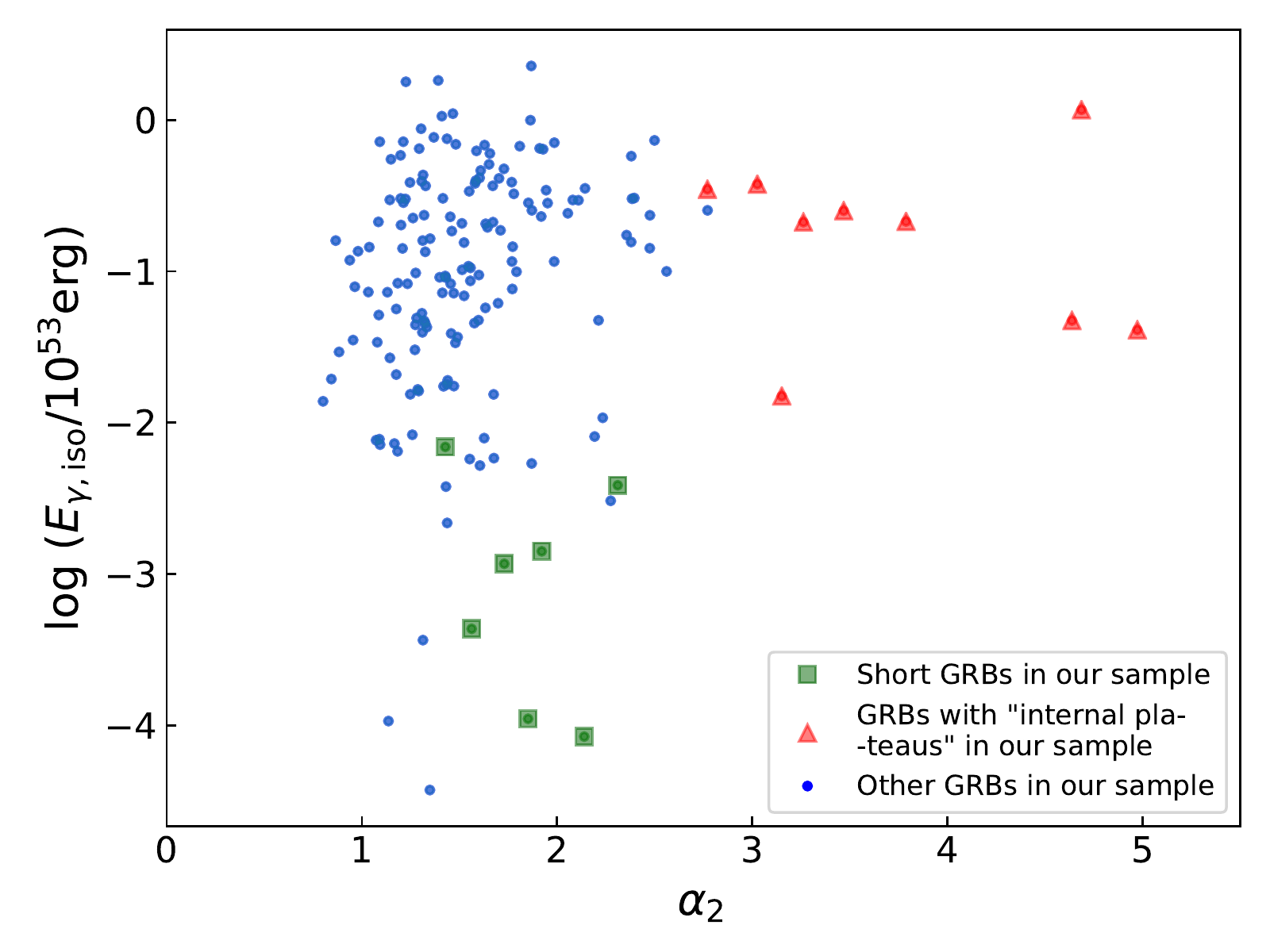}
	}
	
	\subfloat[]{
		\label{fig:2D_L-T90}
		\includegraphics[width=0.5\linewidth]{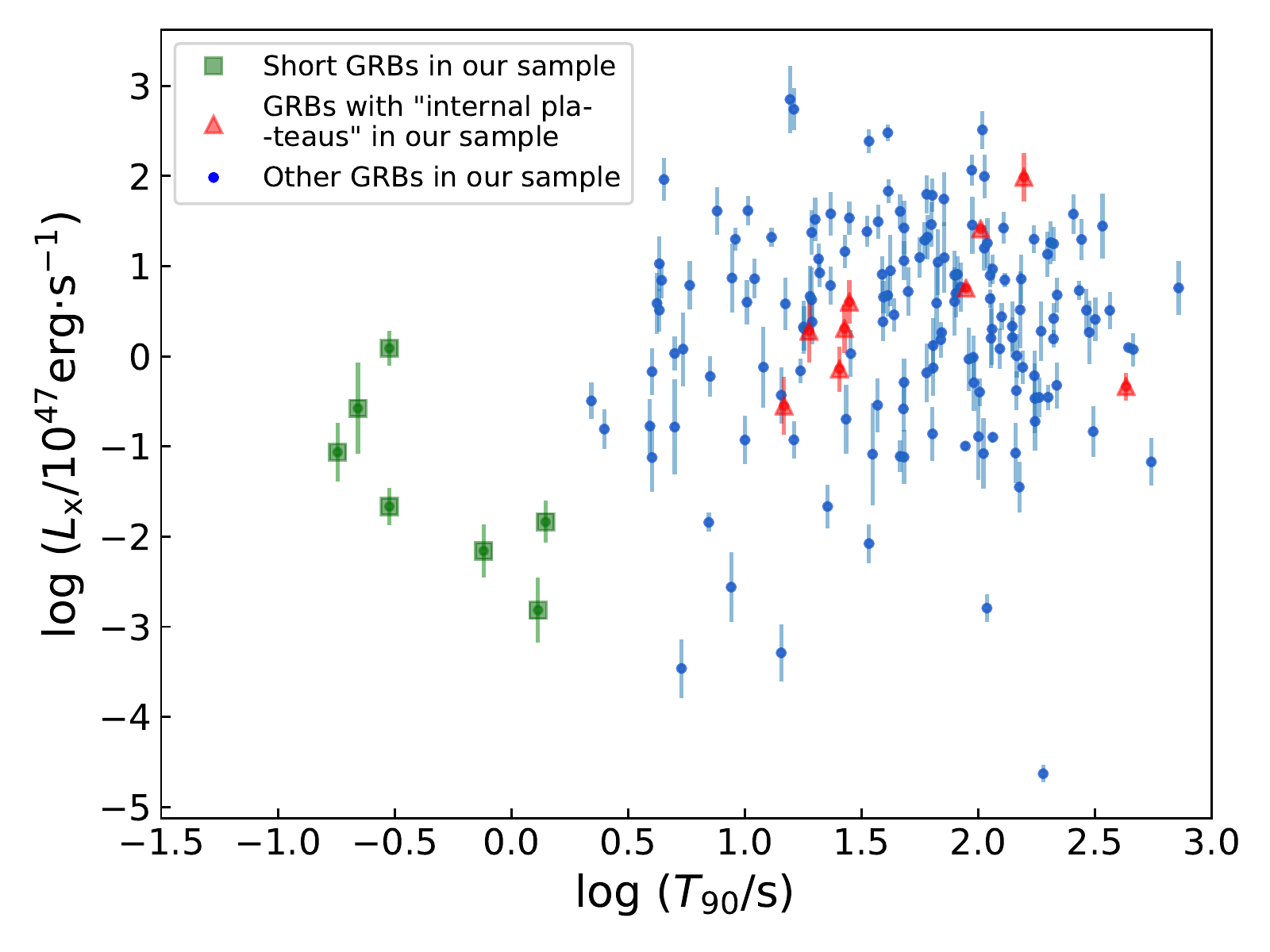}
	}
	\subfloat[]{
		\label{fig:2D_T-T90}
		\includegraphics[width=0.5\linewidth]{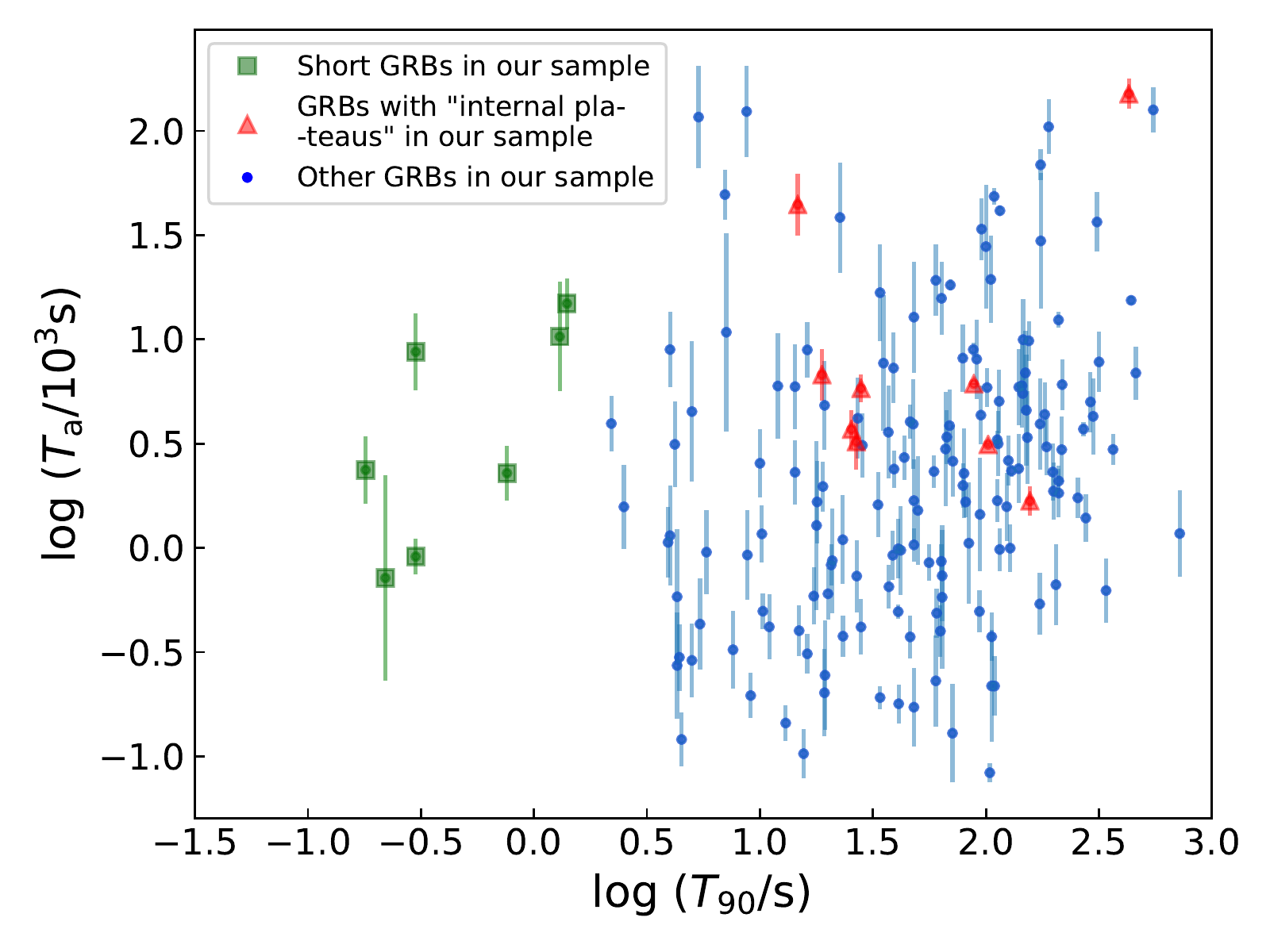}
	}
	\caption{--- Continued}
\end{figure}

\begin{figure}[htbp]
	\centering
	\includegraphics[width=0.8\linewidth]{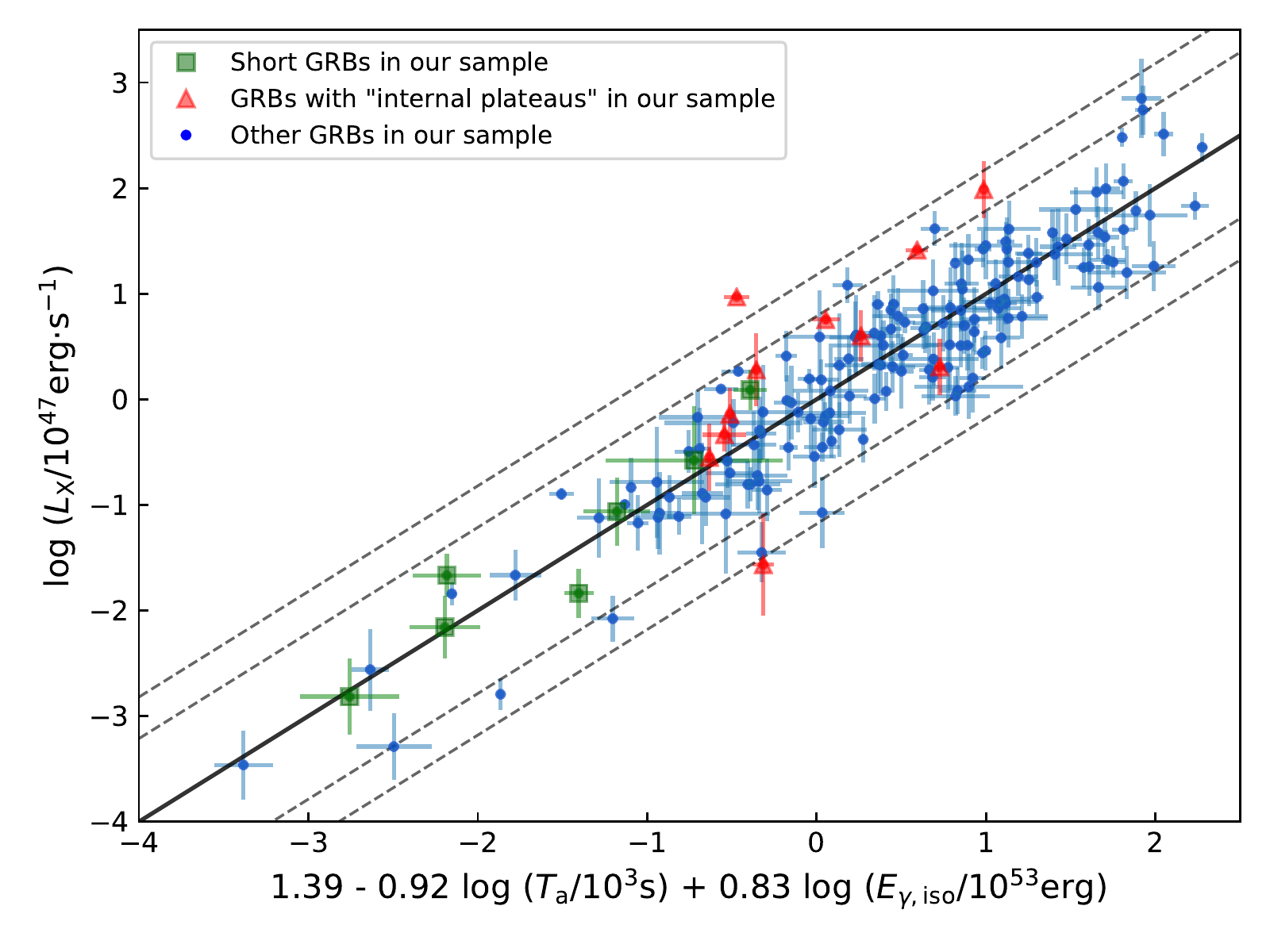}
	\caption{The best-fit L-T-E correlation by using all the 174 GRBs of our sample. Here, $L_{X}$ is the luminosity at the end time of the plateau, $T_{a}$ is the corresponding end time, and $E_{\gamma,\rm{iso}}$ is the isotropic $\gamma$-ray energy of the prompt emission. The black solid lines are the best fit for the observational data points. The black dashed lines are error lines of 2$\sigma$ and 3$\sigma$ confidence intervals.}
	\label{fig:result_fig}
\end{figure}

\begin{longrotatetable}
	\begin{deluxetable*}{lccccccccc}
		\tablecaption{Fitting Results and Some Key Parameters of the 174 GRBs in Our Sample \label{tab:parameter}}
		\tablewidth{700pt}
		\tabletypesize{\scriptsize}
		\tablehead{
			\colhead{GRB Name} & \colhead{T$_{90}^{a}$} &
			\colhead{$z^{a}$} & \colhead{$S$(15-150 keV)$^{a}$} & \colhead{$\alpha_{1}^{b}$} & \colhead{$\alpha_{2}^{b}$} & \colhead{$\log(T_{a}/10^{3})^{b}$} &
			\colhead{$\log(L_{X}/10^{47})^{c}$} & \colhead{$\log(E_{\rm \gamma,iso}'/10^{53})^{d}$} & \colhead{$\log(E_{\rm \gamma,iso}/10^{53})^{e}$} \\
			\colhead{} & \colhead{(s)} & \colhead{} & \colhead{($10^{-7}$ erg/cm$^{2}$)} & \colhead{} & \colhead{} &
			\colhead{(s)} & \colhead{(erg/s)} & \colhead{(erg)} & \colhead{(erg)}
		}
		\startdata
GRB 050315  & 95.6   & 1.949   & 32.2$\pm$1.46 & -0.03$\pm$0.05 & 1.94$\pm$0.25  & 1.53$\pm$0.15  & -0.29$\pm$0.32 & -0.52$\pm$0.02 & -0.46$\pm$0.02 \\
GRB 050319  & 152.5  & 3.24    & 13.1$\pm$1.48 & 0.15$\pm$0.11  & 1.41$\pm$0.17  & 0.53$\pm$0.22  & 0.86$\pm$0.27  & -0.53$\pm$0.05 & -0.52$\pm$0.05 \\
GRB 050401  & 33.3   & 2.9     & 82.2$\pm$3.06 & 0.55$\pm$0.03  & 1.63$\pm$0.28  & 0.21$\pm$0.16  & 1.38$\pm$0.17  & 0.19$\pm$0.02  & -0.17$\pm$0.02 \\
GRB 050416A & 2.5    & 0.6535  & 3.67$\pm$0.37 & 0.32$\pm$0.09  & 1.09$\pm$0.09  & 0.2$\pm$0.2    & -0.81$\pm$0.22 & -2.38$\pm$0.04 & -2.14$\pm$0.04 \\
GRB 050505  & 58.9   & 4.27    & 24.9$\pm$1.79 & 0.06$\pm$0.14  & 1.78$\pm$0.11  & 0.37$\pm$0.08  & 1.29$\pm$0.2   & -0.06$\pm$0.03 & -0.49$\pm$0.03 \\
GRB 050730  & 156.5  & 3.97    & 23.8$\pm$1.52 & 0.23$\pm$0.1   & 2.77$\pm$0.19  & 0.22$\pm$0.07  & 1.99$\pm$0.27  & -0.13$\pm$0.03 & -0.46$\pm$0.03 \\
GRB 050801  & 19.4   & 1.38    & 3.1$\pm$0.48  & 0.44$\pm$0.45  & 1.25$\pm$0.15  & -0.61$\pm$0.26 & 0.38$\pm$0.25  & -1.81$\pm$0.06 & -1.81$\pm$0.06 \\
GRB 050802  & 19     & 1.71    & 20$\pm$1.57   & 0.5$\pm$0.19   & 1.6$\pm$0.08   & 0.29$\pm$0.12  & 0.67$\pm$0.34  & -0.82$\pm$0.03 & -1.02$\pm$0.03 \\
GRB 050803  & 87.9   & 0.422   & 21.5$\pm$1.35 & -0.09$\pm$0.11 & 1.63$\pm$0.05  & 0.95$\pm$0.03  & -1$\pm$0.04    & -2.01$\pm$0.03 & -2.1$\pm$0.03  \\
GRB 050814  & 150.9  & 5.3     & 20.1$\pm$2.2  & 0.06$\pm$0.14  & 1.81$\pm$0.34  & 0.66$\pm$0.27  & 0.52$\pm$0.43  & -0.01$\pm$0.05 & -0.17$\pm$0.05 \\
GRB 050824  & 22.6   & 0.83    & 2.66$\pm$0.52 & 0.05$\pm$0.13  & 1.09$\pm$0.23  & 1.58$\pm$0.27  & -1.67$\pm$0.24 & -2.31$\pm$0.08 & -2.11$\pm$0.08 \\
GRB 050922C & 4.5    & 2.199   & 16.2$\pm$0.54 & 0.61$\pm$0.09  & 1.4$\pm$0.07   & -0.92$\pm$0.13 & 1.96$\pm$0.24  & -0.72$\pm$0.01 & -1.04$\pm$0.01 \\
GRB 051016B & 4      & 0.9364  & 1.7$\pm$0.22  & 0$\pm$0.09     & 1.6$\pm$0.18   & 0.95$\pm$0.18  & -1.12$\pm$0.38 & -2.4$\pm$0.05  & -2.28$\pm$0.05 \\
GRB 051109A & 37.2   & 2.346   & 22$\pm$2.72   & 0.32$\pm$0.08  & 1.31$\pm$0.05  & -0.19$\pm$0.11 & 1.49$\pm$0.19  & -0.54$\pm$0.05 & -0.8$\pm$0.05  \\
GRB 051109B & 14.3   & 0.08    & 2.56$\pm$0.41 & -0.02$\pm$0.15 & 1.35$\pm$0.16  & 0.36$\pm$0.15  & -3.29$\pm$0.32 & -4.43$\pm$0.06 & -4.43$\pm$0.06 \\
GRB 051221A & 1.4    & 0.547   & 11.5$\pm$0.35 & 0$\pm$0.22     & 1.43$\pm$0.11  & 1.17$\pm$0.12  & -1.84$\pm$0.23 & -2.04$\pm$0.01 & -2.16$\pm$0.01 \\
GRB 060108  & 14.3   & 2.03    & 3.69$\pm$0.37 & 0.14$\pm$0.13  & 1.46$\pm$0.22  & 0.77$\pm$0.2   & -0.43$\pm$0.31 & -1.42$\pm$0.04 & -1.41$\pm$0.04 \\
GRB 060115  & 139.6  & 3.53    & 17.1$\pm$1.5  & 0.22$\pm$0.18  & 1.28$\pm$0.17  & 0.38$\pm$0.16  & 0.33$\pm$0.28  & -0.35$\pm$0.04 & -1.01$\pm$0.04 \\
GRB 060116  & 105.9  & 4       & 24.1$\pm$2.61 & 0.63$\pm$0.24  & 1.2$\pm$0.17   & -0.66$\pm$0.27 & 1.2$\pm$0.25   & -0.12$\pm$0.04 & -0.52$\pm$0.04 \\
GRB 060202  & 198.9  & 0.78    & 21.3$\pm$1.65 & -0.08$\pm$0.29 & 0.88$\pm$0.03  & 0.27$\pm$0.14  & -0.45$\pm$0.14 & -1.46$\pm$0.03 & -1.53$\pm$0.03 \\
GRB 060206  & 7.6    & 4.05    & 8.31$\pm$0.42 & 0.22$\pm$0.17  & 1.03$\pm$0.07  & -0.49$\pm$0.19 & 1.61$\pm$0.27  & -0.57$\pm$0.02 & -1.14$\pm$0.02 \\
GRB 060210  & 255    & 3.91    & 76.6$\pm$4.09 & 0.07$\pm$0.16  & 1.47$\pm$0.08  & 0.24$\pm$0.1   & 1.58$\pm$0.22  & 0.37$\pm$0.02  & 0.04$\pm$0.02  \\
GRB 060502A & 28.4   & 1.51    & 23.1$\pm$1.02 & 0.2$\pm$0.11   & 1.18$\pm$0.08  & 0.49$\pm$0.15  & 0.03$\pm$0.26  & -0.86$\pm$0.02 & -1.08$\pm$0.02 \\
GRB 060522  & 71.1   & 5.11    & 11.4$\pm$1.11 & 0.48$\pm$0.35  & 1.32$\pm$0.17  & -0.89$\pm$0.24 & 1.74$\pm$0.3   & -0.28$\pm$0.04 & -0.63$\pm$0.04 \\
GRB 060526  & 298.2  & 3.21    & 12.6$\pm$1.65 & 0.09$\pm$0.17  & 1.85$\pm$0.28  & 0.63$\pm$0.18  & 0.27$\pm$0.35  & -0.55$\pm$0.05 & -0.55$\pm$0.05 \\
GRB 060604  & 95     & 2.1357  & 4.02$\pm$1.06 & 0.04$\pm$0.17  & 1.32$\pm$0.11  & 0.64$\pm$0.14  & -0.01$\pm$0.24 & -1.35$\pm$0.1  & -1.34$\pm$0.1  \\
GRB 060605  & 79.1   & 3.8     & 6.97$\pm$0.9  & 0.17$\pm$0.12  & 2.56$\pm$0.27  & 0.3$\pm$0.11   & 0.9$\pm$0.27   & -0.69$\pm$0.05 & -1$\pm$0.05    \\
GRB 060607A & 102.2  & 3.082   & 25.5$\pm$1.12 & 0.47$\pm$0.02  & 3.47$\pm$0.1   & 0.5$\pm$0.01   & 1.41$\pm$0.02  & -0.28$\pm$0.02 & -0.6$\pm$0.02  \\
GRB 060614  & 108.7  & 0.13    & 204$\pm$3.63  & -0.02$\pm$0.05 & 2.19$\pm$0.13  & 1.69$\pm$0.04  & -2.79$\pm$0.15 & -2.09$\pm$0.01 & -2.09$\pm$0.01 \\
GRB 060707  & 66.2   & 3.43    & 16$\pm$1.51   & 0.13$\pm$0.14  & 1.28$\pm$0.16  & 0.47$\pm$0.27  & 0.59$\pm$0.44  & -0.4$\pm$0.04  & -1.31$\pm$0.04 \\
GRB 060708  & 10.2   & 2.3     & 4.94$\pm$0.37 & 0.1$\pm$0.16   & 1.33$\pm$0.1   & 0.07$\pm$0.14  & 0.6$\pm$0.25   & -1.2$\pm$0.03  & -1.37$\pm$0.03 \\
GRB 060714  & 115    & 2.71    & 28.3$\pm$1.67 & 0.13$\pm$0.15  & 1.31$\pm$0.06  & -0.01$\pm$0.1  & 0.97$\pm$0.16  & -0.32$\pm$0.02 & -0.36$\pm$0.02 \\
GRB 060729  & 115.3  & 0.54    & 26.1$\pm$2.11 & 0.05$\pm$0.04  & 1.44$\pm$0.02  & 1.62$\pm$0.02  & -0.9$\pm$0.06  & -1.7$\pm$0.03  & -1.75$\pm$0.03 \\
GRB 060814  & 145.3  & 0.84    & 146$\pm$2.39  & 0.1$\pm$0.17   & 1.51$\pm$0.08  & 0.74$\pm$0.1   & -0.38$\pm$0.22 & -0.56$\pm$0.01 & -0.68$\pm$0.01 \\
GRB 060906  & 43.5   & 3.685   & 22.1$\pm$1.36 & 0.18$\pm$0.14  & 1.93$\pm$0.4   & 0.43$\pm$0.11  & 0.46$\pm$0.19  & -0.21$\pm$0.03 & -0.19$\pm$0.03 \\
GRB 060908  & 19.3   & 1.8836  & 28$\pm$1.11   & 0.25$\pm$0.19  & 1.56$\pm$0.18  & -0.7$\pm$0.21  & 1.37$\pm$0.25  & -0.6$\pm$0.02  & -1.06$\pm$0.02 \\
GRB 060912A & 5      & 0.937   & 13.5$\pm$0.62 & 0.5$\pm$0.2    & 1.14$\pm$0.06  & -0.54$\pm$0.18 & 0.03$\pm$0.19  & -1.5$\pm$0.02  & -1.57$\pm$0.02 \\
GRB 061021  & 46.2   & 0.3463  & 29.6$\pm$1.01 & 0.25$\pm$0.09  & 1.17$\pm$0.04  & 0.6$\pm$0.08   & -1.11$\pm$0.17 & -2.05$\pm$0.01 & -2.14$\pm$0.01 \\
GRB 061121  & 81.3   & 1.314   & 137$\pm$1.99  & 0.01$\pm$0.05  & 1.58$\pm$0.06  & 0.22$\pm$0.07  & 0.91$\pm$0.2   & -0.2$\pm$0.01  & -0.42$\pm$0.01 \\
GRB 061201  & 0.76   & 0.111   & 3.34$\pm$0.27 & 0.32$\pm$0.16  & 2.14$\pm$0.28  & 0.36$\pm$0.13  & -2.16$\pm$0.3  & -4.02$\pm$0.03 & -4.07$\pm$0.03 \\
GRB 061222A & 71.4   & 2.088   & 79.9$\pm$1.58 & 0.04$\pm$0.07  & 1.7$\pm$0.14   & 0.42$\pm$0.17  & 1.09$\pm$0.39  & -0.07$\pm$0.01 & -0.39$\pm$0.01 \\
GRB 070110  & 88.4   & 2.352   & 16.2$\pm$1.08 & 0.06$\pm$0.06  & 8.95$\pm$0.62  & 0.79$\pm$0.01  & 0.76$\pm$0.04  & -0.67$\pm$0.03 & -0.89$\pm$0.03 \\
GRB 070129  & 460.6  & 2.3384  & 29.8$\pm$2.67 & 0.06$\pm$0.11  & 1.31$\pm$0.11  & 0.84$\pm$0.13  & 0.08$\pm$0.18  & -0.41$\pm$0.04 & -0.4$\pm$0.04  \\
GRB 070208  & 47.7   & 1.165   & 4.45$\pm$1.01 & 0.13$\pm$0.14  & 1.67$\pm$0.42  & 0.59$\pm$0.21  & -0.58$\pm$0.25 & -1.79$\pm$0.09 & -1.81$\pm$0.09 \\
GRB 070306  & 209.5  & 1.497   & 53.8$\pm$2.86 & 0.07$\pm$0.05  & 1.92$\pm$0.1   & 1.09$\pm$0.04  & 0.19$\pm$0.1   & -0.5$\pm$0.02  & -0.64$\pm$0.02 \\
GRB 070506  & 4.3    & 2.31    & 2.1$\pm$0.23  & 0.11$\pm$0.33  & 0.84$\pm$0.36  & -0.23$\pm$0.33 & 0.51$\pm$0.23  & -1.57$\pm$0.05 & -1.71$\pm$0.05 \\
GRB 070508  & 20.9   & 0.82    & 196$\pm$2.73  & 0.38$\pm$0.07  & 1.63$\pm$0.27  & -0.06$\pm$0.25 & 0.93$\pm$0.16  & -0.45$\pm$0.01 & -0.69$\pm$0.01 \\
GRB 070529  & 109.2  & 2.4996  & 25.7$\pm$2.45 & 0.18$\pm$0.21  & 1.35$\pm$0.1   & -0.66$\pm$0.14 & 1.26$\pm$0.28  & -0.42$\pm$0.04 & -0.78$\pm$0.04 \\
GRB 070714B & 64     & 0.92    & 7.2$\pm$0.9   & 0.08$\pm$0.33  & 2.23$\pm$0.62  & -0.13$\pm$0.22 & -0.13$\pm$0.31 & -1.79$\pm$0.05 & -1.97$\pm$0.05 \\
GRB 070721B & 340    & 3.626   & 36$\pm$2      & 0.31$\pm$0.12  & 2.14$\pm$0.25  & -0.2$\pm$0.15  & 1.44$\pm$0.36  & -0.01$\pm$0.02 & -0.45$\pm$0.02 \\
GRB 070809  & 1.3    & 0.22    & 1$\pm$0.1     & 0$\pm$0.15     & 1.85$\pm$0.85  & 1.01$\pm$0.26  & -2.82$\pm$0.36 & -3.93$\pm$0.04 & -3.96$\pm$0.04 \\
GRB 070810A & 11     & 2.17    & 6.9$\pm$0.6   & 0.13$\pm$0.16  & 1.45$\pm$0.15  & -0.38$\pm$0.16 & 0.86$\pm$0.22  & -1.1$\pm$0.04  & -1.08$\pm$0.04 \\
GRB 071020  & 4.2    & 2.145   & 23$\pm$1      & 0.79$\pm$0.07  & 1.43$\pm$0.41  & 0.5$\pm$0.2    & 0.59$\pm$0.34  & -0.59$\pm$0.02 & -1.03$\pm$0.02 \\
GRB 080310  & 365    & 2.43    & 23$\pm$2      & -0.01$\pm$0.08 & 1.73$\pm$0.13  & 0.47$\pm$0.07  & 0.51$\pm$0.21  & -0.49$\pm$0.04 & -0.32$\pm$0.04 \\
GRB 080430  & 16.2   & 0.767   & 12$\pm$1      & 0.07$\pm$0.07  & 1.29$\pm$0.08  & 0.95$\pm$0.13  & -0.93$\pm$0.2  & -1.72$\pm$0.03 & -1.79$\pm$0.03 \\
GRB 080516  & 5.8    & 3.2     & 2.6$\pm$0.4   & 0.08$\pm$0.13  & 1.27$\pm$0.24  & -0.02$\pm$0.2  & 0.79$\pm$0.27  & -1.24$\pm$0.06 & -1.35$\pm$0.06 \\
GRB 080603B & 60     & 2.69    & 24$\pm$1      & 0.13$\pm$0.23  & 1.21$\pm$0.27  & -0.64$\pm$0.22 & 1.8$\pm$0.21   & -0.4$\pm$0.02  & -0.85$\pm$0.02 \\
GRB 080605  & 20     & 1.6398  & 133$\pm$2     & 0.44$\pm$0.08  & 1.77$\pm$0.26  & -0.22$\pm$0.12 & 1.52$\pm$0.24  & -0.04$\pm$0.01 & -0.41$\pm$0.01 \\
GRB 080707  & 27.1   & 1.23    & 5.2$\pm$0.6   & 0.03$\pm$0.13  & 1.42$\pm$0.21  & 0.62$\pm$0.19  & -0.7$\pm$0.38  & -1.68$\pm$0.05 & -1.76$\pm$0.05 \\
GRB 080721  & 16.2   & 2.602   & 120$\pm$10    & 0.37$\pm$0.05  & 1.66$\pm$0.05  & -0.51$\pm$0.09 & 2.74$\pm$0.23  & 0.27$\pm$0.03  & -0.22$\pm$0.03 \\
GRB 080810  & 106    & 3.35    & 46$\pm$2      & 0.08$\pm$0.15  & 1.6$\pm$0.11   & -0.43$\pm$0.12 & 2$\pm$0.24     & 0.04$\pm$0.02  & -0.38$\pm$0.02 \\
GRB 080905B & 128    & 2.374   & 18$\pm$2      & 0.1$\pm$0.15   & 1.46$\pm$0.07  & 0$\pm$0.11     & 1.42$\pm$0.18  & -0.62$\pm$0.05 & -0.73$\pm$0.05 \\
GRB 081007  & 10     & 0.5295  & 7.1$\pm$0.8   & 0.21$\pm$0.11  & 1.18$\pm$0.08  & 0.41$\pm$0.16  & -0.93$\pm$0.27 & -2.28$\pm$0.05 & -2.19$\pm$0.05 \\
GRB 081008  & 185.5  & 1.9685  & 43$\pm$2      & 0.21$\pm$0.14  & 2.08$\pm$0.21  & 0.48$\pm$0.14  & 0.28$\pm$0.33  & -0.38$\pm$0.02 & -0.53$\pm$0.02 \\
GRB 081029  & 270    & 3.8479  & 21$\pm$2      & 0.43$\pm$0.07  & 2.77$\pm$0.3   & 0.57$\pm$0.04  & 0.73$\pm$0.11  & -0.21$\pm$0.04 & -0.6$\pm$0.04  \\
GRB 081221  & 34     & 2.26    & 181$\pm$3     & 0.39$\pm$0.11  & 1.3$\pm$0.02   & -0.72$\pm$0.06 & 2.39$\pm$0.14  & 0.35$\pm$0.01  & -0.06$\pm$0.01 \\
GRB 090113  & 9.1    & 1.7493  & 7.6$\pm$0.4   & -0.04$\pm$0.16 & 1.31$\pm$0.06  & -0.71$\pm$0.11 & 1.3$\pm$0.13   & -1.23$\pm$0.02 & -1.4$\pm$0.02  \\
GRB 090205  & 8.8    & 4.7     & 1.9$\pm$0.3   & -0.05$\pm$0.15 & 1.79$\pm$0.32  & -0.03$\pm$0.22 & 0.87$\pm$0.38  & -1.12$\pm$0.06 & -1$\pm$0.06    \\
GRB 090313  & 79     & 3.375   & 14$\pm$2      & 0.04$\pm$0.42  & 2.11$\pm$0.27  & 0.91$\pm$0.16  & 0.61$\pm$0.38  & -0.47$\pm$0.06 & -0.53$\pm$0.06 \\
GRB 090407  & 310    & 1.4485  & 11$\pm$2      & 0.03$\pm$0.08  & 2.21$\pm$0.27  & 1.56$\pm$0.14  & -0.83$\pm$0.28 & -1.22$\pm$0.07 & -1.32$\pm$0.07 \\
GRB 090418A & 56     & 1.608   & 46$\pm$2      & 0.11$\pm$0.12  & 1.71$\pm$0.1   & -0.07$\pm$0.09 & 1.1$\pm$0.23   & -0.51$\pm$0.02 & -0.73$\pm$0.02 \\
GRB 090423  & 10.3   & 8       & 5.9$\pm$0.4   & -0.14$\pm$0.14 & 1.49$\pm$0.11  & -0.3$\pm$0.08  & 1.62$\pm$0.16  & -0.29$\pm$0.03 & -1.43$\pm$0.03 \\
GRB 090510  & 0.3    & 0.903   & 3.4$\pm$0.4   & 0.62$\pm$0.07  & 2.31$\pm$0.25  & -0.04$\pm$0.09 & 0.09$\pm$0.19  & -2.13$\pm$0.05 & -2.41$\pm$0.05 \\
GRB 090516  & 210    & 4.109   & 90$\pm$6      & 0.07$\pm$0.23  & 1.87$\pm$0.14  & 0.32$\pm$0.07  & 1.25$\pm$0.19  & 0.47$\pm$0.03  & 0.36$\pm$0.03  \\
GRB 090519  & 64     & 3.9     & 12$\pm$1      & 0.2$\pm$0.42   & 1.77$\pm$0.84  & -0.24$\pm$0.34 & 0.12$\pm$0.31  & -0.44$\pm$0.03 & -1.12$\pm$0.03 \\
GRB 090529  & 100    & 2.625   & 6.8$\pm$1.7   & 0.06$\pm$0.18  & 1.55$\pm$0.5   & 1.45$\pm$0.3   & -0.89$\pm$0.48 & -0.97$\pm$0.1  & -0.97$\pm$0.1  \\
GRB 090530  & 48     & 1.266   & 11$\pm$1      & 0.06$\pm$0.12  & 1.08$\pm$0.09  & 0.23$\pm$0.2   & -0.29$\pm$0.26 & -1.33$\pm$0.04 & -1.47$\pm$0.04 \\
GRB 090618  & 113.2  & 0.54    & 1050$\pm$10   & 0.48$\pm$0.06  & 1.59$\pm$0.06  & 0.5$\pm$0.16   & 0.2$\pm$0.33   & -0.1$\pm$0     & -0.2$\pm$0     \\
GRB 090927  & 2.2    & 1.37    & 2$\pm$0.3     & 0$\pm$0.19     & 1.26$\pm$0.11  & 0.6$\pm$0.13   & -0.49$\pm$0.2  & -2$\pm$0.06    & -2.08$\pm$0.06 \\
GRB 091018  & 4.4    & 0.971   & 14$\pm$1      & 0.29$\pm$0.1   & 1.27$\pm$0.06  & -0.53$\pm$0.16 & 0.84$\pm$0.2   & -1.45$\pm$0.03 & -1.52$\pm$0.03 \\
GRB 091029  & 39.2   & 2.752   & 24$\pm$1      & 0.01$\pm$0.09  & 1.2$\pm$0.05   & 0.38$\pm$0.09  & 0.66$\pm$0.18  & -0.38$\pm$0.02 & -0.69$\pm$0.02 \\
GRB 091109A & 48     & 3.5     & 16$\pm$2      & 0.05$\pm$0.35  & 1.04$\pm$0.07  & -0.76$\pm$0.19 & 1.06$\pm$0.21  & -0.39$\pm$0.05 & -0.84$\pm$0.05 \\
GRB 091127  & 7.1    & 0.49    & 90$\pm$3      & 0.6$\pm$0.19   & 1.63$\pm$0.41  & 1.03$\pm$0.48  & -0.22$\pm$0.23 & -1.25$\pm$0.01 & -1.24$\pm$0.01 \\
GRB 091208B & 14.9   & 1.0633  & 33$\pm$2      & 0.11$\pm$0.14  & 1.23$\pm$0.07  & -0.4$\pm$0.12  & 0.58$\pm$0.29  & -1$\pm$0.03    & -1.08$\pm$0.03 \\
GRB 100219A & 18.8   & 4.7     & 3.7$\pm$0.6   & 0.21$\pm$0.26  & 4.64$\pm$1.44  & 0.83$\pm$0.12  & 0.28$\pm$0.35  & -0.83$\pm$0.07 & -1.33$\pm$0.07 \\
GRB 100302A & 17.9   & 4.813   & 3.1$\pm$0.4   & 0.03$\pm$0.15  & 0.96$\pm$0.1   & 0.22$\pm$0.2   & 0.31$\pm$0.25  & -0.89$\pm$0.05 & -1.1$\pm$0.05  \\
GRB 100418A & 7      & 0.6235  & 3.4$\pm$0.5   & -0.16$\pm$0.11 & 1.43$\pm$0.15  & 1.69$\pm$0.12  & -1.84$\pm$0.11 & -2.46$\pm$0.06 & -2.42$\pm$0.06 \\
GRB 100424A & 104    & 2.465   & 15$\pm$1      & -0.23$\pm$0.25 & 2.36$\pm$0.13  & -1.08$\pm$0.05 & 2.51$\pm$0.21  & -0.67$\pm$0.03 & -0.76$\pm$0.03 \\
GRB 100425A & 37     & 1.755   & 4.7$\pm$0.9   & 0.22$\pm$0.13  & 1.18$\pm$0.16  & 0.55$\pm$0.22  & -0.54$\pm$0.3  & -1.43$\pm$0.08 & -1.25$\pm$0.08 \\
GRB 100513A & 84     & 4.772   & 14$\pm$1      & 0.24$\pm$0.24  & 1.14$\pm$0.2   & 0.02$\pm$0.29  & 0.77$\pm$0.28  & -0.24$\pm$0.03 & -0.53$\pm$0.03 \\
GRB 100615A & 39     & 1.398   & 50$\pm$1      & 0.16$\pm$0.07  & 1.45$\pm$0.23  & 0.86$\pm$0.17  & 0.38$\pm$0.22  & -0.59$\pm$0.01 & -0.64$\pm$0.01 \\
GRB 100621A & 63.6   & 0.542   & 210$\pm$0     & 0.47$\pm$0.08  & 1.52$\pm$0.11  & 1.2$\pm$0.18   & -0.86$\pm$0.3  & -0.79$\pm$0    & -0.81$\pm$0    \\
GRB 100704A & 197.5  & 3.6     & 60$\pm$2      & 0.25$\pm$0.09  & 1.41$\pm$0.09  & 0.36$\pm$0.14  & 1.13$\pm$0.25  & 0.2$\pm$0.01   & 0.03$\pm$0.01  \\
GRB 100814A & 174.5  & 1.44    & 90$\pm$2      & 0.37$\pm$0.06  & 2.4$\pm$0.27   & 1.84$\pm$0.07  & -0.47$\pm$0.39 & -0.31$\pm$0.01 & -0.52$\pm$0.01 \\
GRB 100901A & 439    & 1.408   & 21$\pm$3      & 0.01$\pm$0.04  & 1.47$\pm$0.04  & 1.19$\pm$0.02  & 0.1$\pm$0.02   & -0.96$\pm$0.06 & -1.14$\pm$0.06 \\
GRB 100902A & 428.8  & 4.5     & 32$\pm$2      & 0.65$\pm$0.03  & 4.69$\pm$1.59  & 2.18$\pm$0.07  & -0.34$\pm$0.16 & 0.08$\pm$0.03  & 0.07$\pm$0.03  \\
GRB 100906A & 114.4  & 1.727   & 120$\pm$0     & 0.38$\pm$0.13  & 2.5$\pm$0.32   & 0.7$\pm$0.15   & 0.3$\pm$0.39   & -0.04$\pm$0    & -0.13$\pm$0    \\
GRB 101219B & 34     & 0.5519  & 21$\pm$4      & 0.11$\pm$0.24  & 0.8$\pm$0.11   & 1.22$\pm$0.23  & -2.08$\pm$0.22 & -1.77$\pm$0.08 & -1.86$\pm$0.08 \\
GRB 110213A & 48     & 1.46    & 59$\pm$4      & -0.41$\pm$0.26 & 1.95$\pm$0.06  & 0.01$\pm$0.08  & 1.42$\pm$0.3   & -0.48$\pm$0.03 & -0.55$\pm$0.03 \\
GRB 110715A & 13     & 0.82    & 118$\pm$2     & 0.35$\pm$0.1   & 0.98$\pm$0.01  & -0.84$\pm$0.09 & 1.32$\pm$0.1   & -0.67$\pm$0.01 & -0.87$\pm$0.01 \\
GRB 110808A & 48     & 1.348   & 3.3$\pm$0.8   & 0.03$\pm$0.19  & 1.18$\pm$0.23  & 1.11$\pm$0.27  & -1.12$\pm$0.3  & -1.8$\pm$0.09  & -1.68$\pm$0.09 \\
GRB 111008A & 63.46  & 5       & 53$\pm$3      & 0.01$\pm$0.07  & 1.39$\pm$0.07  & -0.06$\pm$0.08 & 1.79$\pm$0.19  & 0.37$\pm$0.02  & 0.26$\pm$0.02  \\
GRB 111123A & 290    & 3.1516  & 73$\pm$3      & 0.24$\pm$0.22  & 1.86$\pm$0.3   & 0.7$\pm$0.14   & 0.51$\pm$0.24  & 0.2$\pm$0.02   & 0$\pm$0.02     \\
GRB 111209A & -      & 0.677   & 360$\pm$10    & 0.33$\pm$0.03  & 15.11$\pm$2.02 & 1.5$\pm$0.06   & -1.57$\pm$0.48 & -0.36$\pm$0.01 & -0.47$\pm$0.01 \\
GRB 111228A & 101.2  & 0.71627 & 85$\pm$2      & 0.14$\pm$0.11  & 1.32$\pm$0.06  & 0.77$\pm$0.09  & -0.4$\pm$0.16  & -0.93$\pm$0.01 & -0.87$\pm$0.01 \\
GRB 111229A & 25.4   & 1.3805  & 3.4$\pm$0.7   & -0.14$\pm$0.09 & 3.15$\pm$1.13  & 0.57$\pm$0.09  & -0.14$\pm$0.25 & -1.77$\pm$0.08 & -1.82$\pm$0.08 \\
GRB 120118B & 23.26  & 2.943   & 18$\pm$1      & -0.17$\pm$0.18 & 1.25$\pm$0.22  & 0.04$\pm$0.21  & 0.79$\pm$0.21  & -0.46$\pm$0.02 & -0.41$\pm$0.02 \\
GRB 120326A & 69.6   & 1.798   & 26$\pm$3      & -0.23$\pm$0.13 & 1.99$\pm$0.1   & 1.26$\pm$0.03  & 0.26$\pm$0.05  & -0.67$\pm$0.05 & -0.93$\pm$0.05 \\
GRB 120327A & 62.9   & 2.813   & 36$\pm$1      & 0.08$\pm$0.17  & 1.55$\pm$0.12  & -0.4$\pm$0.12  & 1.46$\pm$0.25  & -0.19$\pm$0.01 & -0.47$\pm$0.01 \\
GRB 120404A & 38.7   & 2.876   & 16$\pm$1      & 0.05$\pm$0.17  & 2.05$\pm$0.28  & -0.04$\pm$0.12 & 0.91$\pm$0.2   & -0.53$\pm$0.03 & -0.62$\pm$0.03 \\
GRB 120422A & 5.35   & 0.28    & 2.3$\pm$0.4   & 0.16$\pm$0.12  & 1.31$\pm$0.39  & 2.07$\pm$0.25  & -3.46$\pm$0.33 & -3.35$\pm$0.07 & -3.44$\pm$0.07 \\
GRB 120521C & 26.7   & 6       & 11$\pm$1      & 0.07$\pm$0.19  & 3.03$\pm$1.57  & 0.51$\pm$0.13  & 0.31$\pm$0.27  & -0.2$\pm$0.04  & -0.42$\pm$0.04 \\
GRB 120712A & 14.7   & 4.1745  & 18$\pm$1      & -0.04$\pm$0.41 & 3.26$\pm$1.67  & 1.65$\pm$0.15  & -0.55$\pm$0.32 & -0.22$\pm$0.02 & -0.67$\pm$0.02 \\
GRB 120802A & 50     & 3.796   & 19$\pm$3      & 0.1$\pm$0.11   & 0.87$\pm$0.37  & 0.18$\pm$0.26  & 0.72$\pm$0.27  & -0.26$\pm$0.06 & -0.8$\pm$0.06  \\
GRB 120811C & 26.8   & 2.671   & 30$\pm$3      & 0.3$\pm$0.13   & 1.26$\pm$0.17  & -0.14$\pm$0.17 & 1.16$\pm$0.18  & -0.31$\pm$0.04 & -0.65$\pm$0.04 \\
GRB 120922A & 173    & 3.1     & 62$\pm$7      & 0.09$\pm$0.24  & 1.09$\pm$0.06  & -0.27$\pm$0.15 & 1.3$\pm$0.15   & 0.11$\pm$0.05  & -0.14$\pm$0.05 \\
GRB 121024A & 69     & 2.298   & 11$\pm$1      & 0.24$\pm$0.29  & 1.52$\pm$0.28  & 0.59$\pm$0.18  & 0.18$\pm$0.2   & -0.86$\pm$0.04 & -1.16$\pm$0.04 \\
GRB 121128A & 23.3   & 2.2     & 69$\pm$4      & 0.2$\pm$0.19   & 1.67$\pm$0.12  & -0.42$\pm$0.1  & 1.58$\pm$0.25  & -0.09$\pm$0.02 & -0.43$\pm$0.02 \\
GRB 121211A & 182    & 1.023   & 13$\pm$2.67   & 0.09$\pm$0.22  & 1.32$\pm$0.15  & 0.64$\pm$0.16  & -0.46$\pm$0.22 & -1.44$\pm$0.08 & -1.33$\pm$0.08 \\
GRB 130131B & 4.3    & 2.539   & 3.4$\pm$0.4   & 0.23$\pm$0.33  & 1.47$\pm$0.41  & -0.56$\pm$0.26 & 1.03$\pm$0.3   & -1.29$\pm$0.05 & -1.76$\pm$0.05 \\
GRB 130408A & 28     & 3.758   & 23$\pm$4      & 0.28$\pm$0.32  & 3.79$\pm$0.74  & 0.76$\pm$0.07  & 0.6$\pm$0.24   & -0.18$\pm$0.07 & -0.67$\pm$0.07 \\
GRB 130420A & 123.5  & 1.297   & 71$\pm$3      & 0.32$\pm$0.13  & 1.08$\pm$0.07  & 0.2$\pm$0.16   & 0.08$\pm$0.22  & -0.5$\pm$0.02  & -0.67$\pm$0.02 \\
GRB 130511A & 5.43   & 1.3033  & 2.2$\pm$0.4   & 0.38$\pm$0.13  & 1.55$\pm$0.19  & -0.37$\pm$0.22 & 0.08$\pm$0.41  & -2$\pm$0.07    & -2.24$\pm$0.07 \\
GRB 130514A & 204    & 3.6     & 91$\pm$2      & 0.01$\pm$0.36  & 1.23$\pm$0.07  & -0.18$\pm$0.2  & 1.26$\pm$0.24  & 0.39$\pm$0.01  & 0.25$\pm$0.01  \\
GRB 130603B & 0.18   & 0.3564  & 6.3$\pm$0.3   & 0.14$\pm$0.11  & 1.92$\pm$0.23  & 0.37$\pm$0.16  & -1.06$\pm$0.33 & -2.69$\pm$0.02 & -2.85$\pm$0.02 \\
GRB 130606A & 276.58 & 5.913   & 29$\pm$2      & -0.01$\pm$0.19 & 1.91$\pm$0.25  & 0.14$\pm$0.12  & 1.3$\pm$0.23   & 0.22$\pm$0.03  & -0.19$\pm$0.03 \\
GRB 130612A & 4      & 2.006   & 2.3$\pm$0.5   & 0.11$\pm$0.27  & 1.44$\pm$0.34  & 0.06$\pm$0.24  & -0.17$\pm$0.26 & -1.64$\pm$0.09 & -2.66$\pm$0.09 \\
GRB 131030A & 41.1   & 1.293   & 290$\pm$0     & 0.54$\pm$0.12  & 1.21$\pm$0.03  & -0.75$\pm$0.09 & 1.83$\pm$0.13  & 0.11$\pm$0     & -0.14$\pm$0    \\
GRB 131103A & 17.3   & 0.599   & 8.2$\pm$1     & 0.03$\pm$0.19  & 1.07$\pm$0.04  & -0.23$\pm$0.14 & -0.16$\pm$0.14 & -2.11$\pm$0.05 & -2.12$\pm$0.05 \\
GRB 131105A & 112.3  & 1.686   & 71$\pm$5      & 0.15$\pm$0.13  & 1.22$\pm$0.06  & 0.23$\pm$0.1   & 0.64$\pm$0.1   & -0.29$\pm$0.03 & -0.52$\pm$0.03 \\
GRB 140114A & 139.7  & 3       & 32$\pm$1      & 0.03$\pm$0.13  & 1.48$\pm$0.35  & 0.77$\pm$0.18  & 0.21$\pm$0.22  & -0.2$\pm$0.01  & -0.16$\pm$0.01 \\
GRB 140206A & 93.6   & 2.73    & 160$\pm$3     & 0.19$\pm$0.08  & 1.37$\pm$0.05  & -0.3$\pm$0.1   & 2.06$\pm$0.17  & 0.43$\pm$0.01  & -0.11$\pm$0.01 \\
GRB 140213A & 60     & 1.2076  & 120$\pm$0     & 0.7$\pm$0.08   & 1.58$\pm$0.13  & 1.28$\pm$0.17  & -0.18$\pm$0.33 & -0.33$\pm$0    & -0.4$\pm$0     \\
GRB 140304A & 15.6   & 5.283   & 12$\pm$1      & -0.79$\pm$0.67 & 2.38$\pm$0.2   & -0.99$\pm$0.12 & 2.85$\pm$0.38  & -0.24$\pm$0.03 & -0.81$\pm$0.03 \\
GRB 140430A & 173.6  & 1.6     & 11$\pm$2      & 0.03$\pm$0.22  & 1.13$\pm$0.25  & 0.59$\pm$0.22  & -0.22$\pm$0.28 & -1.14$\pm$0.07 & -1.14$\pm$0.07 \\
GRB 140512A & 154.8  & 0.725   & 140$\pm$3     & 0.66$\pm$0.04  & 1.77$\pm$0.14  & 0.99$\pm$0.1   & -0.12$\pm$0.19 & -0.71$\pm$0.01 & -0.84$\pm$0.01 \\
GRB 140518A & 60.5   & 4.707   & 10$\pm$1      & 0.04$\pm$0.15  & 1.7$\pm$0.33   & -0.31$\pm$0.12 & 1.32$\pm$0.24  & -0.39$\pm$0.04 & -1.21$\pm$0.04 \\
GRB 140614A & 720    & 4.233   & 13$\pm$4      & 0.4$\pm$0.18   & 1.64$\pm$0.26  & 0.07$\pm$0.21  & 0.76$\pm$0.3   & -0.35$\pm$0.12 & -0.71$\pm$0.12 \\
GRB 140629A & 42     & 2.275   & 24$\pm$2      & 0.37$\pm$0.12  & 1.87$\pm$0.17  & -0.01$\pm$0.21 & 0.95$\pm$0.39  & -0.52$\pm$0.03 & -0.6$\pm$0.03  \\
GRB 140703A & 67.1   & 3.14    & 39$\pm$3      & 0.41$\pm$0.14  & 2.38$\pm$0.44  & 0.53$\pm$0.13  & 1.04$\pm$0.36  & -0.08$\pm$0.03 & -0.24$\pm$0.03 \\
GRB 140903A & 0.3    & 0.351   & 1.4$\pm$0.1   & 0.03$\pm$0.1   & 1.56$\pm$0.34  & 0.94$\pm$0.19  & -1.67$\pm$0.21 & -3.36$\pm$0.03 & -3.36$\pm$0.03 \\
GRB 141004A & 3.92   & 0.57    & 6.7$\pm$0.3   & 0.24$\pm$0.16  & 1.87$\pm$0.26  & 0.03$\pm$0.17  & -0.77$\pm$0.3  & -2.24$\pm$0.02 & -2.27$\pm$0.02 \\
GRB 141026A & 146    & 3.35    & 13$\pm$1      & 0.03$\pm$0.15  & 1.65$\pm$0.51  & 1$\pm$0.19     & 0.01$\pm$0.24  & -0.51$\pm$0.03 & -0.29$\pm$0.03 \\
GRB 141121A & 549.9  & 1.47    & 53$\pm$4      & 0.06$\pm$0.16  & 2.48$\pm$0.5   & 2.1$\pm$0.11   & -1.17$\pm$0.26 & -0.52$\pm$0.03 & -0.63$\pm$0.03 \\
GRB 150323A & 149.6  & 0.593   & 61$\pm$2      & 0.14$\pm$0.17  & 1.31$\pm$0.22  & 0.84$\pm$0.2   & -1.45$\pm$0.29 & -1.25$\pm$0.01 & -1.28$\pm$0.01 \\
GRB 150403A & 40.9   & 2.06    & 170$\pm$3     & 0.29$\pm$0.03  & 1.44$\pm$0.02  & -0.31$\pm$0.03 & 2.48$\pm$0.09  & 0.25$\pm$0.01  & -0.12$\pm$0.01 \\
GRB 150423A & 0.22   & 1.394   & 0.63$\pm$0.1  & 0.08$\pm$0.23  & 1.73$\pm$0.75  & -0.14$\pm$0.49 & -0.58$\pm$0.51 & -2.49$\pm$0.06 & -2.93$\pm$0.06 \\
GRB 150424A & 91     & 3       & 15$\pm$1      & 0.35$\pm$0.09  & 1.51$\pm$0.19  & 0.91$\pm$0.19  & -0.03$\pm$0.35 & -0.53$\pm$0.03 & -0.99$\pm$0.03 \\
GRB 150910A & 112.2  & 1.359   & 48$\pm$4      & 0.51$\pm$0.07  & 2.47$\pm$0.11  & 0.52$\pm$0.04  & 0.9$\pm$0.13   & -0.63$\pm$0.03 & -0.85$\pm$0.03 \\
GRB 151027A & 129.69 & 0.81    & 78$\pm$2      & 0.02$\pm$0.05  & 1.77$\pm$0.05  & 0.37$\pm$0.03  & 0.85$\pm$0.08  & -0.86$\pm$0.01 & -0.93$\pm$0.01 \\
GRB 151027B & 80     & 4.063   & 15$\pm$3      & 0.02$\pm$0.18  & 1.33$\pm$0.19  & 0.36$\pm$0.21  & 0.7$\pm$0.26   & -0.32$\pm$0.08 & -0.44$\pm$0.08 \\
GRB 151112A & 19.32  & 4.1     & 9.4$\pm$1.2   & -0.01$\pm$0.13 & 1.67$\pm$0.3   & 0.68$\pm$0.21  & 0.63$\pm$0.36  & -0.51$\pm$0.05 & -0.67$\pm$0.05 \\
GRB 151215A & 17.8   & 2.59    & 3.1$\pm$0.7   & 0.21$\pm$0.22  & 1.6$\pm$0.67   & 0.11$\pm$0.4   & 0.32$\pm$0.29  & -1.32$\pm$0.09 & -1.32$\pm$0.09 \\
GRB 160121A & 12     & 1.96    & 6.1$\pm$0.5   & 0.02$\pm$0.11  & 1.58$\pm$0.61  & 0.78$\pm$0.25  & -0.12$\pm$0.45 & -1.23$\pm$0.03 & -1.34$\pm$0.03 \\
GRB 160227A & 316.5  & 2.38    & 31$\pm$2      & 0.04$\pm$0.1   & 1.43$\pm$0.14  & 0.89$\pm$0.14  & 0.41$\pm$0.24  & -0.38$\pm$0.03 & -1.04$\pm$0.03 \\
GRB 160303A & 5      & 2.3     & 1.5$\pm$0.3   & 0$\pm$0.21     & 1.68$\pm$0.76  & 0.65$\pm$0.34  & -0.78$\pm$0.53 & -1.72$\pm$0.08 & -2.23$\pm$0.08 \\
GRB 160314A & 8.73   & 0.726   & 2.8$\pm$0.4   & 0.14$\pm$0.28  & 2.27$\pm$1.13  & 2.09$\pm$0.22  & -2.56$\pm$0.39 & -2.4$\pm$0.06  & -2.52$\pm$0.06 \\
GRB 160327A & 28     & 4.99    & 14$\pm$1      & -0.22$\pm$0.31 & 1.61$\pm$0.15  & -0.38$\pm$0.13 & 1.53$\pm$0.18  & -0.21$\pm$0.03 & -0.33$\pm$0.03 \\
GRB 160804A & 144.2  & 0.736   & 114$\pm$3     & 0.26$\pm$0.17  & 0.94$\pm$0.12  & 0.78$\pm$0.2   & -1.07$\pm$0.34 & -0.78$\pm$0.01 & -0.93$\pm$0.01 \\
GRB 161108A & 105.1  & 1.159   & 11$\pm$1      & 0.24$\pm$0.12  & 0.95$\pm$0.19  & 1.29$\pm$0.21  & -1.08$\pm$0.39 & -1.4$\pm$0.04  & -1.45$\pm$0.04 \\
GRB 161117A & 125.7  & 1.549   & 200$\pm$0     & 0.1$\pm$0.43   & 1.2$\pm$0.06   & 0.42$\pm$0.12  & 0.44$\pm$0.15  & 0.1$\pm$0      & -0.23$\pm$0    \\
GRB 170113A & 20.66  & 1.968   & 6.7$\pm$0.7   & 0.08$\pm$0.12  & 1.29$\pm$0.06  & -0.08$\pm$0.1  & 1.08$\pm$0.17  & -1.19$\pm$0.04 & -1.78$\pm$0.04 \\
GRB 170202A & 46.2   & 3.65    & 33$\pm$1      & -0.42$\pm$0.28 & 1.15$\pm$0.05  & -0.43$\pm$0.1  & 1.61$\pm$0.19  & -0.05$\pm$0.01 & -0.26$\pm$0.01 \\
GRB 170519A & 216.4  & 0.818   & 11$\pm$2      & 0.03$\pm$0.17  & 1.44$\pm$0.19  & 0.47$\pm$0.16  & -0.32$\pm$0.26 & -1.7$\pm$0.07  & -1.72$\pm$0.07 \\
GRB 170607A & -      & 0.557   & 75$\pm$3      & 0.03$\pm$0.1   & 1.09$\pm$0.07  & 0.89$\pm$0.1   & -0.8$\pm$0.16  & -1.21$\pm$0.02 & -1.29$\pm$0.02 \\
GRB 170705A & 217.3  & 2.01    & 95$\pm$3      & 0.29$\pm$0.07  & 1.29$\pm$0.09  & 0.78$\pm$0.12  & 0.68$\pm$0.2   & -0.02$\pm$0.01 & -0.19$\pm$0.01 \\
GRB 170714A & -      & 0.793   & 28$\pm$3      & 0.54$\pm$0.01  & 4.97$\pm$0.16  & 0.89$\pm$0.01  & 0.97$\pm$0.03  & -1.33$\pm$0.04 & -1.39$\pm$0.04 \\
GRB 171205A & 189.4  & 0.0368  & 36$\pm$3      & -0.05$\pm$0.09 & 1.14$\pm$0.15  & 2.02$\pm$0.13  & -4.63$\pm$0.1  & -3.96$\pm$0.03 & -3.97$\pm$0.03 \\
GRB 171222A & 174.8  & 2.409   & 19$\pm$2      & 0.02$\pm$0.13  & 1.21$\pm$0.42  & 1.47$\pm$0.33  & -0.72$\pm$0.33 & -0.58$\pm$0.04 & -0.55$\pm$0.04 \\
GRB 180115A & 40.9   & 2.487   & 7.6$\pm$1.1   & 0.15$\pm$0.14  & 1.41$\pm$0.13  & 0$\pm$0.14     & 0.68$\pm$0.24  & -0.96$\pm$0.06 & -1.14$\pm$0.06 \\
GRB 180325A & 94.1   & 2.25    & 65$\pm$2      & 0.45$\pm$0.11  & 2.38$\pm$0.69  & 0.16$\pm$0.27  & 1.46$\pm$0.32  & -0.1$\pm$0.01  & -0.52$\pm$0.01 \\
GRB 180329B & 210    & 1.998   & 33$\pm$3      & 0.03$\pm$0.13  & 1.56$\pm$0.16  & 0.26$\pm$0.12  & 0.42$\pm$0.17  & -0.49$\pm$0.04 & -0.98$\pm$0.04 \\
GRB 180404A & 35.2   & 1       & 13$\pm$1      & 0.02$\pm$0.13  & 1.48$\pm$0.38  & 0.89$\pm$0.33  & -1.09$\pm$0.57 & -1.46$\pm$0.03 & -1.47$\pm$0.03 \\
GRB 180720B & -      & 0.654   & 860$\pm$10    & 0.67$\pm$0.09  & 1.99$\pm$0.45  & 1.33$\pm$0.17  & 0.32$\pm$0.49  & -0.01$\pm$0.01 & -0.15$\pm$0.01
		\enddata
		\tablenotetext{a}{Taken from the \emph{Swift} GRB Table at \url{https://swift.gsfc.nasa.gov/archive/grb\_table.html/}.}
		\tablenotetext{b}{Fitting results by using Equation (\ref{eq:smo}), carried out through a Markov chain Monte Carlo (MCMC) algorithm}
		\tablenotetext{c}{Calculated from Equation (\ref{eq:lx}).}
		\tablenotetext{d}{Calculated from Equation (\ref{eq:ex}).}
		\tablenotetext{e}{Calculated from Equation (\ref{eq:corr_ex}).}
	\end{deluxetable*}
\end{longrotatetable}

\begin{table}[h!]
	\renewcommand{\thetable}{\arabic{table}}
	\centering
	\caption{The Bromberg criteria test on the ``internal plateau'' GRBs.
        Except for two GRBs with $T_{90}$ unavailable (GRBs 111209A and 170714A), all other events are
        explicitly identified as long GRBs. }
	\label{tab:fNC}
	\begin{tabular}{ccccc}
		\tablewidth{0pt}
		\hline
		GRB name & $T_{90}$(s) & PL & $f_{NC}$ & Type \\
		\hline
		GRB 050730  & 156.5 & 1.53 & 0.08 ($<50\%$) & Long \\
		GRB 060607A & 102.2 & 1.47 & 0.10 ($<50\%$) & Long \\	
		GRB 070110  & 88.4 & 1.58 & 0.10 ($<50\%$) & Long \\
		GRB 100219A & 18.8 & 1.34 & 0.20 ($<50\%$) & Long \\
		GRB 100902A & 428.8 & 1.98 & 0.06 ($<50\%$) & Long \\
		GRB 111209A & -- & 1.48 & -- & -- \\
		GRB 111229A & 25.4 & 1.85 & 0.17 ($<50\%$) & Long \\
		GRB 120521C & 26.7 & 1.73 & 0.17 ($<50\%$) & Long \\
		GRB 120712A & 14.7 & 1.36 & 0.23 ($<50\%$) & Long \\
		GRB 130408A & 28.0 & 1.28 & 0.17 ($<50\%$) & Long \\
		GRB 170714A & -- & 1.76 & -- & -- \\
		\hline
	\end{tabular}
\end{table}

\begin{table}[h!]
	\renewcommand{\thetable}{\arabic{table}}
	\centering
	\caption{Kendall's Tau test for a few pairs of parameters. The coefficients indicate that no
        significant correlations existed between these parameters. }
	\label{tab:kendalltau}
	\begin{tabular}{ccc}
		\hline
		& $\tau$ & $p$ \\
		\hline
		log$T_{a}$ - $z$ & -0.24 & 2.55e-6 \\
		log$E_{\gamma,{\rm iso}}$ - log$T_{a}$ & -0.11 & 0.038 \\
		$\alpha_{1}$ - $\alpha_{2}$ & 0.03 & 0.57 \\
		log$L_{X}$ - $\alpha_{1}$ & 0.11 & 0.024 \\
		log$T_{a}$ - $\alpha_{1}$ & -0.11 & 0.028 \\
		log$E_{\gamma,{\rm iso}}$ - $\alpha_{1}$ & 0.12 & 0.019 \\
		log$L_{X}$ - $\alpha_{2}$ & 0.04 & 0.41 \\
		log$T_{a}$ - $\alpha_{2}$ & 0.16 & 0.002 \\
		log$E_{\gamma,{\rm iso}}$ - $\alpha_{2}$ & 0.13 & 0.014 \\
		log$L_{X}$ - log$T_{90}$ & 0.06 & 0.22 \\
		log$T_{a}$ - log$T_{90}$ & 0.17 & 0.0008 \\
		\hline
	\end{tabular}
\end{table}

\end{document}